{{\PassOptionsToPackage
{spanish,english} {babel}
}}

\documentclass{aa}  
\usepackage{babel}
\usepackage[nottoc]{tocbibind} % add references to table of context
\usepackage{graphicx}
\usepackage{csvsimple}
\usepackage{txfonts}
\usepackage{lscape}
\usepackage{natbib}
\usepackage[utf8]{inputenc}
\usepackage[T1]{fontenc}
\bibpunct{(}{)}{;}{a}{}{,}
\usepackage[colorlinks, citecolor=blue]{hyperref}
\usepackage[toc,page]{appendix}
\usepackage{booktabs,caption}
\usepackage[flushleft]{threeparttable}

\newcommand{\hii}{\relax \ifmmode {\mbox H\,{\scshape ii}}\else
H\,{\scshape ii}\fi}
\newcommand{\mi}{\relax \ifmmode {\mu{\mbox m}}\else $\mu$m\fi}
\newcommand{\ha}{\relax \ifmmode {\mbox H}\alpha\else H$\alpha$\fi}
\newcommand{\hb}{\relax \ifmmode {\mbox H}\beta\else H$\beta$\fi}

\newcommand{\sii}{\relax \ifmmode {\mbox S\,{\scshape ii}}\else
S\,{\scshape ii}\fi}
\newcommand{\siii}{\relax \ifmmode {\mbox S\,{\scshape iii}}\else
S\,{\scshape iii}\fi}
\newcommand{\nii}{\relax \ifmmode {\mbox N\,{\scshape ii}}\else
N\,{\scshape ii}\fi}
\newcommand{\oi}{\relax \ifmmode {\mbox O\,{\scshape i}}\else
O\,{\scshape i}\fi}
\newcommand{\oii}{\relax \ifmmode {\mbox O\,{\scshape ii}}\else
O\,{\scshape ii}\fi}
\newcommand{\oiii}{\relax \ifmmode {\mbox O\,{\scshape iii}}\else
O\,{\scshape iii}\fi}
\newcommand{\neiii}{\relax \ifmmode {\mbox Ne\,{\scshape iii}}\else
Ne\,{\scshape iii}\fi}

\newcommand{\rdostres}{\relax \ifmmode {\,\mbox{R}}_{\rm 23}\else
\,\mbox{R}$_{\rm 23}$\fi}

\begin{document}

\selectlanguage{english}

\title{A MUSE/VLT spatially resolved study of the emission structure of Green Pea galaxies}

\selectlanguage{spanish}

\author{A. Arroyo-Polonio\inst{1},
        J. Iglesias-P\'{a}ramo\inst{1},\inst{2},
        C. Kehrig\inst{1},
        J.M. V\'{i}lchez\inst{1},
        R. Amor\'{i}n\inst{3},\inst{4},
        I. Breda\inst{1},
        E. P\'{e}rez-Montero\inst{1},
        B. P\'{e}rez-D\'{i}az\inst{1}
        \and
        M. Hayes\inst{5}
        }

\institute{Instituto de Astrof\'{i}sica de Andaluc\'{i}a, CSIC, Apartado de correos 3004, 18080 Granada, Spain 
           %\email{aarroyo@iaa.es}
           \and
           Centro Astron\'{o}mico Hispano en Andaluc\'{i}a, Observatorio de Calar Alto, Sierra de los Filabres, 04550 G\'{e}rgal, Spain
           \and
           Instituto de Investigaci\'{o}n Multidisciplinar en Ciencia y Tecnolog\'{i}a, Universidad de La Serena, Ra\'{u}l Bitr\'{a}n 1305, La Serena, Chile
           \and
           Departamento de Astronom\'{i}a, Universidad de La Serena, Av. Juan Cisternas 1200 Norte, La Serena, Chile
           \and
           Department of Astronomy, Stockholm University, AlbaNova University Centre, 106 91 Stockholm, Sweden
           }

\selectlanguage{english}

\date{Received 20 February 2023 / Accepted: 08 June 2023}

\abstract
{Green Pea galaxies (GPs) present among the most intense starbursts known in the nearby Universe. These galaxies are regarded as local analogs of high-redshift galaxies, making them a  benchmark in the understanding of the star formation processes and the galactic evolution in the early Universe.
In this work, we performed an integral field spectroscopic (IFS) study for a set of 24 GPs to investigate the interplay between its ionized interstellar medium (ISM) and the massive star formation that these galaxies present.
Observations were taken in the optical spectral range ($\lambda 4750 \AA - \lambda 9350 \AA$) with the MUSE spectrograph attached to the $8.2 \ \mathrm{m}$ telescope VLT.
Spatial extension criteria were employed to verify which GPs are spatially resolved in the MUSE data cubes.
We created and analyzed maps of spatially distributed emission lines (at different stages of excitation), continuum emission, and properties of the ionized ISM (e.g., ionization structure indicators, physical-chemical conditions, dust extinction).
We also took advantage of our IFS data to produce integrated spectra of selected galactic regions in order to study their physical-chemical conditions.
Maps of relevant emission lines and emission line ratios show that higher-excitation gas is preferentially located in the center of the galaxy, where the starburst is present. 
The continuum maps, with an average angular extent of 4'', exhibit more complex structures than the emission line maps. However, the $[O\textsc{iii}]{\lambda}5007\AA$ emission line maps tend to extend beyond the continuum images (the average angular extent is 5.5''), indicating the presence of low surface brightness ionized gas in the outer parts of the galaxies.
$H \alpha / H \beta$, $[S\textsc{ii}] / H \alpha,$ and $[O\textsc{i}] / H \alpha$ maps trace low-extinction, optically thin regions. The line ratios $[O\textsc{iii}]/H\beta$ and $[N\textsc{ii}]/H\alpha$ span extensive ranges, with values varying from 0.5 dex to 0.9 dex and from -1.7 dex to -0.8 dex, respectively. 
Regarding the integrated spectra, the line ratios were fit to derive physical properties including the electron densities $n_e = 30-530 \ \mathrm{cm^{-3}}$, and, in six GPs with a measurable $[O\textsc{iii}]{\lambda}4363\AA$ line, electron temperatures of
$T_e = 11500\mathrm{K}-15500 \ \mathrm{K}$, so the direct method was applied in these objects to retrieve metallicities $12+\log (O/H)\simeq 8$.
We found the presence of the high-ionizing nebular $He\textsc{ii} \lambda 4686 \AA$ line in three
GPs, where two of them present among the highest sSFR values (> $8\times 10^8 yr^{-1}$) in this sample. Non-Wolf-Rayet (WR) features are detected in these galaxy spectra.

%Maps of relevant emission lines and emission line ratios show that higher-excitation gas is preferentially located in the center of the galaxy where the starburst is present.
%Maps of the continuum have more complex structures than emission line maps. Nonetheless, the $[O\textsc{iii}]{\lambda}5007\AA$ emission line map tend to present more extension that continuum images, indicating the presence of low surface brightness ionized gas in the outer part of the galaxies.
%$H \alpha / H \beta$, $[S\textsc{ii}] / H \alpha$ and $[O\textsc{i}] / H \alpha$ maps trace low extinction, optically thin regions, indicating channels where Ly-continuum photons can escape.

}


\titlerunning{Green Pea galaxies seen with MUSE/VLT}
\authorrunning{A. Arroyo-Polonio et al.}
\maketitle
\section{Introduction}

The cosmic dawn (6 $\lesssim$ z $\lesssim$ 10) marks a major phase transition of the Universe, during which the “first light” [metal-free stars (or the so-called PopIII-stars) and the subsequent formation of numerous low-mass, extremely metal-poor galaxies] appeared, putting an end to the dark ages. The details of the reionization history reflects the nature of these first sources, which is just starting to be constrained with the James Webb Space Telescope (JWST) \citep[e.g.,][]{curtis2022spectroscopy,robertson2022discovery,naidu2022two}. However, exactly when and how the Universe was reionized remains one of the most important questions of modern astrophysics, and that will be explored during the next decades.
%Still it is subject of considerable observational and theoretical efforts (e.g., \cite{bromm2013formation}, \cite{fialkov2014observable} $\&$ \cite{visbal2017maximum}).

%In recent years, Hubble Space Telescope (HST) deep field surveys (e.g.,, HUDF, CANDELS-deep) have improved the statistics of galaxies at z > 6. Now, JWST is studying the sources identified at this epoch. At z > 6, however, observing the UV photons (10 nm < $\lambda$ < 130 nm) gets progressively more difficult due to the increasing opacity of the intergalactic medium (IGM) \citep{dijkstra2011detectability}.
%Even JWST can study in detail (i.e., spectroscopically) only the brightest sources, which corresponds to high-mass galaxies (M > $10^{10} M_\odot$, \cite{windhorst2006jwst}), rather than the more common lower-mass galaxies that are expected to reionize the Universe \citep{bouwens2015reionization}, for which only the integrated properties can be derived.
An immediately accessible approach to better understand the first sources is to identify galaxies at lower redshifts with properties similar to galaxies in the very early Universe \citep[e.g.,][]{schaerer2022first,chen2023jwst}. Among these local analogs, we find the objects known as GPs, which are a subset of galaxies inside the extreme emission line galaxies (EELGs) set \citep[e.g.,][]{perez2021extreme,breda2022characterisation,iglesias2022minijpas}.
GPs are compact starbursts commonly found at redshift z $\in (0.112, 0.360)$, corresponding to $2.5 - 4.3 \ \mathrm{Gyr}$ ago. The upper size limit of these galaxies is $5 \ \mathrm{kpc}$ in Hubble Space Telescope (HST) images \citep{yang2017lyalpha}. The JWST has already shown the similarity between primeval galaxies ($z \sim 8$) and GPs \citep{JWST_Peas}. The three high-z galaxies presented in \cite{JWST_Peas} are all strong line emitters, with spectra reminiscent of nearby GPs. They also show compact morphologies typical in GPs. So, without any doubt, GPs are excellent local analogs of high-redshift galaxies. % {\bf CK delete: these galaxies are the local analogs to primeval galaxies.}
 
%The study of galactic evolution is really challenging. It is essential for this task to observe the very first galaxies that once populate the Universe, to see how it starts. But to observe these galaxies is really difficult due to it's distance and hence faintness, nowadays only around 20 galaxies have been observed at z$>$7 DE DONDE SALE ESTO?. Here, one option to keep on with the study is to take closer galaxies that somehow look similar to those more distant, these galaxies are the analogs to primeval galaxies.

 %GPs are commonly found at redshift z $\in$ (0.112, 0.360), that correspond to 2.5 - 4.3 Gyr past from now. 
 On average, a GP galaxy has a stellar mass of  $\mathrm{M}_\star = 1 \times 10^9 \ \mathrm{M_\odot}$ and a star formation rate (SFR) of $10 \ \mathrm{M_\odot/yr}$ \citep{cardamone2009galaxy,izotov2011green},  and thus its mass-doubling timescale easily goes below $100 \ \mathrm{Myr}$. In this way, GPs present a very strong starburst similar to in high-redshift galaxies \citep{lofthouse2017local}. %Furthermore, GPs are purely star-ionizing sources with an extreme level of ionization as BPT diagrams shows \citep{erb2016high}. These properties gives GPs the title of local analogs of primeval galaxies
 Furthermore, different studies suggest hot massive stars as the main excitation source in GPs, resulting in a highly 
ionized ISM \citep[e.g.,][]{jaskot2013origin}.
%{\bf CK: Furthermore, different studies suggest hot massive stars as the main excitation source in GPs producing a high level gas ionization (REF). YO NO HABLARIA DE BPTs AQUI.}

The spectra of GPs  present optical emission lines with high equivalent width (EW) (up to $2000\ \mathrm{\r{A}}$), where $[O\textsc{iii}]{\lambda}5007\AA$ is the most intense line. 
If fact, the high intensity of this line brought to the discovery of these galaxies in Galaxy Zoo \citep{lintott2008galaxy}. The gas metallicity of GPs is low in comparison to regular star-forming galaxies ($12+\log(O/H)=7.6 \ to \ 8.4$) \citep[e.g.,][]{amorin2010oxygen}, but it rarely goes below $12+\log(O/H)=7.6$. 

%In order to have extremely low metallicities a necessary (but not sufficient) condition for that is that the $H\alpha$ line has to be more intense that the $[O\textsc{iii}]{\lambda}5007\AA$ line \citep{kojima2020extremely}. So the most remarkable spectroscopic feature of GPs, wich is the high intensity in the $[O\textsc{iii}]{\lambda}5007\AA$ makes these galaxies unable to reach metallicities of primeval galaxies. This one is the main difference between GPs and high redshift galaxies. In this way, GPs are more suitable to be the local analogs of galaxies in the cosmic noon (1 $\lesssim$ z $\lesssim$ 3), that correspond to 2.1 - 5.9 Gyr after the big bang. Time where the Universe presented galaxies with the most potents star-burst and metallicities similar to GPs \citep{wang2020census}. Nevertheless, GPs constitute one of the best laboratories in the low-redshift (or Local) Universe to analyze its first epochs.

Compared to normal galaxies, GPs are unusual (i.e.,  on average,
there are 2 GPs per square degree brighter than 20.5 mag) and tend to be isolated (i.e., they reside in regions $2/3$ less dense than regular galaxies \citep{cardamone2009galaxy}).
%GPs are unusual objects considering their scardity, in average there are 2 GPs per square degree brighter than 20.5 mag \citep{cardamone2009galaxy}. 
%%This fact in addition to the low depletion time leads to the conclusion that these galaxies are experimenting a rare episode of his life and somehow they have such a big yet futile star-burst. 
%They tend to be isolated galaxies, residing in regions two-thirds less dense than normal galaxies \citep{cardamone2009galaxy}. 
However, the possible existence of interactions in GPs
with other close neighbors is discussed in \cite{laufman2022triggering}, which concluded that interactions are unrelated to the strong star formation bursts present in these galaxies.
As a consequence of their very high SFR, almost all GPs
are Ly$\alpha$ emitters, where a significant fraction (between 2$\%$ and 70$\%$) of their Ly$\alpha$ photons escape into the IGM \citep{yang2016green,yang2017lyalpha,jaskot2017kinematics,yang2017lyalphaa,henry2018lyalpha,mckinney2019neutral,kim2020importance,hayes2023spectral}, with at least $\sim 10 \%$ of them being Lyman continuum (LyC) leakers \citep{yang2017lyalpha}. These facts reinforce GPs as local analogs to high-redshift galaxies, making GPs essential to understanding the process of reionization in the early Universe.

The spatially resolved analysis from IFS, in comparison to integrated single-aperture and long-slit spectroscopy, can largely improve 
our understanding of the warm ISM conditions in different systems as demonstrated in the literature
%The importance of integral field spectroscopy (IFS), in comparison to single-aperture/long-slit spectroscopy, for improving our understanding of the warm ISM conditions in different systems has been demonstrated in the literature 
\citep[e.g.,][]{kehrig2008interplay,cairos2009mapping,james2010vlt,monreal20112d,papaderos2013nebular,kehrig2015extended,bae2017limited,herenz2017vlt}.
However, not many spatially resolved properties of GPs have been studied so far (see \cite{lofthouse2017local}). 
In this work, we present the spatially resolved properties (e.g., excitation, extinction, ionization) of a sample of 24 GPs observed with the MUSE integral field unit (IFU) spectrograph. A detailed bidimensional spectroscopic study of our set of GPs is relevant to shed light on the ISM properties from galaxies in the intermediate/high-z Universe. 
%In this way, we investigate the spatially resolved properties of the ionized gas in the ISM: excitation, ionization source, extinction, physical conditions and chemical abundances.
%Moreover, we use the IFU data to create continuum maps, showing that GPs have a richer structure in the continuum rather than in the emission lines. Finally, taking advantage of the IFU data, we derive the integrated spectrum  for the whole galaxy in each GP, to study physical-chemical characteristics using their prominent emission lines (e.g.,, SFR, metallicity, electron temperature and density). In the integrated spectra we can see that for 3 GPs the emission line $He\textsc{ii} \lambda 4686 \AA$. This line is an indicator of a really extreme ionization \citep{Carol_HeII}.

This paper is organized as follows. In Section \ref{seccion_observaciones}, we report observations and data reduction. Flux measurements and a spatially resolved structure are presented in Section \ref{seccion_resuelta}. In Section \ref{seccion_espectros_integrados}, we present the integrated properties from each GP as a whole. Finally, Section \ref{seccion_conclusiones} summarizes the main conclusions derived from this work. Throughout this paper, we use physical distances and assume flat $\Lambda CDM$ cosmology with $H_0 = 70 \ \mathrm{km \ s^{-1} \ Mpc^{-1}}$, $\Omega_m = 0.3$ and $\Omega_\Lambda = 0.7$.

\section{Observations}
\label{seccion_observaciones}

\subsection{MUSE data}

We studied a sample of GPs observed with MUSE \citep{bacon2010muse} at the Very Large Telescope (VLT; ESO Paranal Observatory, Chile). MUSE is a panoramic IFS, which, operating in its wide field mode (WFM), provides a field of view (FoV) of $1' \times 1'$ with a spatial sampling of $0.2''$ and a FWHM spatial resolution of $0.3'' - 0.4 ''$.
%With the AOF, the MUSE WFM is supported by ground layer adaptive optics (GLAO) through four artificial laser guide stars and the adaptive optics system GALACSI (ground atmospheric optics for spectroscopic imaging). 
The data were obtained in nominal mode (wavelength range $\lambda4750\AA-\lambda9350\AA$) with a spectral sampling of about $1.07 \ \mathrm{\AA pix^{-1}}$ and an average resolving power of $R \sim 3000$.
The selection criteria corresponding to the observations consist in all the GPs presented in \cite{cardamone2009galaxy} at a declination $<+20$, so there is visibility from the Paranal observatory.
The program ID corresponding to the observations of these galaxies is $0102.B-0480(A)$ (PI: Hayes, Matthew).

We retrieved the fully reduced data cubes from the ESO archive. The data reduction was performed with MUSE Instrument Pipeline v. 1.6.1 with default parameters \citep{weilbacher2020data}, which consists of the standard procedures of bias subtraction, flat fielding, sky subtraction, wavelength calibration, and flux calibration.

%with the IP $Hayes , Matthew$. The title of the proposal is "Triggering the most extreme starburst galaxies" and the abstract: We propose an unbiased census of the environments of the most extreme star forming galaxies in the low-z Universe. The requested sample of `Green Pea' galaxies ($z\approx 0.25$) are close analogs of the faint galaxies that dominate high-z samples and likely reionized the Universe. However a satisfactory picture of how such intense star formation was initiated remains completely illusive. Star formation must either be internally or externally triggered, and any test requires knowledge of the local environment -- current catalogues have neither the sensitivity nor spectroscopic completeness for this. The proposed MUSE observations will provide a complete census of the Green Peas' environment within 200~kpc, and will identify galaxies to limits 100 times fainter than the main object. This will tightly constrain the triggering mechanism of the starburst -- if the external triggering scenario holds, the counterparts will be found; the alternative is that triggering must be internal.\ref{table:galaxy_stuff}

\begin{figure}[h!]
  \resizebox{\hsize}{!}{\includegraphics{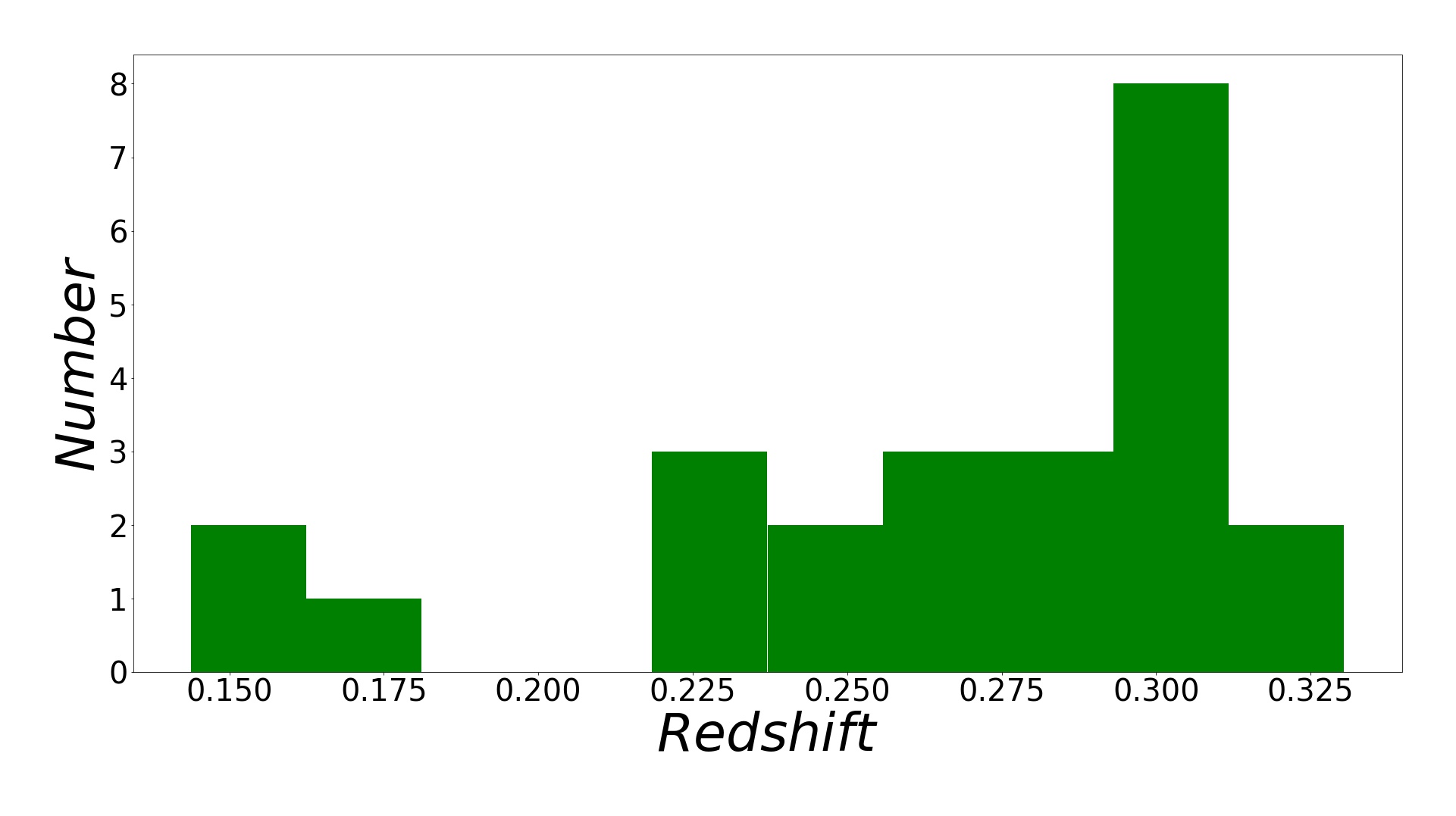}}
  \caption{Redshift histogram of the 24 GPs in our sample.}
  \label{Redshift}
\end{figure}

The sample consists of 24 GPs. It is an unbiased representative set of all GP galaxies: stellar masses, redshifts, metallicities, SFR, $[O\textsc{iii}]5007$ EWs, and line ratios of our galaxies spanning the ranges typical of GPs \citep[e.g.,][]{cardamone2009galaxy,amorin2010oxygen}.
The names, positions, redshifts, and information about the observations of the galaxies are in Table \ref{table:galaxy_stuff}. A histogram of the redshift
of the GPs in our sample can be seen in Fig. \ref{Redshift}.

\begin{table*}[h]                                  % used for centering table
\caption{Name, position, redshift, and observation data. The names shown in the first column are the ones adopted in this work. SDSS name refers to the name of each galaxy in SDSS. Exposure time, seeing and date are taken from the ESO archive. Positions and redshifts are taken from the cube header. Each seeing is referring to the corresponding night of observation.} %Redshifts have been calculated using the position of the most prominent lines ie. $H\alpha$ and $[O\textsc{iii}]$. } 
\label{table:galaxy_stuff}
\centering   
\begin{tabular*}{\linewidth}{l @{\extracolsep{\fill}}c c c c c c c c}       % centered columns (4 columns)
\hline\hline   
Name & SDSS name & R.A.  & Dec  & Redshift & Exp. time & Seeing  & Date \\   
 &  & (deg) & (deg) &  & (s) &  ('') &  \\
\hline

  \begin{filecontents*}{Tablas/GPss.csv}
Name,Ra J2000.0,Dec J2000.0,   Redshift,  Exp. time,      Seeing,  Date, SDSS name,    Name
GP01,146.242675357,-0.762983127,300,5600,0.73,25 Jan 2020, J094458.22-004545.4, GP01
GP02,195.546538444,-0.087455295,225,2800,0.94,23 Jan 2020, J130211.15-000516.4, GP02
GP04,351.413396148,0.751926756,276,2800,1.06,6 Oct 2019,  J232539.23+004507.2, GP03
GP05,50.687443658,0.745180948,304,2800,0.81,26 Feb 2020, J032244.89+004442.3, GP04
GP06,22.292355138,14.993034053,280,2800,0.93,28 Oct 2019, J012910.15+145934.6, GP05
GP07,45.839213815,-7.989384947,164,2800,0.67,17 Nov 2019, J030321.41-075923.2, GP06
GP08,51.556553152,-6.587214591,162,2800,1.33,18 Nov 2019, J032613.62-063512.5, GP07
GP09,54.949309218,-7.427966919,260,4900,0.83,30 Dec 2019, J033947.79-072541.2, GP08
GP10,164.319294819,2.535426310,302,2800,0.45,26 Jan 2020, J105716.72+023207.0, GP09
GP11,191.097707648,2.261384896,239,2800,1.01,23 Jan 2020, J124423.37+021540.4, GP10
GP12,130.570286219,3.634824172,219,3200,1.21,24 Jan 2020, J084216.95+033806.6, GP11
GP13,236.788327547,3.604116133,231,2800,1.11,20 Feb 2020, J154709.10+033614.0, GP12
GP18,339.396355427,13.612971936,293,2800,1.68,7 Nov 2019,  J223735.05+133647.0, GP13
GP27,204.299314389,-2.434471896,273,2800,1.77,11 Feb 2020, J133711.88-022605.4, GP14
GP28,220.630417561,-2.164615960,293,2800,1.14,11 Feb 2020, J144231.37-020952.0, GP15
GP30,157.912232701,7.265419384,252,2800,0.41,26 Jan 2020, J103138.93+071556.5, GP16
GP47,192.144500401,12.566958312,263,2800,0.68,28 Jan 2020, J124834.63+123402.9, GP17
GP48,241.152494936,8.333018461,312,2800,1.45,20 Feb 2020, J160436.66+081959.1, GP18
GP49,239.858998751,8.688757585,297,2800,0.54,21 Feb 2020, J155925.97+084119.1, GP19
GP50,152.988016342,13.139831495,143,2800,0.69,28 Jan 2020, J101157.08+130822.0, GP20
GP63,249.330323649,14.650662530,292,2800,1.78,26 Feb 2020, J163719.30+143904.9, GP21
GP75,141.384719535,14.053597370,301,2800,0.64,28 Jan 2020, J092532.36+140313.1, GP22
GP76,243.276808967,9.496738576,299,2800,1.33,14 Mar 2020, J161306.31+092949.1, GP23
GP77,121.325130188,9.425853403,330,2800,1.12,24 Jan 2020, J080518.06+092533.3, GP24
\end{filecontents*}
\csvreader[tabular = c]{Tablas/GPss.csv}{}{\csvcolix} 
& \begin{filecontents*}{Tablas/GPs.csv}
Name,Ra J2000.0,Dec J2000.0,   Redshift,  Exp. time,      Seeing,  Date, SDSS name,    Name
GP01,146.242675357,-0.762983127,0.300,5600,0.73,25 Jan 2020, J094458.22-004545.4, GP01
GP02,195.546538444,-0.087455295,0.225,2800,0.94,23 Jan 2020, J130211.15-000516.4, GP02
GP04,351.413396148,0.751926756, 0.276,2800,1.06,6 Oct 2019,  J232539.23+004507.2, GP03
GP05,50.687443658,0.745180948,  0.304,2800,0.81,26 Feb 2020, J032244.89+004442.3, GP04
GP06,22.292355138,14.993034053, 0.280,2800,0.93,28 Oct 2019, J012910.15+145934.6, GP05
GP07,45.839213815,-7.989384947, 0.164,2800,0.67,17 Nov 2019, J030321.41-075923.2, GP06
GP08,51.556553152,-6.587214591, 0.162,2800,1.33,18 Nov 2019, J032613.62-063512.5, GP07
GP09,54.949309218,-7.427966919, 0.260,4900,0.83,30 Dec 2019, J033947.79-072541.2, GP08
GP10,164.319294819,2.535426310, 0.302,2800,0.45,26 Jan 2020, J105716.72+023207.0, GP09
GP11,191.097707648,2.261384896, 0.239,2800,1.01,23 Jan 2020, J124423.37+021540.4, GP10
GP12,130.570286219,3.634824172, 0.219,3200,1.21,24 Jan 2020, J084216.95+033806.6, GP11
GP13,236.788327547,3.604116133, 0.231,2800,1.11,20 Feb 2020, J154709.10+033614.0, GP12
GP18,339.396355427,13.612971936,0.293,2800,1.68,7 Nov 2019,  J223735.05+133647.0, GP13
GP27,204.299314389,-2.434471896,0.273,2800,1.77,11 Feb 2020, J133711.88-022605.4, GP14
GP28,220.630417561,-2.164615960,0.293,2800,1.14,11 Feb 2020, J144231.37-020952.0, GP15
GP30,157.912232701,7.265419384, 0.252,2800,0.41,26 Jan 2020, J103138.93+071556.5, GP16
GP47,192.144500401,12.566958312,0.263,2800,0.68,28 Jan 2020, J124834.63+123402.9, GP17
GP48,241.152494936,8.333018461, 0.312,2800,1.45,20 Feb 2020, J160436.66+081959.1, GP18
GP49,239.858998751,8.688757585, 0.297,2800,0.54,21 Feb 2020, J155925.97+084119.1, GP18
GP50,152.988016342,13.139831495,0.143,2800,0.69,28 Jan 2020, J101157.08+130822.0, GP20
GP63,249.330323649,14.650662530,0.292,2800,1.78,26 Feb 2020, J163719.30+143904.9, GP21
GP75,141.384719535,14.053597370,0.301,2800,0.64,28 Jan 2020, J092532.36+140313.1, GP22
GP76,243.276808967,9.496738576, 0.299,2800,1.33,14 Mar 2020, J161306.31+092949.1, GP23
GP77,121.325130188,9.425853403, 0.330,2800,1.12,24 Jan 2020, J080518.06+092533.3, GP24
\end{filecontents*}
\csvreader[tabular = c]{Tablas/GPs.csv}{}{\csvcolviii}
& \begin{filecontents*}{Tablas/GPs_holi.csv}
Name,Ra J2000.0,Dec J2000.0,   Redshift,  Exp. time,      Seeing,  Date, SDSS name,    Name
GP01,146.242675,-0.762983,300,5600,0.73,25 Jan 2020,J094458.22-004545.4,GP01
GP02,195.546538,-0.087455,225,2800,0.94,23 Jan 2020,J130211.15-000516.4,GP02
GP04,351.413396,0.751927,276,2800,1.06,6 Oct 2019,J232539.23+004507.2,GP03
GP05,50.687444,0.745181,304,2800,0.81,26 Feb 2020,J032244.89+004442.3,GP04
GP06,22.292355,14.993034,280,2800,0.93,28 Oct 2019,J012910.15+145934.6,GP05
GP07,45.839214,-7.989385,164,2800,0.67,17 Nov 2019,J030321.41-075923.2,GP06
GP08,51.556553,-6.587215,162,2800,1.33,18 Nov 2019,J032613.62-063512.5,GP07
GP09,54.949309,-7.427967,260,4900,0.83,30 Dec 2019,J033947.79-072541.2,GP08
GP10,164.319295,2.535426,302,2800,0.45,26 Jan 2020,J105716.72+023207.0,GP09
GP11,191.097708,2.261385,239,2800,1.01,23 Jan 2020,J124423.37+021540.4,GP10
GP12,130.570286,3.634824,219,3200,1.21,24 Jan 2020,J084216.95+033806.6,GP11
GP13,236.788328,3.604116,231,2800,1.11,20 Feb 2020,J154709.10+033614.0,GP12
GP18,339.396355,13.612971,293,2800,1.68,7 Nov 2019,J223735.05+133647.0,GP13
GP27,204.299314,-2.434471,273,2800,1.77,11 Feb 2020,J133711.88-022605.4,GP14
GP28,220.630418,-2.164616,293,2800,1.14,11 Feb 2020,J144231.37-020952.0,GP15
GP30,157.912233,7.265419,252,2800,0.41,26 Jan 2020,J103138.93+071556.5,GP16
GP47,192.144500,12.566958,263,2800,0.68,28 Jan 2020,J124834.63+123402.9,GP17
GP48,241.152495,8.333018,312,2800,1.45,20 Feb 2020,J160436.66+081959.1,GP18
GP49,239.858999,8.688758,297,2800,0.54,21 Feb 2020,J155925.97+084119.1,GP18
GP50,152.988016,13.139831,143,2800,0.69,28 Jan 2020,J101157.08+130822.0,GP20
GP63,249.330324,14.650662,292,2800,1.78,26 Feb 2020,J163719.30+143904.9,GP21
GP75,141.384720,14.053597,301,2800,0.64,28 Jan 2020,J092532.36+140313.1,GP22
GP76,243.276809,9.496739,299,2800,1.33,14 Mar 2020,J161306.31+092949.1,GP23
GP77,121.325130,9.425853,330,2800,1.12,24 Jan 2020,J080518.06+092533.3,GP24
\end{filecontents*}
\csvreader[tabular = c]{Tablas/GPs_holi.csv}{}{\csvcolii}
& \csvreader[tabular = c]{Tablas/GPs_holi.csv}{}{\csvcoliii}
& \csvreader[tabular = c]{Tablas/GPs.csv}{}{\csvcoliv}
& \csvreader[tabular = c]{Tablas/GPs.csv}{}{\csvcolv}
& \csvreader[tabular = c]{Tablas/GPs.csv}{}{\csvcolvi}
& \csvreader[tabular = c]{Tablas/GPs.csv}{}{\csvcolvii}\\
\hline
                            
\end{tabular*}
\end{table*}

%una columna

From a first inspection of the data cubes, our GPs appear compact and almost point-like except for a few of them. In order to ensure that we can spatially resolve these galaxies,
%have enough spatial resolution to resolve these galaxies, 
the FWHM of all the sources within the FoV has been measured in white light \footnote{Corresponding to the integrated light in the entire range of wavelength ($\lambda 4750 \AA - \lambda 9350 \AA$).}.
We checked the objects in the FoV that are stars according to SDSS DR16 \citep{SDSS_DR16}.
Then, we defined the stellar-like FWHM ($\mathrm{FWHM}_\star$) as the median stellar FWHM, and, in case there are no SDSS stars in the FoV, we defined the $\mathrm{FWHM}_\star$ as the FWHM of the least extended object within the FoV.
We considered a GP to be extended if its FWHM ($\mathrm{FWHM}_{GP}$) follows this condition:
$$ \mathrm{FWHM}_{GP} > \mathrm{FWHM}_\star + 0.1 ''. $$After applying this criterion, 12 GPs are considered to be extended out of 24.
The $\mathrm{FWHM}_\star$ and the $\mathrm{FWHM}_{GP}$
along with the number of stars in the FoV 
can be seen in Table \ref{table:extension}.

The previous analysis allows us to determine which galaxies present a
%slightly 
resolved core.
Nevertheless, thanks to the very high VLT/MUSE sensitivity, we can also check the extended nature of these galaxies in  low surface brightness.
%Nevertheless, MUSE sensitivity would allow us to see the low surface brightness regions of these galaxies which are expected to be more extended. 
In order to prove it, an analogous analysis of the previous one is realized but using instead the full width at $\tfrac{1}{10}$ of the maximum ($\mathrm{FW}\tfrac{1}{10}\mathrm{M}$) and the full width at $\frac{1}{100}$ of the maximum ($\mathrm{FW}\frac{1}{100}\mathrm{M}$). These parameters trace the extension of the objects at lower surface brightness.
We consider that a GP is extended in the low surface brightness regions if one of the following conditions is satisfied:
$$ \mathrm{FW}\tfrac{1}{10}\mathrm{M}_{GP} > \mathrm{FW}\tfrac{1}{10}\mathrm{M}_\star + 0.3 '', $$$$ \mathrm{FW}\tfrac{1}{100}\mathrm{M}_{GP} > \mathrm{FW}\tfrac{1}{100}\mathrm{M}_\star + 0.5 ''. $$After applying this criterion, 11 GPs are considered to be resolved in the low surface brightness regions out of 24.
The $\mathrm{FW}\frac{1}{10}\mathrm{M}$ and $\mathrm{FW}\frac{1}{100}\mathrm{M}$ of the stellar-like sources and the GPs 
%along with the number of stars in the FoV 
can be seen in Table \ref{table:extension}.

\begin{table*}[h]                                  % used for centering table
\caption{Extension of stellar-like sources and GPs. Column (1): Name of the galaxy; bold text indicates that the galaxy satisfies the extension criteria. Column (2): Number of SDSS stars in the FoV. Column (3): FWHM of stellar-like sources. Column (4): $\mathrm{FW}\frac{1}{10}\mathrm{M}$ of stellar-like sources. Column (5): $\mathrm{FW}\frac{1}{100}\mathrm{M}$ of stellar-like sources. Column (6): FWHM of the GP. Column (7): $\mathrm{FW}\frac{1}{10}\mathrm{M}$ of the GP. Column (8): $\mathrm{FW}\frac{1}{100}\mathrm{M}$ of of the GP.} %Redshifts have been calculated using the position of the most prominent lines ie. $H\alpha$ and $[O\textsc{iii}]$. } 
\label{table:extension}
\centering   
\begin{tabular*}{\linewidth}{l @{\extracolsep{\fill}}c c c c c c c c}       % centered columns (4 columns)
\hline\hline   
Name & Number & FWHM$_\star$  & $\mathrm{FW}\frac{1}{10}\mathrm{M}$$_\star$   & $\mathrm{FW}\frac{1}{100}\mathrm{M}$$_\star$   & FWHM$_{GP}$ & $\mathrm{FW}\frac{1}{10}\mathrm{M}$$_{GP}$ & $\mathrm{FW}\frac{1}{100}\mathrm{M}$$_{GP}$   \\   
 & of stars & ('') & ('') & ('') & ('') &  ('') & ('')  \\
 (1) & (2) & (3) & (4) & (5) & (6) & (7) & (8)  \\
\hline
%\csvreader[tabular = c]{Tablas/EXTENSION_MADRE.csv}{}{\boldify{\csvcoli} & \csvcolii & \csvcoliii & \csvcoliv & \csvcolv & \csvcolvi & \csvcolvii & \csvcolviii}

   \begin{filecontents*}{Tablas/EXTENSION_MADRE.csv}
Nombre,Numero_estrellas,FWHM_star-like (arcsec),FW1/10M_star-like (arcsec),FW1/100M_star-like (arcsec),FWHM_GP (arcsec),FW1/10M_GP (arcsec),FW1/100M_GP (arcsec),Extension,Rayleight (x10^5 arcsec/AA)
\textbf{GP01},1,0.825,1.82,3.498,1.011,2.44,5.489, CORE MEDIA LOW,4.592
GP02,0,1.232,2.261,3.229,1.232,2.261,3.229,,5.155
GP03,0,1.04,2.31,4.485,1.04,2.31,4.485,,7.072
\textbf{GP04},1,0.821,1.782,3.325,0.902,1.972,3.725,,4.953
\textbf{GP05},1,0.862,1.852,3.397,1.105,2.371,4.342, CORE MEDIA LOW,4.549
\textbf{GP06},0,0.692,1.265,1.797,0.795,1.648,2.844,Likely CORE MEDIA LOW,2.174
\textbf{GP07},1,1.338,3.476,8.435,1.559,3.76,8.435, CORE,12.017
\textbf{GP08},1,0.792,1.735,3.291,0.94,1.998,3.604, CORE,5.602
GP09,1,0.924,1.794,2.792,0.936,2.038,3.82, LOW,4.055
\textbf{GP10},3,0.977,2.095,3.813,1.116,2.54,5.148, CORE MEDIA LOW,5.359
\textbf{GP11},1,1.126,2.681,4.857,1.257,2.681,4.857, CORE,6.335
GP12,3,1.309,2.711,4.22,1.395,2.711,4.22,,7.235
\textbf{GP13},1,1.313,3.026,6.265,1.529,3.355,6.383, CORE MEDIA,7.657
\textbf{GP14},1,1.269,2.707,4.906,1.44,3.042,5.425, CORE MEDIA LOW,7.151
\textbf{GP15},1,1.275,2.718,4.921,1.345,2.917,5.44, LOW,6.839
\textbf{GP16},1,0.696,1.339,2.049,0.76,1.656,3.118, MEDIA LOW,2.611
\textbf{GP17},1,0.768,1.601,2.791,0.878,1.759,2.873, CORE,2.734
GP18,1,1.643,3.335,5.564,1.643,3.335,5.564,,11.553
GP19,2,0.86,1.813,3.22,0.926,1.934,3.382,,2.074
GP20,0,0.903,1.839,3.088,0.903,1.839,3.088,,5.601
\textbf{GP21},4,1.546,3.39,6.36,1.705,3.675,6.782, CORE,6.5
GP22,2,0.717,1.491,2.421,0.747,1.491,2.421,,3.896
\textbf{GP23},4,1.144,2.545,4.961,1.324,2.941,5.707, CORE MEDIA LOW,5.78
\textbf{GP24},2,1.064,2.179,3.7,1.1,2.374,4.387, LOW,5.479
\end{filecontents*}
\csvreader[tabular = c]{Tablas/EXTENSION_MADRE.csv}{}{\csvcoli}
&  \csvreader[tabular = c]{Tablas/EXTENSION_MADRE.csv}{}{\csvcolii}
&  \csvreader[tabular = c]{Tablas/EXTENSION_MADRE.csv}{}{\csvcoliii}
&  \csvreader[tabular = c]{Tablas/EXTENSION_MADRE.csv}{}{\csvcoliv}
&  \csvreader[tabular = c]{Tablas/EXTENSION_MADRE.csv}{}{\csvcolv}
&  \csvreader[tabular = c]{Tablas/EXTENSION_MADRE.csv}{}{\csvcolvi}
&  \csvreader[tabular = c]{Tablas/EXTENSION_MADRE.csv}{}{\csvcolvii}
&  \csvreader[tabular = c]{Tablas/EXTENSION_MADRE.csv}{}{\csvcolviii}

 \\
\hline
                            
\end{tabular*}
\end{table*}

Due to the redshift distribution in our sample of GPs, the spectral coverage of MUSE provides us with most of the emission lines in the optical wavelengths. In particular, for all GPs we have all the emission lines from $H\delta$ to $[S\textsc{ii}]{\lambda}6731\AA.$   For 13 of them we can reach the $[O\textsc{ii}]{\lambda}3727\AA$  line.

%In this work we \textbf{divide} emission line maps at substantially different wavelengths \textbf{
%(e.g., $H\alpha$ / $H\beta$ and $[O\textsc{iii}]5007\AA$ / $[O\textsc{ii}]3727\AA,3729\AA$)}, in the limit of spatial resolution.
%Rayleigh scattering of the atmosphere would spread the light from a source depending on wavelength, thus 
%likely affecting the extension of different emission line maps.

Rayleigh scattering of the atmosphere would spread the light from a source depending on wavelength, and the extension of the emission line maps at substantially different wavelengths (e.g., $H\alpha$ vs $H\beta$ and $[O\textsc{iii}]5007\AA$ vs. $[O\textsc{ii}]3727\AA,3729\AA$)) might be affected.
Aiming to correct by this effect, a detailed analysis was performed sampling the FWHM for all the sources (within the FoV of each GP) along the entire wavelength range, showing that blue images in the cubes have a lower spatial resolution than red images. We derived a decrease in the FWHM (as wavelength increases) of the sources ranging from $2.25\times 10^{-5} \ \mathrm{arcsec}  \ \AA^{-1} $ to $1.15\times 10^{-4} \ \mathrm{arcsec} \ \AA^{-1}$.
This information allowed us to apply the appropriate gaussian kernel to red images to effectively compare them with the blue images.

The MUSE pipeline corrects the data for the effect of atmospheric differential refraction. As a sanity check, we measured the center of all the objects detected in the cubes in the spectral ranges of [4900,5100]$\AA$ and [8950,9150]$\AA$, finding a mean difference of only $0.07''$ \citep{weilbacher2020data}.
%Best and worst conditions are found for GP06 and GP18 respectively (see color Table \ref{table:galaxy_stuff}).

%The evidence of interaction between the GP galaxies and other objects is discussed in \cite{adelantaos}. 4 GPs have a companion within the FoV of the cube. Nevertheless the companions are likely unrelated to the bursts in these galaxies \citep{adelantaos}. 

%In this work for the analysis we have made a cut in the cubes around the center GP of $15 '' \times 15 ''$ covering from $(60 Kpc)^2$ to $(130 Kpc)^2$.

%There are no evidence of interactions between the GP galaxies and other objects. No other source in the field of view of the cube have a measurable redshift similar to the redshift of the central GP. That's why in this work we have focused only in the GP. In order to do so we have make a cut in the cubes around the center GP of $15 '' \times 15 ''$ with covers from $(60 Kpc)^2$ to $(130 Kpc)^2$ and is enough to see the whole galaxy.

%In order to ensure the absence of Atmospheric Differential Refraction in the cubes, a comparison between the center of the objects in the red part of the spectra ($8950-9150Å$) and the blue part of the spectra ($4900-5100Å$) is made. Leading to differences in distance of $0.07''$. This indicates that this effect is well corrected \citep{weilbacher2020data}.
 %word isolated

%\subsection{COMPLEMENTARY DATA}
\subsection{SDSS spectra}

 We retrieved SDSS-DR16 integrated spectra for all the GPs \citep{SDSS_DR16}.
%The SDSS spectrograph has 640 fibers per plate. 
The diameter of the SDSS fiber is 3''. The wavelength coverage is 3800-9200Å. The spectral resolution is 1500 at 3800Å and 2500 at 9000Å.
The pixel spacing       log wavelength is $10^{-4}$ dex and the exposure times are in the range of $2800\ \mathrm{s}-8500 \ \mathrm{s}$.

%The MUSE integrated spectra (see Section \ref{seccion_espectros_integrados}) and the SDSS spectra were compared. The line fluxes measured in both cases are within the margins of error. Additionally, the spectral coverage of SDSS extends to shorter wavelengths than MUSE, enabling us to obtain the flux of the $[O\textsc{ii}]$ lines for each of our galaxies.

\section{Flux measurements and spatially resolved structure}
\label{seccion_resuelta}

In this section, we describe the methodology followed to measure the flux of the emission lines in the galaxies. Additionally, we present the spatially resolved maps generated from the spaxel-to-spaxel measurements of fluxes and continuum along with the so-called BPT diagrams \citep{baldwin1981classification,kewley2006host}.

\subsection{Emission line and continuum maps}

For each spaxel, all the emission lines were fit to a gaussian profile, which gives us the flux of each line. The errors in the fluxes were calculated using the bootstrap method \citep{efron1985bootstrap}. We combined the line fluxes with the position of the fibers in the sky to create the maps of emission lines presented in this paper.

\begin{figure*}[h!]
\centering
   \includegraphics[width=17cm]{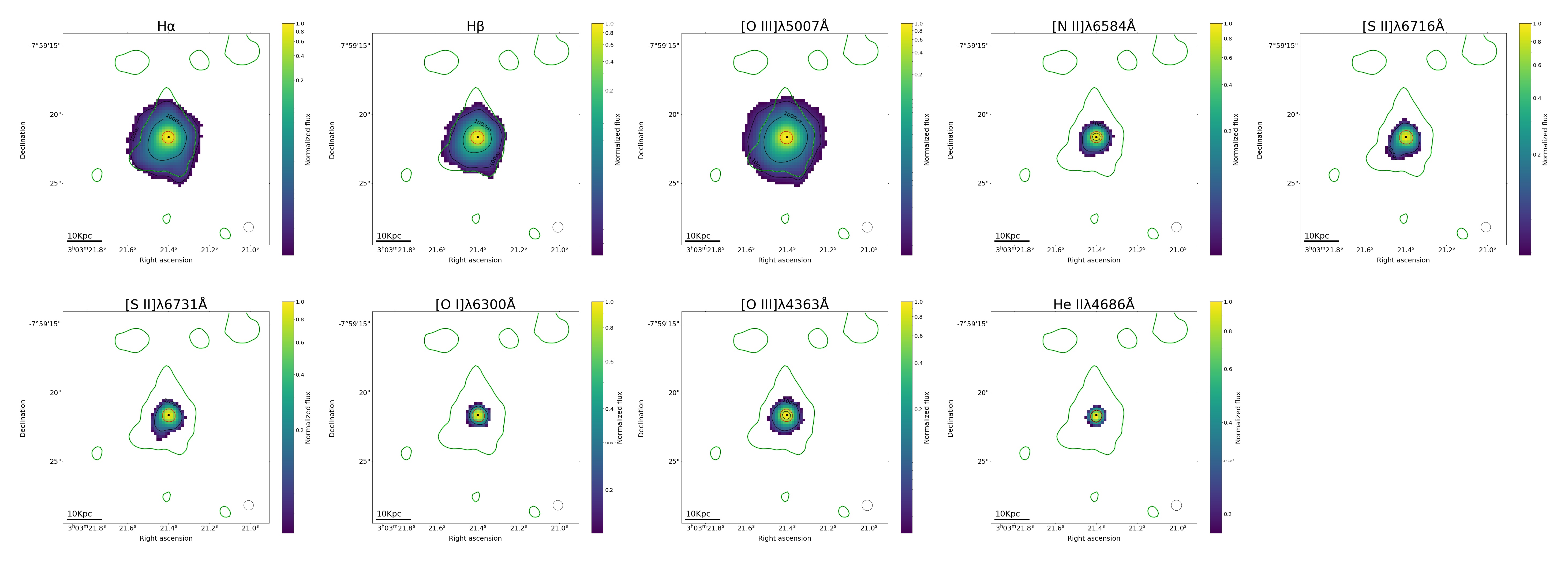}
     \caption{GP06 emission line maps. The black line in the bottom left corresponds to a distance of 10 kpc. The circle in the bottom right is the representation of the seeing. The peak in the $H \alpha$ emission is represented by the black point in the middle. The green and black contour represents the 3 $\sigma_{sky}$ level of the continuum map corresponding to the same galaxy. All maps presented in this work have the previously mentioned features except for the continuum maps, which do not present the continuum contour. Black contours are regions with different $signal = k \times \sigma_{sky}$ levels (with $k = 10,100$), and all spaxels represented in all maps are above 3 $\sigma_{sky}$. The red contour indicates the FWHM of the map. The same contours are shown in each emission line map and continuum map in this study.}
     \label{GP06_maps}
\end{figure*}

As an example, the emission line maps of the galaxy
GP06 are shown in Fig. \ref{GP06_maps}. Among the galaxy sample, GP06 is the one that presents the most complex
structure in these maps, where even in dim lines like $[S\textsc{ii}]6716\AA$ and $[S\textsc{ii}]6731\AA$ a small bump is seen in the south-western position with respect to the central burst. Furthermore, this galaxy presents the brightest nebular $HeII4686\AA$ emission (also detected in GP20 and GP15) in our sample. This line indicates the presence of very high energy photons (E > 4 Ry) that strongly ionize the gas and raise the electron temperature \citep[e.g.,][]{kehrig2015extended}. 
The emission line maps corresponding to the rest of the
galaxies are presented in Appendix \ref{appendix:emission line maps}.
All emission lines in our set of GPs peak in the center of the galaxy. This indicates that the total brightness of the galaxies is dominated by a compact region where the ionization sources are present. This region is slightly resolved for only a few GPs (e.g., GP13; see Fig. \ref{13})
The only four GPs that present non-circular symmetry in     the low surface brightness regions    are GP06, GP13, GP07, and marginally GP01 (see Figs. \ref{GP06_maps}, \ref{13}, \ref{07}, and \ref{1}). The most extended structures in the maps are the ones corresponding to the $[O\textsc{iii}]5007\AA$ line and, in lesser terms, to the $H\alpha$ line. The high intensity of these lines allows us to trace the low surface brightness regions corresponding to the ionized gas in the outer parts of the galaxies.
Dimmer emission lines (e.g., $H\beta$, $[N\textsc{ii}]6584\AA$, $[S\textsc{ii}]6716\AA$, $[S\textsc{ii}]6731\AA$, $[O\textsc{i}]6300\AA,$ and $[O\textsc{iii}]4363\AA$) present less extension and more circular symmetry.

%dos columnas

%CONTINUO

\begin{figure}[h!]
  \resizebox{\hsize}{!}{\includegraphics{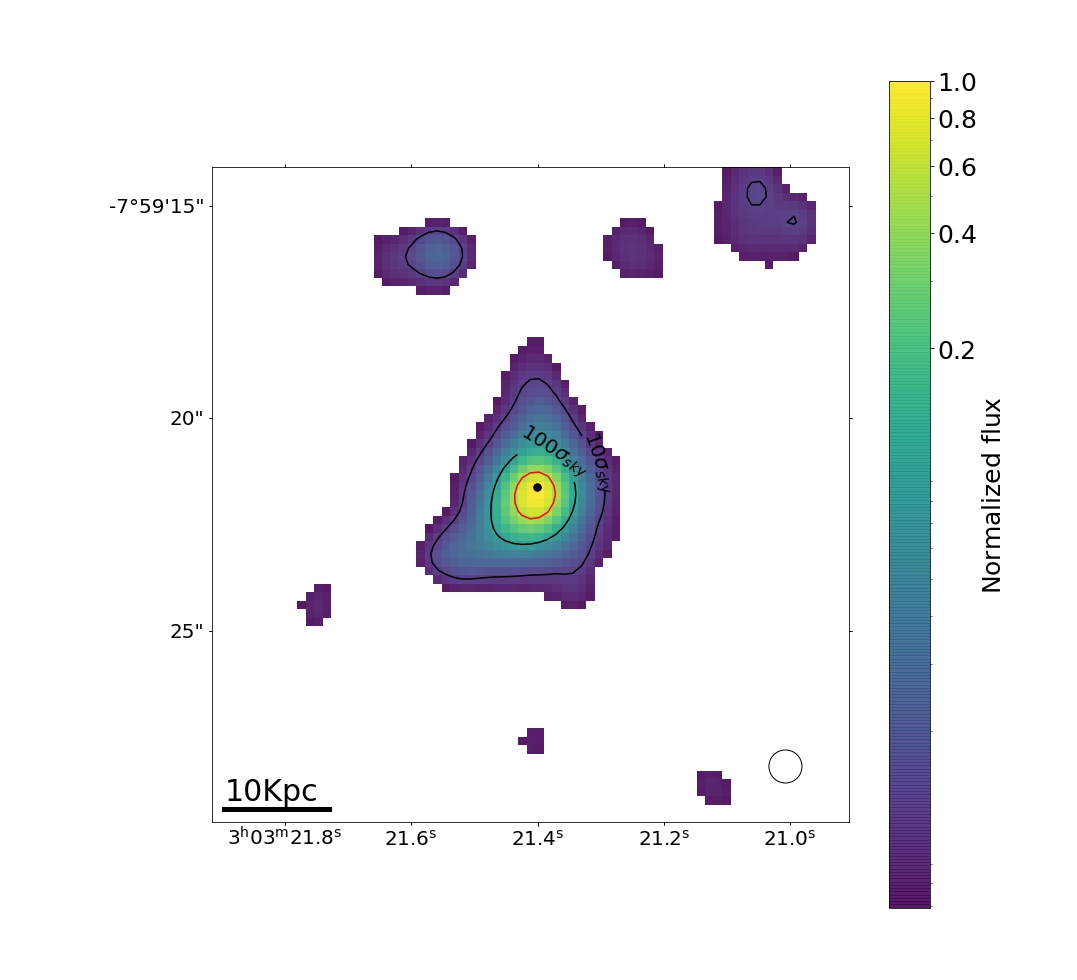}}
  \caption{Continuum map of GP06. Integration is done between 4154Å and 7026 in rest frame.}
  \label{continuo_GP15}
\end{figure}

To make the continuum maps, we selected a rest frame spectral range common to all galaxies to integrate the continuum. Since each galaxy has a different redshift, they present a different rest frame spectral range. So, we can take the intersection of all these to define our rest frame spectral range for defining the continuum. This is indeed between 4154Å and 7026Å, very close to the spectral range of the human eye. %Hereafter we will call it continuum light. 
To make these maps we removed the lines in the spectrum of each spaxel. Then, we integrated the spectrum in our selected rest-frame spectral range.
In
Fig. \ref{continuo_GP15}, we present the continuum map corresponding to GP06.
The continuum maps of all GPs are in Fig. \ref{continuo}. 
As seen in these maps, all galaxies  present a richer structure than in the emission line maps. One of the reasons behind this is the fact that we are collecting more light, since the rest frame spectral range is $2872\AA$ wide. In order to collect the same amount of light from a line, its EW must be similar, while their values are up to $2000\AA$. Continuum maps show bumps attached to the central part of the galaxy in many cases (e.g., GP03, GO06, GP13, GP15). Such disturbed morphologies of the continuum could indicate a spread in the underlying stellar population due to recent mergers \citep{lofthouse2017local}.

%una columna

Even if we are collecting more light in the continuum than in the emission lines, the galaxies show a more compact appearance in the continuum maps than the $[O\textsc{iii}]5007$ emission line maps. This possibly indicates that we are tracing the ionized gas in the outer parts of the galaxies, since the underlying stellar population is less extended than the ionized gas.
In particular, GP06 presents two regions in the north east and north west (with respect to the $H\alpha$ peak), where the $[O\textsc{iii}]5007$ emission line extends much further (up to 10kpc) than the continuum (Figs. \ref{GP06_maps} $\&$ \ref{continuo_GP15}). In these galactic regions, there is almost no extinction (H$\alpha /$ H$ \beta \sim 2.86$), and the gas has higher excitation ($\log([O\textsc{iii}]\lambda5007/H\beta) \sim 0.7$) in comparison to the other outer regions where $\log([O\textsc{iii}]\lambda5007/H\beta) \sim 0.45$ (see Section \ref{seccion_resuelta}, Figs. \ref{ExtinctionGP06} and \ref{combine}). Similar results were found for the galaxy SBS 0335–052E where $[O\textsc{iii}]5007$ and H$\alpha$ filaments with low H$\alpha /$ H$ \beta$ ratio were discovered \citep{2023A&A...670A.121H}. These observations point to the fact that these regions can be channels where LyC photons can escape.
In particular, for the case of GP06 it is found that this galaxy presents an optically thick, neutral outflow along the line of sight (LOS) \citep{jaskot2014linking}, making the escape of high-energy photons difficult in this particular direction (with $f_{esc} (Ly\alpha) = 0.05$ as found in \cite{jaskot2017kinematics}). However, this galaxy displays one of the highest ratios of $[O\textsc{iii}]$ to $[O\textsc{ii}]$ in this study ($[O\textsc{iii}]$ / $[O\textsc{ii}]$ = 6.56). This suggests that there could be potential escape paths for ionizing photons in other directions \citep{izotov2022lyman}.

%These maps present a richer structure than the emission line maps. One of the reasons behind this is the fact that we are collecting more light. The rest frame spectral range is 2872Å wide, so in order to collect the same amount of light from a line its EW must be similar. The EWs of all lines in all the GPs is color $2000\AA$. Anyway, the structure of the continuum shows bumps attached to the central part of the galaxy in many cases. Such disturbed morphologies of the continuum could indicate a spread in the underlying stellar population due to recent mergers \citep{lofthouse2017local}.

\subsection{Emission line ratio maps}
\label{subseccion_emission_line_ratio_maps}

\begin{figure}[h!]
  \resizebox{\hsize}{!}{\includegraphics{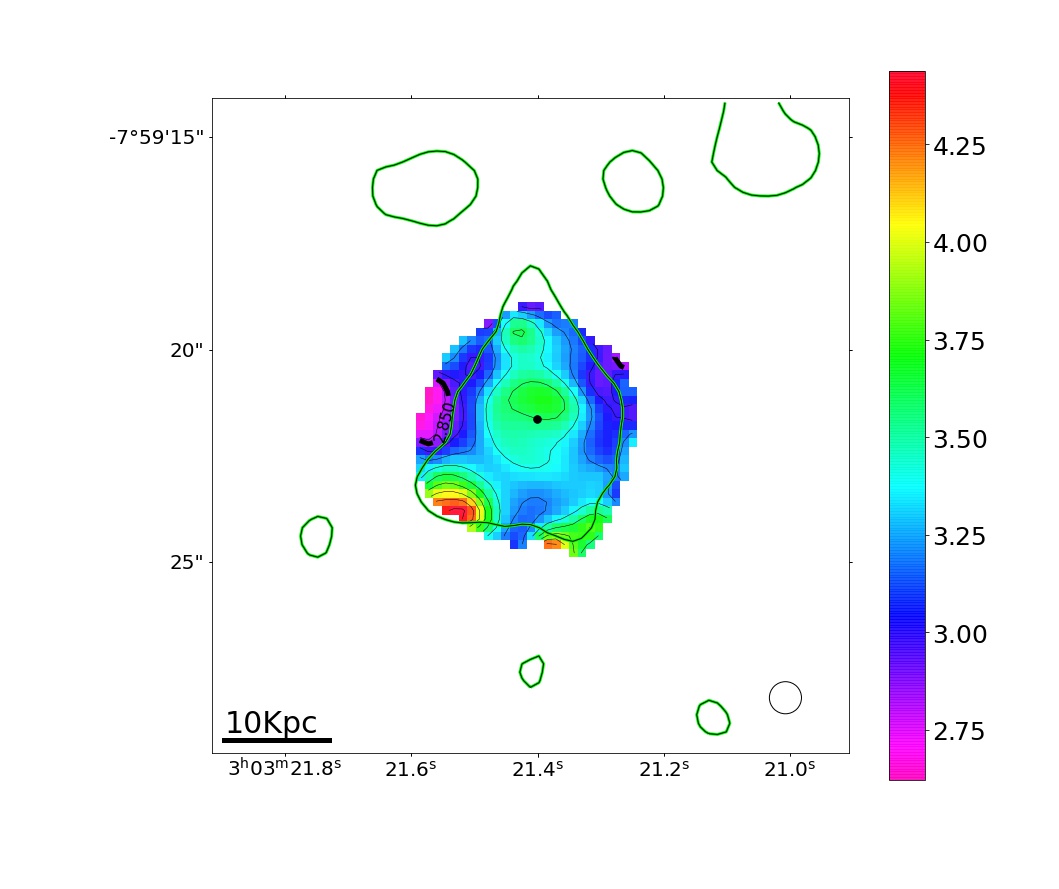}}
  \caption{$H\alpha$/$H\beta$ map of GP06 before reddening correction.}
  \label{ExtinctionGP06}
\end{figure}

Here, we present maps of some of the relevant line ratios to study the ionization structure of the gas in our set of GPs.
These line ratios are corrected for reddening using the corresponding c(H$\beta$) for each spaxel; c(H$\beta$) was computed from the ratio of the measured-to-theoretical $H\alpha$/$H\beta$ assuming the reddening law of \cite{cardelli1989relationship}, and case B recombination with the electron temperature $T_e = 10^4 \ \mathrm{K}$ and electron density $n_e = 100 \ \mathrm{cm^{-3,}}$ which give an intrinsic value of $H\alpha$/$H\beta$= 2.86. 
In Fig. \ref{ExtinctionGP06}, we show the uncorrected $H\alpha / H\beta$ map for GP06.
The $H\alpha / H\beta$ map is a tracer of dusty regions, and the zones with higher extinction are likely to present dust that acts like a wall for high-energy photons \citep{weingartner2006photoelectric}. Hence, if there is an escape of LyC photons, the preferred direction would be the one that presents lower extinction.
The emission line ratio maps for all GPs are provided in Appendix \ref{appendix:cocient maps}.

%una columna

%Examples of maps corresponding to the most relevant line ratios which are indicators of the ionization structure for ionized gaseous nebulae are displayed in Figs. \ref{IonizationGP06}, \ref{oiii_hbeta_GP06}, \ref{nii_halpha_GP06}, \ref{sii_halpha_GP06} $\&$ \ref{oi_halpha_GP06}. These line ratios are corrected for reddening using the corresponding $c(H\beta)$ for each spaxel. The same maps for all 24 GPs can be seen in Appendix \ref{appendix:cocient maps}.

\begin{figure}[h!]
  \resizebox{\hsize}{!}{\includegraphics{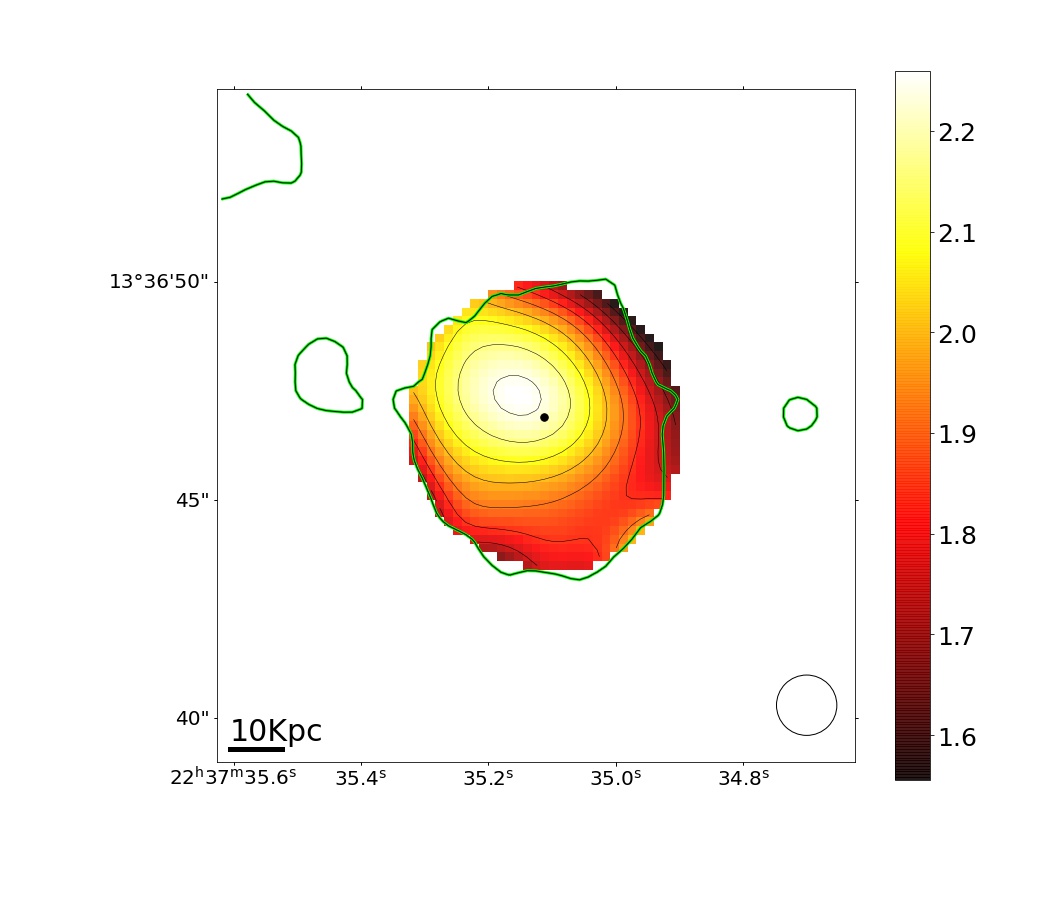}}
  \caption{$[O\textsc{iii}]5007/([O\textsc{ii}]3727+[O\textsc{ii}]3729)$ map of GP13.}
  \label{IonizationGP06}
\end{figure}

The MUSE spectral coverage allows us to observe the $[O\textsc{ii}]\lambda\lambda3727,3729$ lines for 13 of our GPs due to their redshifts, 
so we can analyze the map of the $[O\textsc{iii}]\lambda5007$ / $[O\textsc{ii}]\lambda\lambda3727,3729$ ($[O\textsc{iii}]$ / $[O\textsc{ii}]$)  line ratio (see Appendix \ref{appendix:cocient maps}), which traces the ionization of the gas, in these galaxies. As an example, in Fig. \ref{IonizationGP06} we show the map of this ratio for GP13, which is the most extended galaxy in our sample.
High values of $[O\textsc{iii}]$ / $[O\textsc{ii}]$ trace galactic regions with high ionization. 
These regions tend to be near the center of the galaxy where the $H\alpha$ peaks and the main ionizing sources 
are also located. 
The highest values of $[O\textsc{iii}]$ / $[O\textsc{ii}]$ are found in GP22 and GP15 (see Figs. \ref{GP22_color} and \ref{GP15_color}), reaching a value of $6.5$. 

%una columna
%The ionization maps of all GPs are in Fig. \ref{Ionization}. 

\begin{figure*}[h!]
\centering
  \includegraphics[width=17cm]{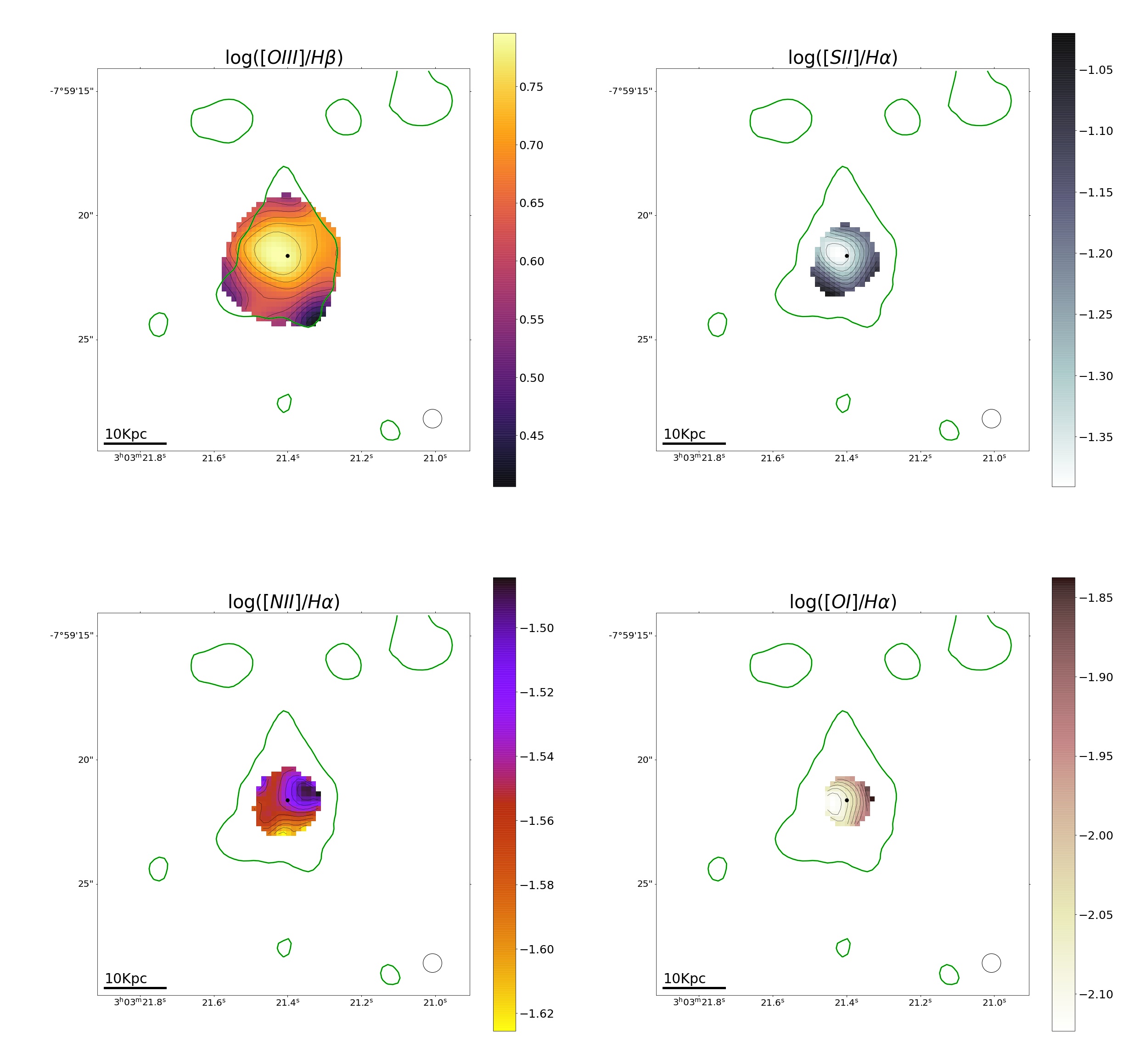}
  \caption{BPT maps corresponding to GP06.}
  \label{combine}
\end{figure*}

The ratios used in the BPT diagrams are 
$[O\textsc{iii}]\lambda5007/H\beta$ ($[O\textsc{iii}]/H\beta$), 
$[N\textsc{ii}]\lambda6584/H\alpha$ ($[N\textsc{ii}]/H\alpha$),
$[S\textsc{ii}]\lambda\lambda6716,6731/H\alpha$ ($[S\textsc{ii}]/H\alpha$), and
$[O\textsc{i}]\lambda6300/H\alpha$ ($[O\textsc{i}]/H\alpha$).
Examples of maps corresponding to these line ratios 
are displayed in Fig. \ref{combine}.

High values of $[O\textsc{iii}]/H\beta$ correspond to the areas of ionized gas with relatively higher excitation.  
The $[O\textsc{iii}]/H\beta$ maps do not show significant spatial variations (with a maximum difference of < 0.1 dex) for any of the GPs except for GP06, indicating low spatial gradients in gas excitation. 
For this galaxy, the spatial variation of $[O\textsc{iii}]/H\beta$ goes up to 0.4 dex, with a maximum very close to the center of H$\alpha$ emission (see Fig. \ref{combine}). The outer zones of this galaxy with lower $[O\textsc{iii}]/H\beta$ ($[O\textsc{iii}]/H\beta \sim 0.45$) present high extinction ($H\alpha / H\beta \sim 4.25$) (see north, south-east and south-west regions in Figs \ref{ExtinctionGP06} and  \ref{combine}). As the photons reach the outer parts from the central starburst, they tend to be absorbed, so this radiation becomes softer as it goes through dusty regions.
GP20 presents the highest value of $[O\textsc{iii}]/H\beta$, reaching $\log([O\textsc{iii}]/H\beta) = 0.91$ (see Fig. \ref{GP20_color}). This galaxy also exhibits the most prominent emission lines (see \ref{table:emission line fluxes}).

Low values of $[N\textsc{ii}]/H\alpha$ in the maps indicate regions with higher gas excitation and trace low metallicity zones \citep{pettini2004iii}.
The $[N\textsc{ii}]/H\alpha$ maps do not present much spatial variance, with a maximum difference of 0.14 dex for GP08 (see Fig. \ref{GP08_color}). Indicating low spatial variations in gas excitation and metallicity.
The lowest values in $[N\textsc{ii}]/H\alpha$ are found in GP06, GP15, and GP20 (see Figs. \ref{GP06_color}, \ref{GP15_color} and \ref{GP20_color}). These three galaxies are
also the ones that show nebular $HeII4686\AA$ emission, which indicates the
presence of hard ionizing sources.

% Metallicity is traced by the $[N\textsc{ii}]/H\alpha$ maps. If we translate this ratio into metallicity by using the relation in \cite{pettini2004iii},
% %12 + \log (O/H) = 8.90 + 0.57 \times ([N\textsc{ii}]/H\alpha)$$,
% the maximal spatial difference in metallicity does not exceed 0.1 dex. While the error in the integrated metallicity calculation (see Section \ref{seccion_espectros_integrados}) is in the range: 0.02 - 0.2 dex. So metallicity gradients are pretty mild or even non-existence for most GPs. Only in some cases like GP06 and GP20 we can see metallicity gradients above the error. Nonetheless, it is not possible to see a clear radial gradient, since %(ie. for the set of galaxies we can not say if the metallicity tend to increase or decrease towards the center). 
% each galaxy present his own pattern.

$[S\textsc{ii}]/H\alpha$ maps  trace the opacity of the column of gas (i.e., gas in the line of sight in each spaxel) \citep{Pellegrini_2012}. $[O\textsc{i}]/H\alpha$ peaks in the ionization front (edge of ionization-bounded regions). Low values on both coefficients indicate thin columns of gas (and as well, high gas excitation) where LyC photons are likely to escape \citep{paswan2022}. We can see that most GPs present a blister-type morphology, which is characterized by an optically thin galactic center, and as we go to the outer parts (around 5-10kpc) the medium tends to become optically thick. (see Appendix \ref{appendix:cocient maps}). 
An extremely low $[S\textsc{ii}]/H\alpha$ value could indicate a hole from which photons can escape \citep{wang2021low}, being GP06, GP15, and GP20 the most extreme cases again (see Figs. \ref{GP06_color}, \ref{GP15_color}, and \ref{GP20_color}).

% %una columna
% \begin{figure}[h!]
%   \resizebox{\hsize}{!}{\includegraphics{Fotos/Mapas_sueltos/ejemplo_OIII_Hbeta.jpg}}
%   \caption{log($[O\textsc{iii}]5007/H\beta$) map of GP06.}
%   \label{oiii_hbeta_GP06}
% \end{figure}
% 
% 
% %una columna
% \begin{figure}[h!]
%   \resizebox{\hsize}{!}{\includegraphics{Fotos/Mapas_sueltos/Ejemplo_NII_Halpha.jpg}}
%   \caption{log($[N\textsc{ii}]/H \alpha$) map of GP06.}
%   \label{nii_halpha_GP06}
% \end{figure}
% 
% 
% %una columna
% \begin{figure}[h!]
%   \resizebox{\hsize}{!}{\includegraphics{Fotos/Mapas_sueltos/Ejemplo_SII_Halpha.jpg}}
%   \caption{$\log([S\textsc{ii}]/H\alpha$) map of GP06.}
%   \label{sii_halpha_GP06}
% \end{figure}
% 
% 
% %una columna
% \begin{figure}[h!]
%   \resizebox{\hsize}{!}{\includegraphics{Fotos/Mapas_sueltos/Ejemplo_OI_Halpha.jpg}}
%   \caption{$\log([O\textsc{i}]6300/H\alpha$) map of GP06.}
%   \label{oi_halpha_GP06}
% \end{figure}

\begin{figure*}[h!]
\centering
  \includegraphics[width=17cm]{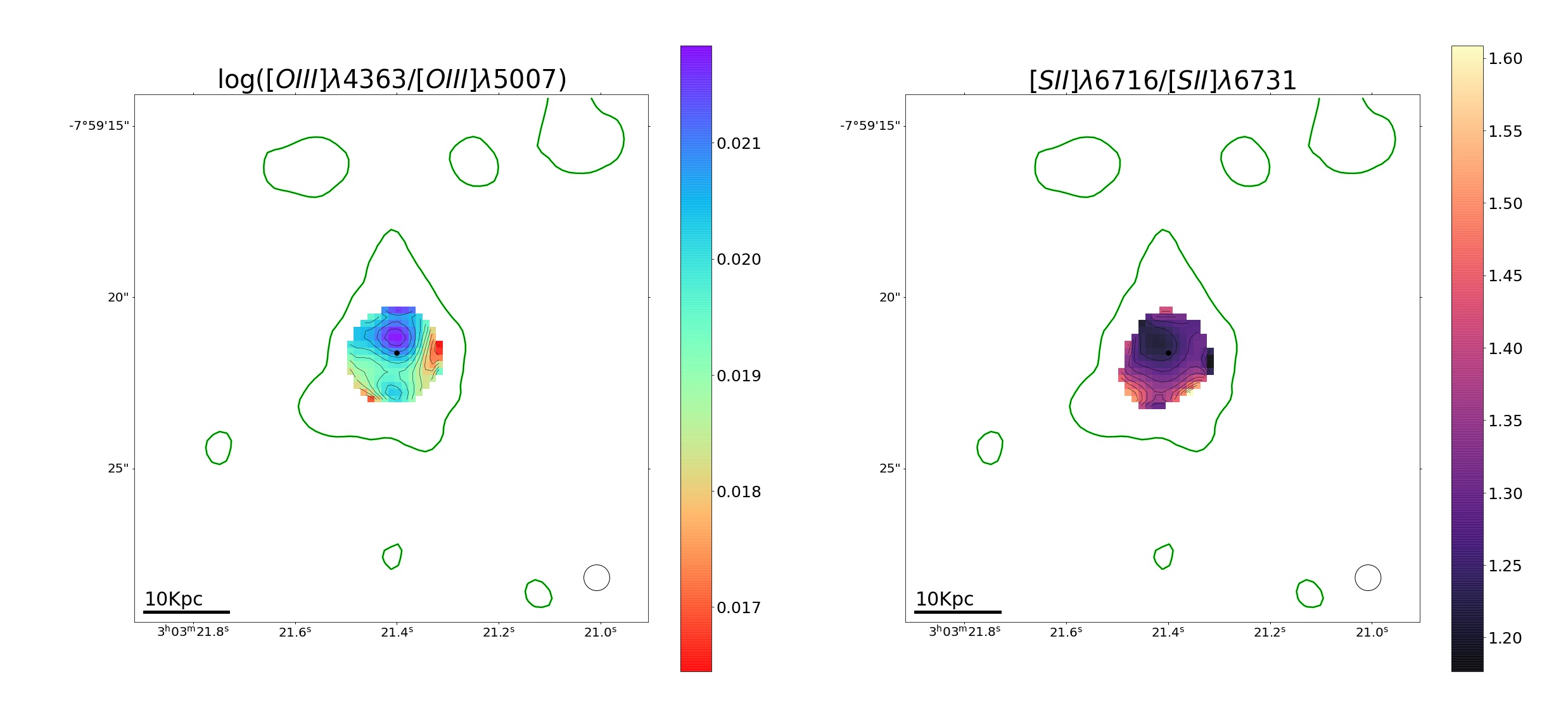}
  \caption{Maps corresponding to GP06.}
  \label{combine_2}
\end{figure*}

The last two maps presented here are $[S\textsc{ii}]\lambda6716/[S\textsc{ii}]\lambda6731$ and $[O\textsc{iii}]\lambda4363/[O\textsc{iii}]\lambda5007$.  
Examples of maps corresponding to these line ratios
are displayed in Fig. \ref{combine_2}.
These emission line ratios give us information about the electron density and electron temperature, respectively \citep[e.g.,][]{Enrique_metodo_directo}. 
Higher $[S\textsc{ii}]\lambda6716/[S\textsc{ii}]\lambda6731$ values correspond to lower electron densities, whereas higher $[O\textsc{iii}]\lambda4363/[O\textsc{iii}]\lambda5007$ values correspond to higher electron temperatures.
The maps corresponding to these two line ratios do not present clear radial variances, while
they show a wide variety of morphologies despite the low spatial extension of these dim lines (see Appendix \ref{appendix:cocient maps}).
In the particular case of GP06, the values of the
$[S\textsc{ii}]\lambda6716/[S\textsc{ii}]\lambda6731$ ratio are higher where the gas shows lower excitation
(as traced by $[O\textsc{iii}]/H\beta$ and $[S\textsc{ii}]/H\alpha$) and higher extinction regions (traced by
$H\alpha$/$H\beta$). This indicates that for this galaxy the electron density tends to
a decrease in the regions with higher dust content and lower gas excitation.
Moreover, the highest values for $[O\textsc{iii}]\lambda4363/[O\textsc{iii}]\lambda5007$ (line-ratio indicator
of the electron temperature) are found in the galaxies with $HeII4686\AA$
emission (see Figs. \ref{GP06_color}, \ref{GP15_color} and \ref{GP20_color}), which reinforces the existence of a
harder ionizing radiation field in these objects \citep[see, e.g.,][]{kehrig2016spatially}.

%dos columnas

We present the radial profiles of the emission line ratios previously mentioned. By plotting the emission line ratio as a function of radial distance, we effectively visualize the radial variations in the line ratios across the galaxy. This representation facilitates the identification of radial trends and provides a comprehensive understanding of how these line ratios evolve as we move away from the galactic center.

To obtain the radial profiles, we utilized a technique that integrates circular crowns centered around the peak of H$\alpha$ emission. The code calculates the average value of the emission line ratio within each crown and plots it against the radial distance from the center. The radial profiles of H$\alpha$/H$\beta$ are shown in Fig. \ref{radial_Halpha_Hbeta}. The rest of the radial profiles are in Appendix \ref{apendix_radial_profiles}.

%dos columnas
\begin{figure*}[h!]
\centering
  \includegraphics[width=17cm]{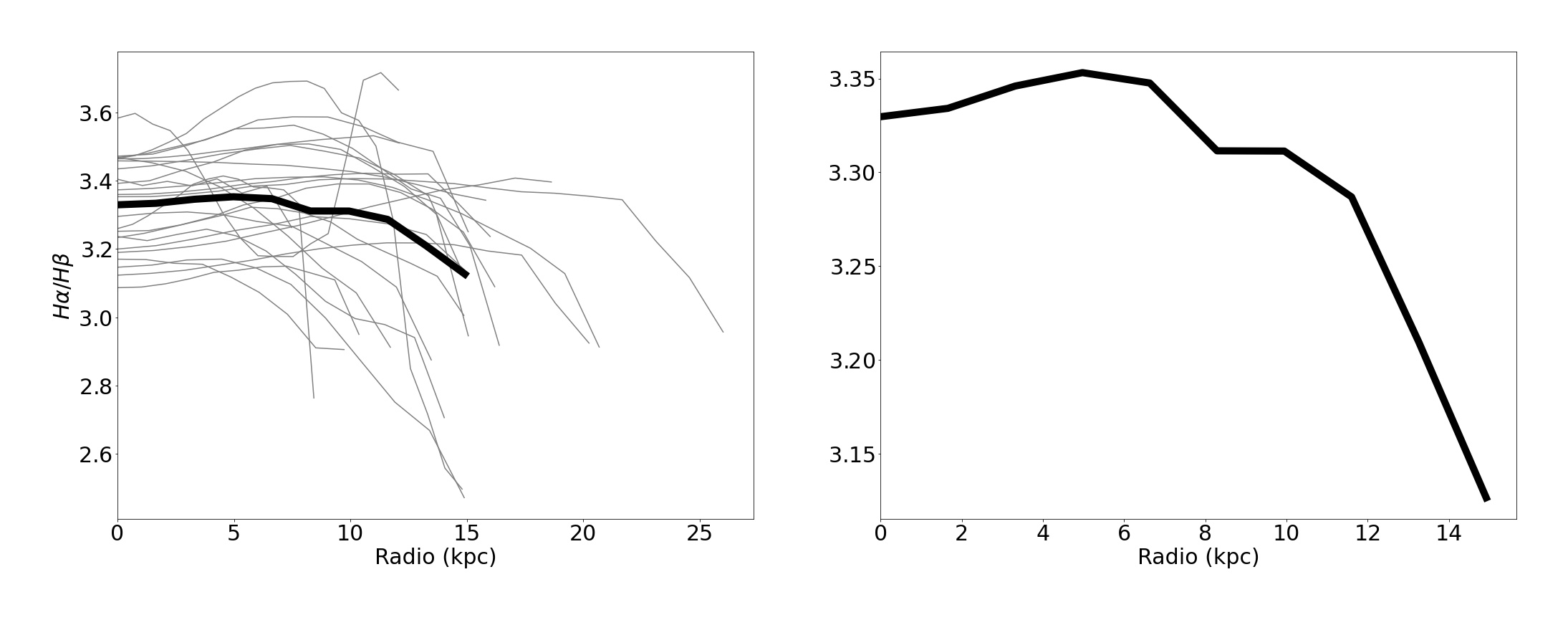}
  \caption{Radial profile of H$\alpha$/H$\beta$. The left image shows the profiles of all GPs with available information about the lines in gray. In black it is represented the mean of all galaxies and extends up to a radius where only 30$\%$ of the galaxies present values at larger radii. On the right is a zoomed-in image of the mean profile. The same display applies to all line ratios presented in the appendix.}
  \label{radial_Halpha_Hbeta}
\end{figure*}

One notable observation is that the overall radial tendencies of the emission line ratio profiles are similar across the entire set of galaxies. However, there are substantial variations in the absolute values of the ratios between different galaxies, typically on the order of 0.5 dex. In contrast, within each individual galaxy, the radial change in the emission line ratios is relatively small, typically less than 0.1 dex.
This small radial variation within each galaxy could potentially be attributed to the low spatial resolution of our observations.

Despite the low radial change within each galaxy, the mean radial profiles of all galaxies present clear tendencies for the various emission line ratios studied. These profiles provide valuable insights into the ionization state within the galaxies.

We observe a consistent decrease in the $[O\textsc{iii}]$ / $[O\textsc{ii}]$ and $[O\textsc{iii}]5007/H\beta$ ratios and an increasing trend in the $[S\textsc{ii}]/H\alpha$, $[O\textsc{i}]/H\alpha$, and $[N\textsc{ii}]/H\alpha$ ratios as we move away from the galactic center. This implies that the highest level of ionization and density-bounded tracers are predominantly concentrated in the central regions of the galaxies. Still, the mean variation in the $[O\textsc{iii}]5007/H\beta$ and $[N\textsc{ii}]/H\alpha$ ratios are really low (0.03 dex and 0.015 dex, respectively), indicating that the radial changes in these emission line ratios are nearly imperceptible.

Furthermore, the radial tendency in the H$\alpha$/H$\beta$ ratio indicates a decreasing trend as we move toward the outer parts of the galaxies. This suggests a relatively higher level of extinction or enhanced dust attenuation toward the central regions, resulting in a higher H$\alpha$/H$\beta$ ratio compared to the outskirts.

In contrast, the radial changes in the $[S\textsc{ii}]\lambda6716/[S\textsc{ii}]\lambda6731$ and $[O\textsc{iii}]\lambda4363/[O\textsc{iii}]\lambda5007$ ratios are small (on the order of the previously mentioned $[N\textsc{ii}]/H\alpha$ ratio) and do not exhibit a clear radial trend. These ratios may be less sensitive to radial variations or they are influenced by other factors such as excitation conditions or observational uncertainties.

Overall, the mean radial profiles of the emission line ratios consistently reveal the spatial variations in the ionization state within the galaxies. The observed trends support the notion that the central regions of the galaxies exhibit higher ionization levels and greater density-bounded tracers, while the outer parts experience lower ionization conditions.

\subsection{BPT diagrams}

\begin{figure*}[h!]
\centering
   \includegraphics[width=18.5cm]{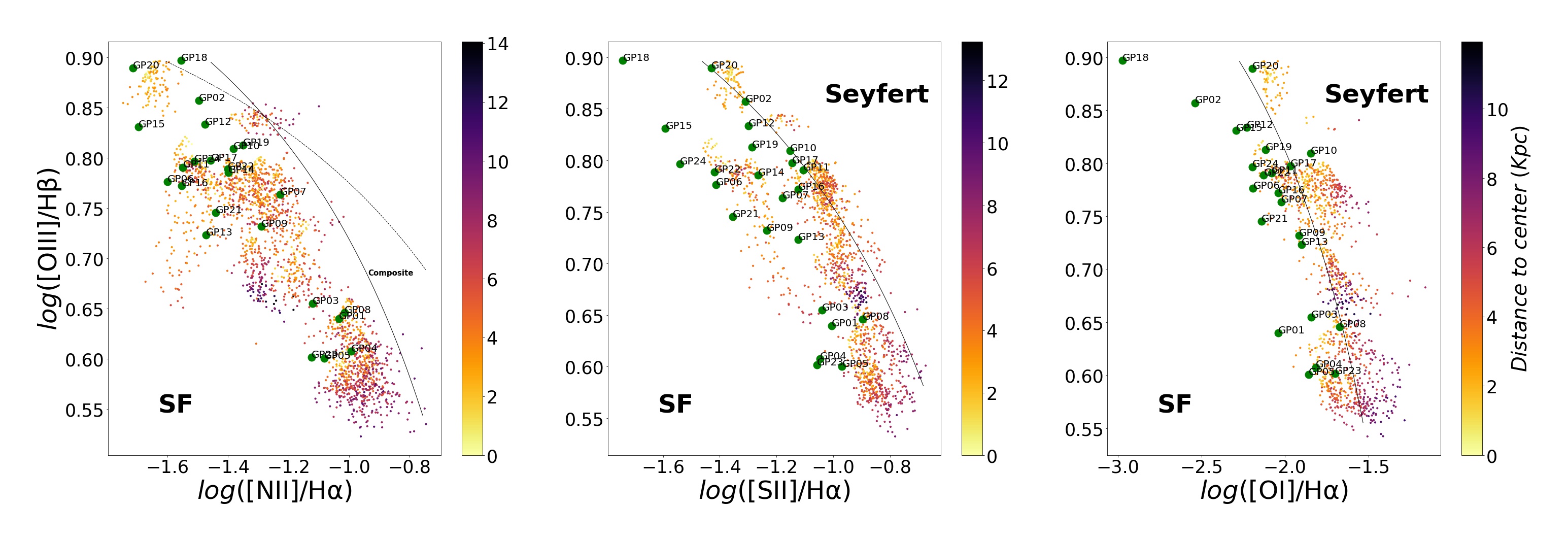}
     \caption{BPT diagram. Small dots in the color scale (from yellow to dark purple) correspond to spaxels in each galaxy. Big green dots correspond to the integrated spectra of the galaxy. Next to the green dots we can see the name of each galaxy. The color of the small dots corresponds to the distance to the center of the galaxy. The closer to the center the more yellow they are, and farther away from the center they become darker. The lines that delimitate each region are taken from \cite{lineaBPT1} and \cite{lineaBPT2}.}
     \label{BPT}
\end{figure*}

The BPT diagrams for all our GPs are shown in Fig. \ref{BPT} on a spaxel-by-spaxel basis; there, we can see the line ratios $[O\textsc{iii}]5007/H\beta$ versus $[N\textsc{ii}]/H\alpha$, $[S\textsc{ii}]/H\alpha,$ and $[O\textsc{i}]/H\alpha $plotted. Small dots correspond to measurements from individual spaxels for each galaxy. Big green dots show the line ratios derived from the integrated spectra (see Section \ref{seccion_espectros_integrados}), these values are shown in Table \ref{table:gas_properties}.
%The aperture of integration is define by the strongest emission line (ie. $[O\textsc{iii}]$). 
%The aperture of integration is define by the faintest emission line (ie. $[N\textsc{ii}]$ in the left, $[S\textsc{ii}]$ in the middle and $[O\textsc{i}]$ in the right). %using the same procedure as section 4.

Overall, our GPs fall in the general locus of SF objects according to the spectral classification scheme proposed by \cite{baldwin1981classification} and \cite{kewley2006host}. This suggests that photoionization from hot massive stars is the dominant excitation mechanism within these galaxies. In particular, GPs are located in the top left part of the diagram where the most extreme galaxies reside  (i.e., lower metallicity and higher excitation of the ionized gas). Previous studies confirm this trend for GPs \citep[e.g.,][]{cardamone2009galaxy,jaskot2013origin}. Furthermore, GPs share with EELGs the same region in the $[O\textsc{iii}]5007/H\beta$ versus $[N\textsc{ii}]/H\alpha$ diagram as \cite{perez2021extreme} showed.

%BH

Furthermore, there is no presence of the $Fe[X]\lambda 6374\AA$ line in any GP. This line is a tracer of black hole (BH) activity. Luminosities of this line on the order of $10^{36}-10^{39} \ \mathrm{erg/s}$ correspond to the presence of BHs with a mass of $\sim 10^5 \ \mathrm{M_\odot}$ \citep{molina2021sample}. Nevertheless, due to the distance of GPs, it is not possible to measure lines with luminosities $<<10^{40} \ \mathrm{erg/s}$. Such limitations lead us to conclude that GPs do not present actively accreting BHs with a mass $>>10^5 \ \mathrm{M_\odot}$.

In addition, we used a method for estimating BH masses involving the use of a BH mass -stellar mass relation- which states that it is nearly independent of redshift \citep{BH_STELLAR}. Nevertheless, our range of stellar masses, which covers from $10^{8.3} \ \mathrm{M_\odot}$ to $10^{10}  \ \mathrm{M_\odot}$ in the low end, is well below the predictive capacity of the relation. Thus, we are only using it for the GPs with higher masses ($\sim 10^{10}  \ \mathrm{M_\odot}$); for these galaxies, the expected BH mass is no greater than $10^{6.7}  \ \mathrm{M_\odot}$.

This estimation is clearly above $10^5 \ \mathrm{M_\odot}$ solar masses, which suggests that if these galaxies do contain BHs, these BHs have a low mass compared to the total stellar mass of the galaxy. It is important to note, however, that with the use of the \cite{BH_STELLAR} relation we cannot accurately predict BH masses in galaxies with $\mathrm{M_\odot} < 10^{9.8}$. That is why it is so difficult  to retrieve any information from the BHs in the low stellar mass GPs. Further observations and analysis are needed to confirm the presence and measure the exact masses of any BHs in these galaxies.

%end BH

In the BPT diagrams, the distance from the center of the galaxy as a parameter is also represented. We can see a common tendency for all galaxies. As we go closer to the center of the galaxy, it shows a higher $[O\textsc{iii}]5007/H\beta$ ratio (i.e., higher excitation close to the center) and lower $[S\textsc{ii}]/H\alpha$ and $[O\textsc{i}]/H\alpha$ ratios (i.e., the center of the galaxies are optically thinner, and neutral oxygen is more abundant in the outer part). Regarding the $[N\textsc{ii}]/H\alpha$ ratio, it is almost independent of distance, possibly indicating that metallicity gradients are low. 
All these results are in agreement with the ones presented in the previous section (Section
\ref{subseccion_emission_line_ratio_maps}).

\section{Properties of GPs from integrated spectra}
\label{seccion_espectros_integrados}

\begin{figure*}[h!]
\centering
   \includegraphics[width=17cm]{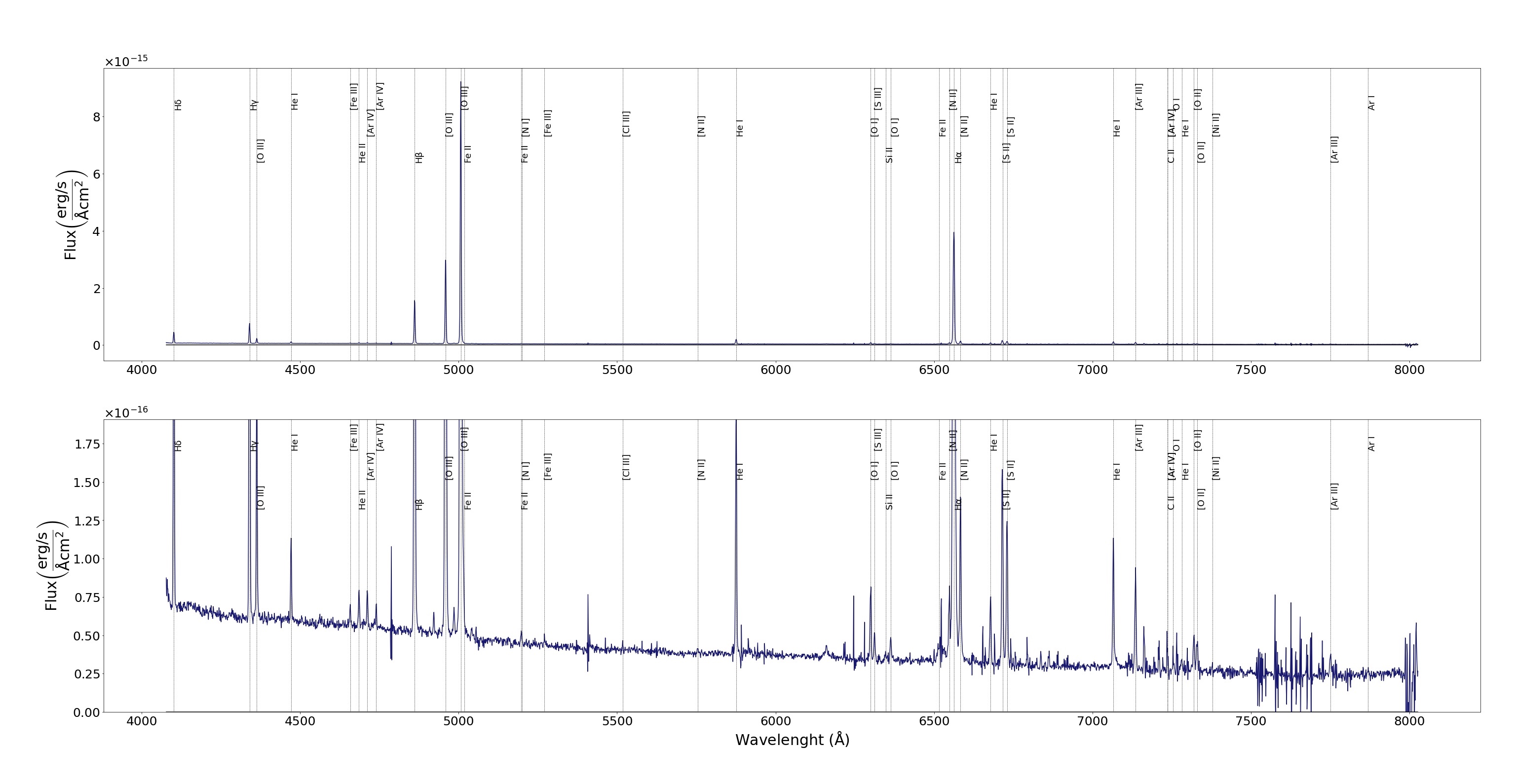}
     \caption{Integrated spectrum of GP06. The bottom image is a zoomed-in view of the y-scale of the top image.}
     \label{espectro_GP06}
\end{figure*}

We also took advantage of our IFS data to produce the 1D spectra of selected galaxy regions. To do so, for each GP, we added the flux in all spaxels for which the $H\alpha$ flux measurements present a signal-to-sigma sky ratio greater than three ($S/ \sigma_{sky} >3$). As an example, the integrated spectrum of the galaxy GP06 can be seen in Fig. \ref{espectro_GP06}. The integrated spectra of all GPs are shown in figure \ref{espectros}. We did not find in any of the spectra features of WR stars.

%dos columnas

We used the
integrated spectrum of each GP to identify and measure the most relevant emission lines.
The same procedure as the one used in Section \ref{seccion_resuelta} was used to calculate emission line fluxes, errors, and extinction correction. 
Tables in \ref{a1} lists all the data of the emission lines.
To check the wellness in the flux measurement, the MUSE integrated spectra and the SDSS spectra were compared, retrieving that, on average, the ratio between the fluxes per line is $1.003 \pm 0.038$.
Additionally, the spectral coverage of SDSS extends to shorter wavelengths than MUSE, enabling us to obtain the flux of the $[O\textsc{ii}]$ line for the 11 GPs that do not present this line in MUSE.
We used the MUSE fluxes in all other cases.

\subsection{SFR}

By means of the Kennicutt relation
\citep{kennicutt1998global}, we derived the SFR through the $H\alpha$ luminosity ($H\alpha$ luminosity was retrieved using luminosity distances from the UCLA cosmology calculator \footnote{See https://astro.ucla.edu/~wright/CosmoCalc.html.}) in the following way: $$SFR(M_{\odot}/yr) =1.26 \times  10^{41}   L_{H\alpha}   (erg/s).$$
Most of the stellar masses ($\mathrm{M}_\star$) were reproduced from \cite{izotov2011green}; nevertheless, \cite{izotov2011green} did not calculate the mass for all the galaxies in our sample, and so the \cite{cardamone2009galaxy} value was used instead. \cite{izotov2011green} recalculated the masses of the GPs and obtained systematically lower values. Their values are lower because in fitting the SED, they subtracted the contribution from gaseous continuum emission. No errors were listed in the original studies.
The corresponding specific SFRs were derived as follows: $sSFR=SFR/\mathrm{M}_\star$.

%With the $H\alpha$ luminosity we can derive the star formation rate SFR through the Kennicutt relation
%$$SFR(M_{\odot}/yr) =1.26 \times  10^{41}   L_{H\alpha}   (erg/s)$$ \citep{kennicutt1998global}. The SFR along with the stellar mass calculated in \cite{cardamone2009galaxy} will allow us to estimate the specific star formation rate (sSFR).

\begin{figure}[h!]
  \resizebox{\hsize}{!}{\includegraphics{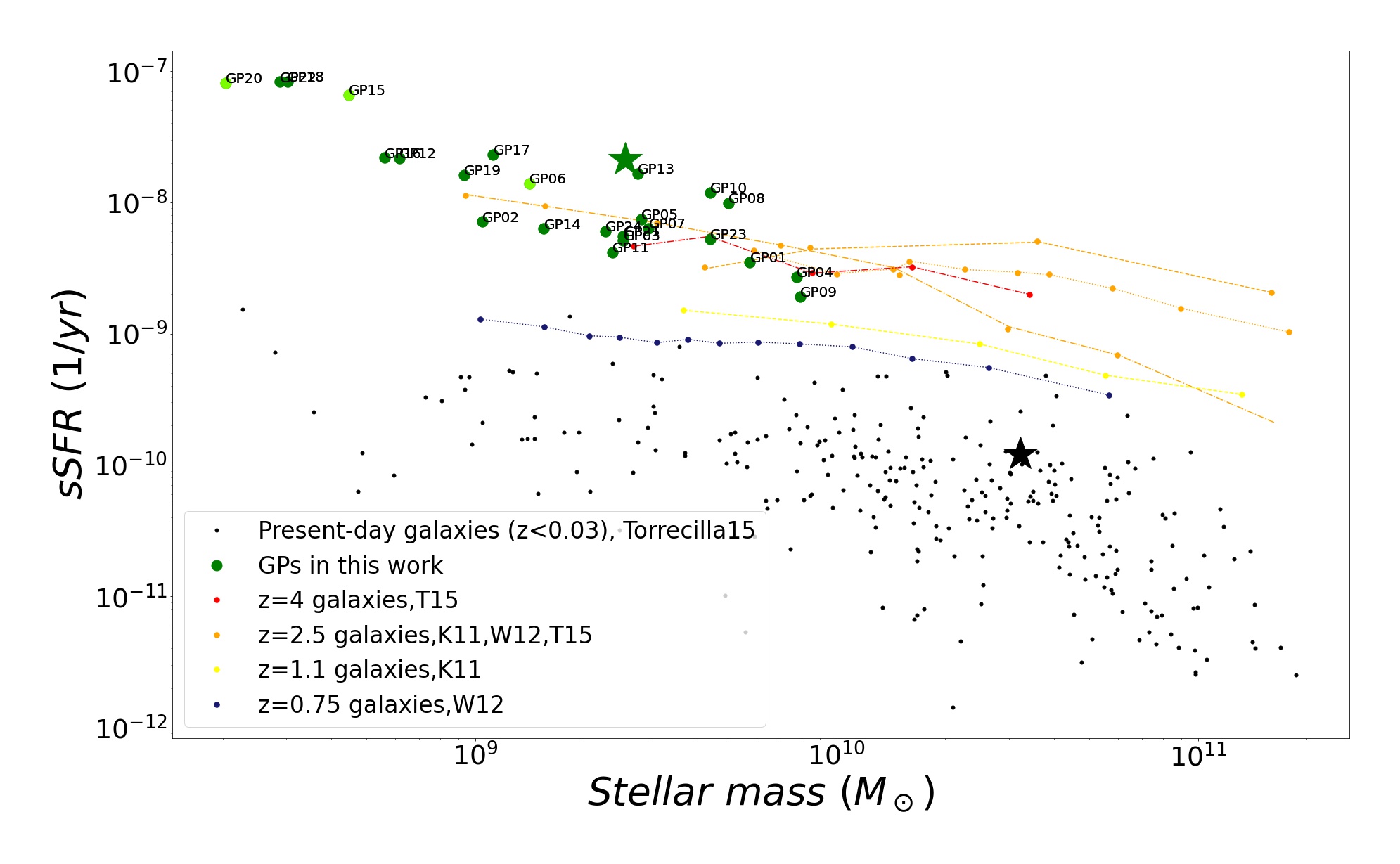}}
  \caption{Stellar mass versus sSFR. Green points are the GPs presented in this work, light green are GPs with HeII emission. The green star represents the median of the GPs, corresponding to a stellar mass of $2.6\times 10^9 \ \mathrm{M_\odot}$, an sSFR of $12 \ \mathrm{Gyr^{-1}} $ , and thus a mass-doubling timescale of $47 \ \mathrm{Myr}$. Black points represent a set of local galaxies (z<0.03) from the Califa survey \citep{catalan2015star} (Torrecilla15). The black star represents the median of this sample, it corresponds to a stellar mass of $3.2\times 10^{10} \ \mathrm{M_\odot}$, a sSFR of $0.12 \ \mathrm{Gyr^{-1}} $ and thus a mass-doubling timescale of $8.3 \ \mathrm{Gyr}$, which is on the order of the age of the Universe. We also show the sSFR-mass relation at a variety of redshifts from \cite{tasca2015evolving} (dash-dot line, T15), \cite{karim2011star} (dashed line, K11), and \cite{whitaker2012star} (dotted line, W12).}
  \label{SFR}
\end{figure}

In Fig. \ref{SFR}, we display sSFR versus stellar mass. Here, GPs occupy a different region in the diagram than present-day galaxies (galaxies at z<0.05) \citep{catalan2015star}. GPs have lower masses and  much higher sSFRs. In fact, GPs share the same space in this diagram with high-redshift galaxies $\mathrm{z} = 1.1-4$.
The mass-doubling timescale of GPs is on average 2 dex lower than that of present-day galaxies, reaching down to $\sim 15 \ \mathrm{Myr}$ for GP20, GP22, GP18, and GP15.

The depletion timescale, which indicates the starburst duration, is on the order of the mass-doubling timescale (which is the inverse of the sSFR) only if the mass of available hydrogen for creating new stars ($\mathrm{M}_{HII}$) is on the order of the $\mathrm{M}_\star$. Nevertheless, if $\mathrm{M}_{HII} > \mathrm{M}_\star$ the starburst duration is larger than the mass-doubling timescale. If we consider that $\mathrm{M}_{HII} \simeq \mathrm{M}_\star$, this could indicate 
 that GPs are short-lived events, since they will not be able to support such an incredibly high SFR for a long time. 
The mechanisms that would stop the SFR are mainly the exhaustion of the gas that fuels star formation and the stellar feedback via supernovae \citep{amorin2012complex}.
Observationally, we have very little information on whether GPs
will immediately quench or not.  However, if they continue forming stars,
they would quite rapidly build stellar mass and increase the stellar
luminosity in the blue, and the EW of $[O\textsc{iii}]5007\AA$ will decrease.
Thus, they would no longer be selected as GPs, given the effective
criterion of having high EW in $[O\textsc{iii}]5007\AA$. Using this kind of argument one could state that they are the
analogs of the early phases of galaxies that reionized the Universe.

The GP15 galaxy is identified as an LyC leaker exhibiting an escape fraction for LyC, $f_{esc}(LyC) = 0.08$ \citep{izotov2016detection} and for Ly$\alpha$, $f_{esc}(Ly\alpha) = 0.51$ \citep{jaskot2017kinematics}. This aligns with the understanding that the most intense starburst events typically generate a higher quantity of ionizing photons. Furthermore, this galaxy present among the most extremes line ratios favoring escape of ionizing photons (high $[O\textsc{iii}]$ / $[O\textsc{ii}]$ and low $[S\textsc{ii}]/H\alpha$ and $[N\textsc{ii}]/H\alpha$) see Table. \ref{table:gas_properties}.

%una columna

\subsection{Electron density, temperature, and abundances of the ionized gas}

In this section, we discuss how we derived chemical abundances of the ionized gas and electron temperatures ($T_e$) and densities ($n_e$). The emission line $[O\textsc{iii}]4363\AA$ is essential to calculate electron temperature and hence abundances. There are six GPs (out of the 24 GPs analyzed in this work) in which this line can be detected (i.e., the line is above the three-sigma detection limit).
For this subset of galaxies, we derived the $T_e$ values of $[O\textsc{iii}]$ using the $[O\textsc{iii}]\lambda4363/[O\textsc{iii}]\lambda4959,5007$ line ratio
and the values of $T_e$ corresponding to $[O\textsc{ii}]$ from the empirical relation between $[O\textsc{ii}]$ and $[O\textsc{iii}]$ electron temperatures given by \cite{campbell}.
We obtained the electron densities, from the $[S\textsc{ii}]\lambda6716/[S\textsc{ii}]\lambda6731$ line ratio.
The oxygen ionic abundance ratios, $O^+ /H^+$ and $O^{2+} /H^+$ , were derived from the $[O\textsc{ii}]\lambda3727$ and $[O\textsc{iii}]\lambda\lambda 4959,5007$ lines, respectively, using the corresponding electron temperatures.
The total oxygen abundance is assumed to be $O/H = O^+ /H^+ + O^{2+} /H^+$. 
The nitrogen ionic abundance ratio, $N^+ /H^+$ , was calculated 
using the $[N\textsc{ii}]\lambda6584$ emission line and assuming $T_e [N\textsc{ii}] \sim T_e [O\textsc{ii}]$; the $N/O$ abundance ratio was computed under the assumption that $N/O = N^+/O^+$, based on the similarity of the ionization potentials of the ions involved.
All this was computed by implementing the Pyneb code \citep{luridiana2015pyneb}.
We calculated the final errors in the derived quantities by error propagation and taking into account errors in flux measurements.
Furthermore, for the subset of galaxies that do not present the $[O\textsc{iii}]4363\AA$ line, we used the HII-CHI-Mistry code \citep{perez2014code}, which calculates oxygen and nitrogen over oxygen abundances (and errors) without this line.
The values corresponding to all these gas properties are shown in Table \ref{table:gas_properties}.

\begin{figure}[h!]
   \resizebox{\hsize}{!}{\includegraphics{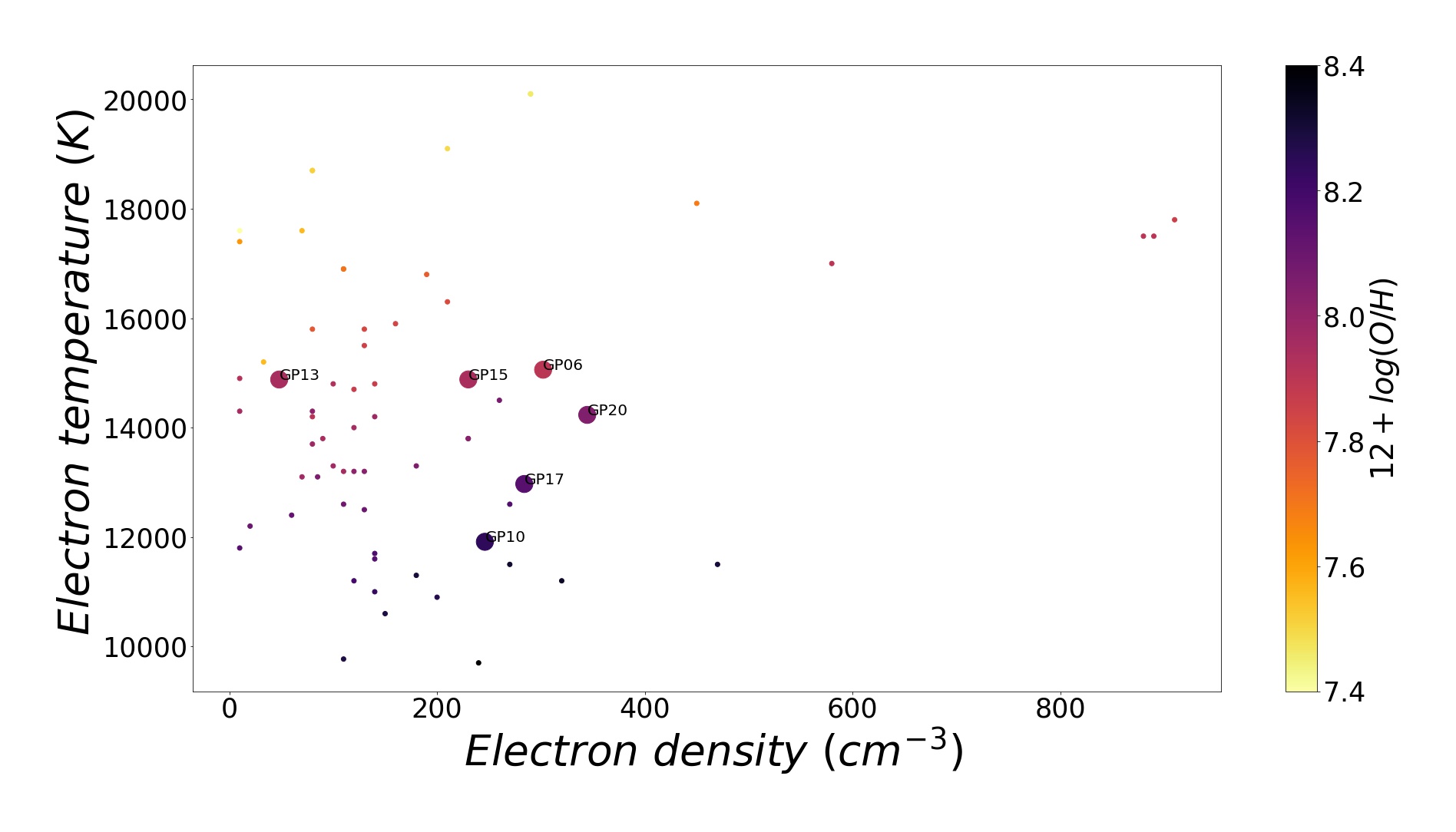}}
  \caption{Electron density, electron temperature, and metallicity representation. Big labeled dots correspond to the six GPs where the direct method can be well applied. The rest of the small dots correspond to all galaxies selected from the NASA-Sloan Atlas with W($\lambda$5007) > 1000$\AA$ and 37 COS Legacy Archive Spectroscopic SurveY galaxies.}
  \label{Electrones}
\end{figure}

In Fig. \ref{Electrones}, we can see the electron density, electron temperature, and metallicity for the six GPs with the $[O\textsc{iii}]4363\AA$ line measured. 
The same properties are also represented for 35
galaxies selected from the NASA-Sloan Atlas \footnote{http://www.nsatlas.org/} with W($\lambda$5007) > 1000$\AA$ and 37 galaxies from the COS Legacy Archive Spectroscopic SurveY \citep{berg2022cos} galaxies, which are the closest local analogs (z<0.18) of high-redshift galaxies in the epoch of reionization. The derivation of gas parameters in these galaxies was done by \cite{peng2021}.
The metallicity of our set of GPs is low and in agreement with previous results (i.e., $12+\log(O/H) = 7.6-8.4$) \citep[e.g.,][]{amorin2010oxygen}. For this subset of GPs with $[O\textsc{iii}]4363\AA$ emission, the electron temperature ranges from $11500 \ \mathrm{K}$ to $15500 \ \mathrm{K}$ and the electron density ranges from $30 \ \mathrm{cm^{-3}}$ to $400 \ \mathrm{cm^{-3}}$.
%Lower electron density and higher electron temperature correspond to lower metallicities, this result is consistent with the theory (photoionization models and empiric result).

%una columna

\begin{figure}[h!]
  \resizebox{\hsize}{!}{\includegraphics{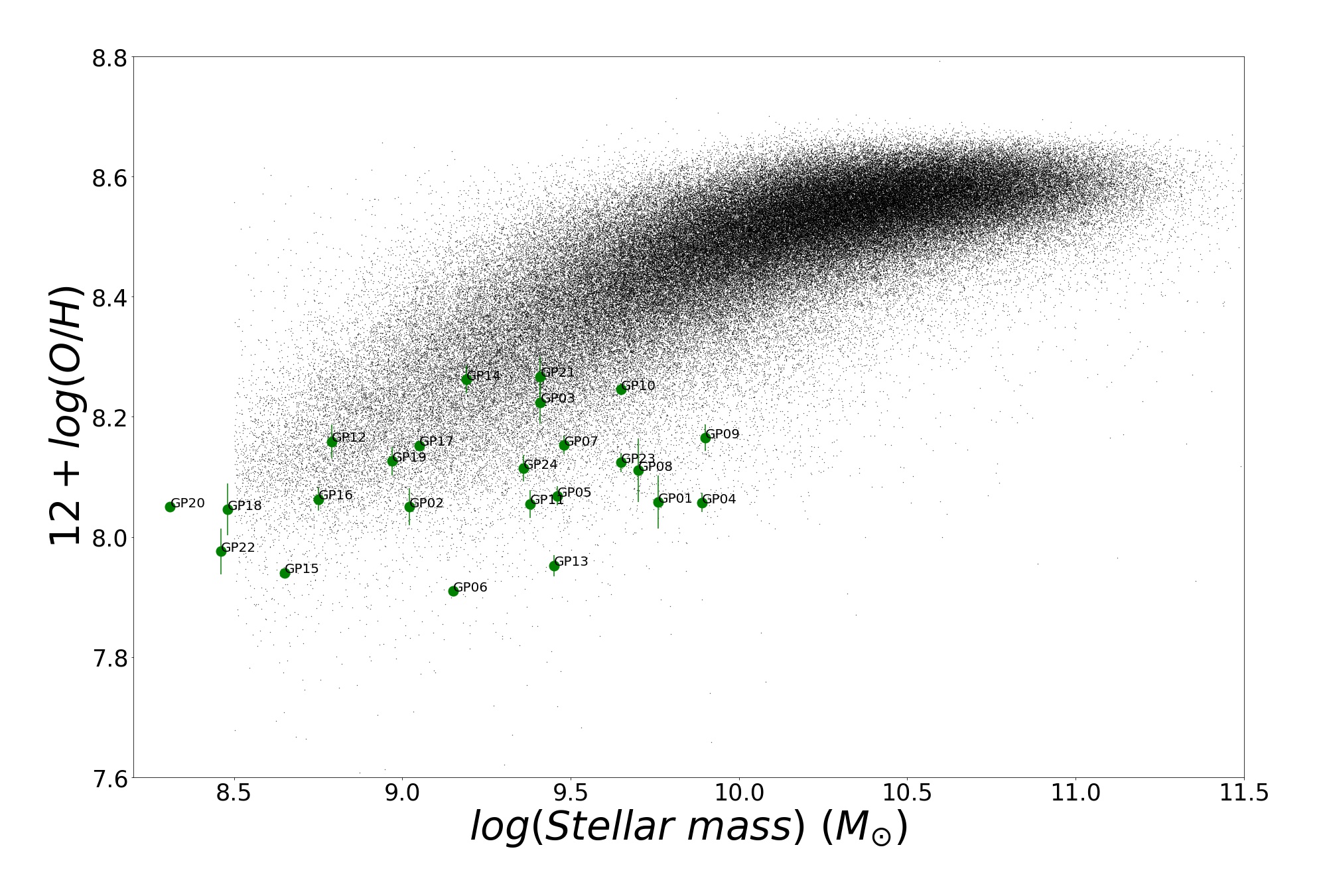}}
  \caption{Oxygen abundance versus stellar mass. GPs are the green points. Black points correspond to $\sim 200000$ galaxies from \cite{Salva}.} 
  \label{Metalicidad}
\end{figure}

\begin{figure}[h!]
  \resizebox{\hsize}{!}{\includegraphics{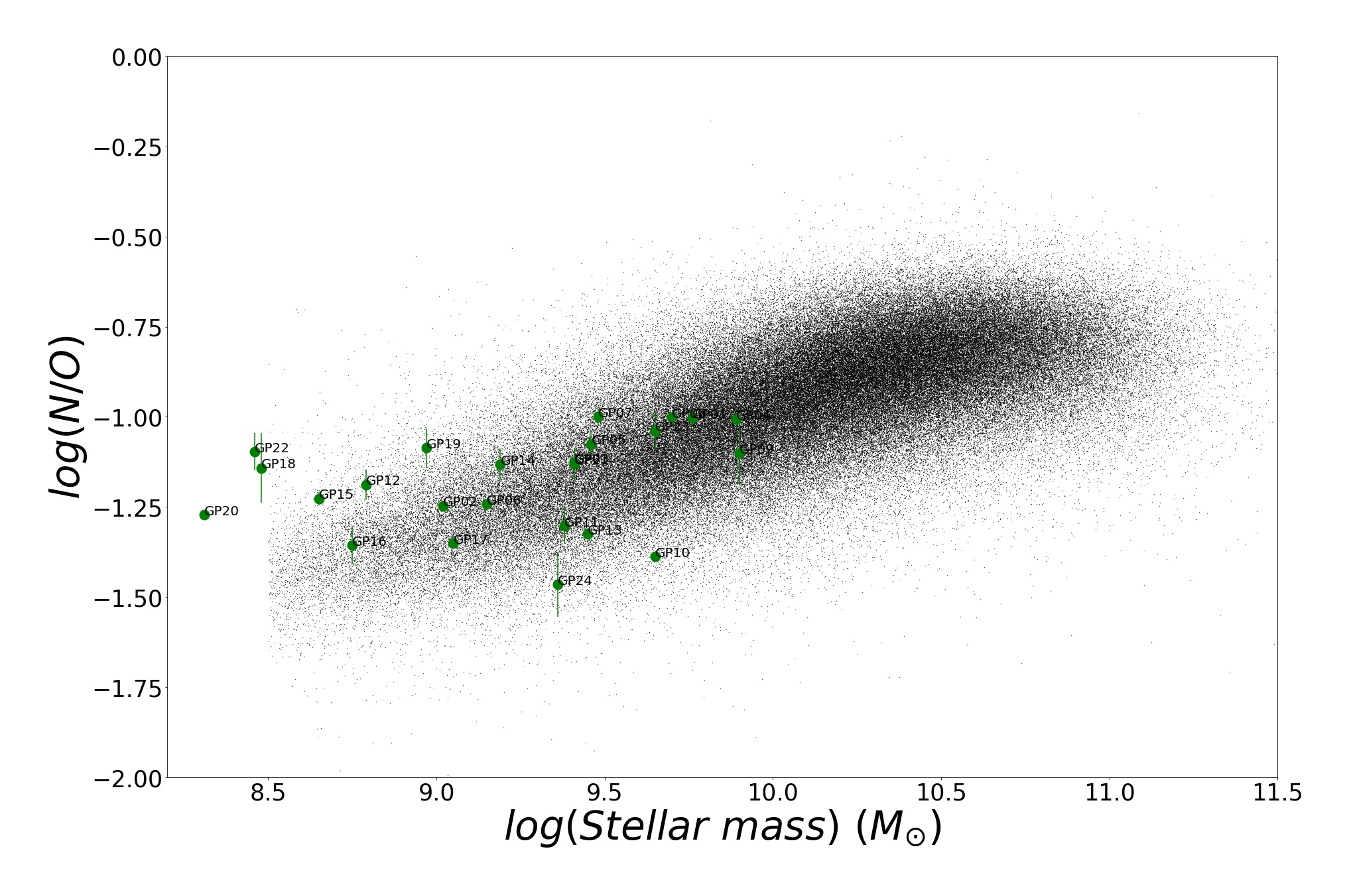}}
  \caption{Nitrogen over oxygen versus stellar mass. GPs are the green points. Black points correspond to $\sim 200000$ galaxies from \cite{Salva}.}
  \label{Nitrogeno}
\end{figure}

\begin{figure}[h!]
  \resizebox{\hsize}{!}{\includegraphics{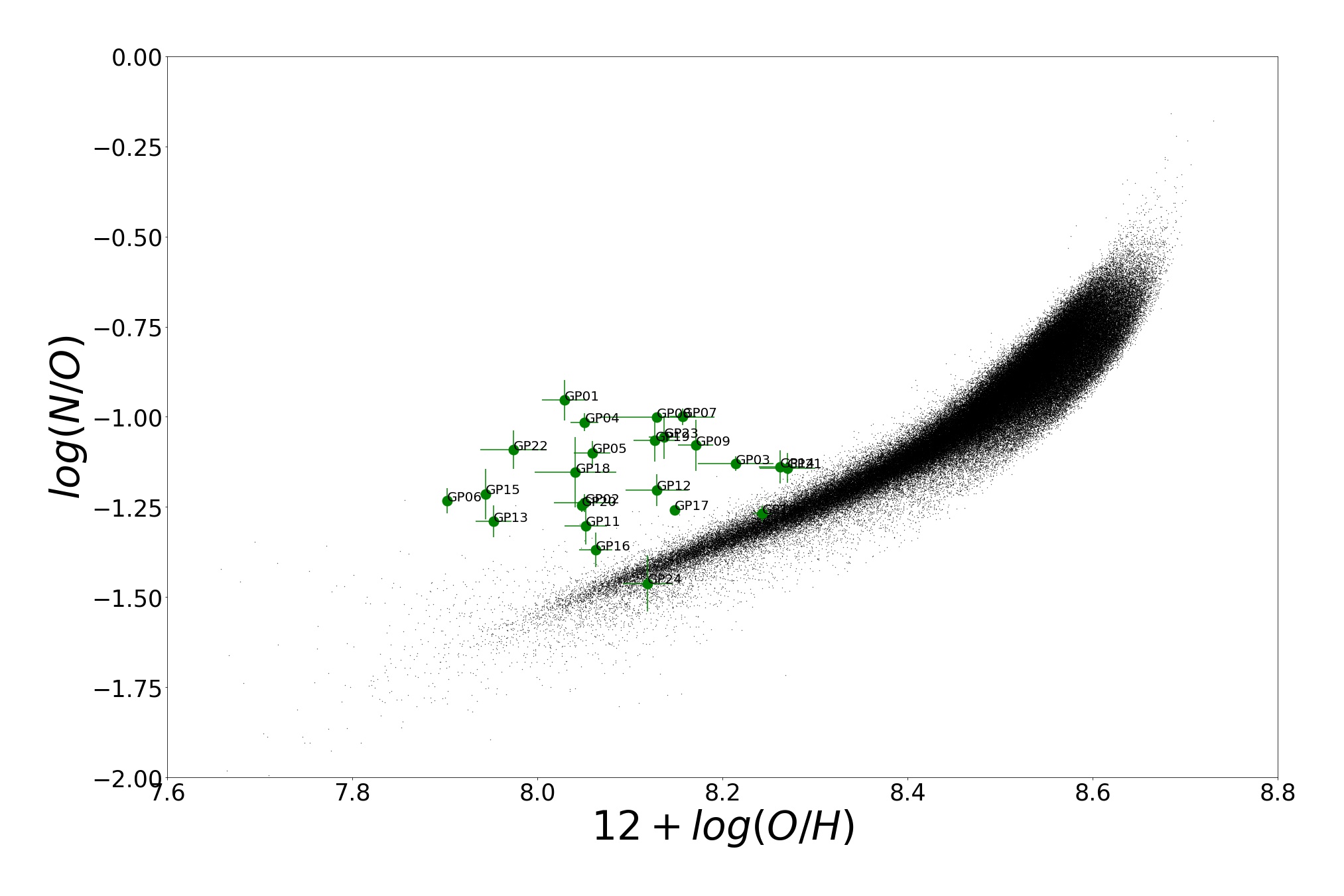}}
  \caption{Nitrogen over oxygen versus oxygen abundance. GPs are the green points. Black points correspond to $\sim 200000$ galaxies from \cite{Salva}.}
  \label{Oxigeno y nitrogeno}
\end{figure}

Figures \ref{Metalicidad}, \ref{Nitrogeno}, $\&$ \ref{Oxigeno y nitrogeno} display the  oxygen abundance versus stellar mass, nitrogen over oxygen versus stellar mass, and nitrogen over oxygen versus oxygen abundance, respectively. 
A comparison is made between $\sim 200000$ star-forming galaxies (SFGs) from SDSS \citep{Salva} and the GPs. 
For a given stellar mass, GPs show lower oxygen abundances than the bulk of the SDSS galaxies. This is in accordance with the result from \cite{amorin2010oxygen}, where GPs follow a relation between mass and metallicity parallel to the one defined by the SDSS SFGs, but it is offset by $\sim 0.3$ dex to lower metallicities. 
The nitrogen over oxygen (N/O) ratio (Fig. \ref{Nitrogeno}) of GPs follows the tendency of SDSS galaxies, except for the low-mass end, where GPs present a higher N/O (see GP22, GP18 and GP20 in Fig. \ref{Nitrogeno}). This ratio ranges from $\log(N/O)=-1.5,-0.85$.

Despite the lack of metals in GP galaxies, the metal ratio (N/O) is mostly conserved. One scenario that can explain this would be a massive and recent inflow of metal-poor gas (basically neutral hydrogen clouds) from the HI halo/reservoir of the galaxy. This accretion could dilute the O/H keeping the N/O unaltered \citep{koppen2005effects}. The fact that their N/O is in most cases the expected for the stellar masses of these galaxies supports this scenario \citep{amorin2010oxygen}.

The proposed scenario of a massive inflow of pristine gas that both sustains low metallicity and increases the SFR is plausible. However, this raises a complex issue regarding the presence of such a process at low redshifts. The origins of this unenriched gas remain an area of inquiry. Nonetheless, empirical data provides insight with evidence from extremely metal-poor galaxies proximal to us (z=0.03) that exhibit substantial amounts of neutral hydrogen, resembling a halo \citep[e.g.,][]{lequeux1980hi,2023A&A...670A.121H}. This suggests that similar circumstances could exist for GPs, thereby supporting the proposed scenario.

Chemical evolution model predictions \citep[e.g.,][]{molla2006nitrogen,vincenzo2016nitrogen} suggest that the transition between primary and secondary N dominance in the N/O versus O/H plane depends on the SF history, particularly on the star formation efficiency (SFE). A "bursty" galaxy, that is, one having experienced a very recent starburst, will see a quicker increase in N/O than a galaxy with a smoother and longer SF history. Another factor that could elevate the N/O at low metallicity is the initial mass function (IMF), where a higher fraction of massive stars could enhance primary N production at low metallicity. However, we observe that the N/O - stellar mass relation generally holds. For galaxies deviating from this relation (see GP22, GP18, and GP20 in Fig. \ref{Nitrogeno}), adjusting the IMF could be a viable solution.

%At least HST or JWST images would be needed in order to confirm this.
%GPs are metal poor galaxies, with metallicities between 7.6-8.4. Nevertheless, their metallicity rarely is color 7.6 and this is because of the high $[O\textsc{iii}]5007\AA / H\alpha$ ratio. According to \citep{kojima2020extremely} extremely metal poor galaxies have a low $[O\textsc{iii}]5007\AA / H\alpha$ ratio, with is usually color 1. GPs have among the highest $[O\textsc{iii}]5007\AA / H\alpha$ ratios where the typical metallicity for this ratios is 8.

%The nitrogen over oxygen ratio (Fig. \ref{Nitrogeno}) of GPs follows the tendency of SDSS galaxies. So despite of the lack of metals in GP galaxies the metal ratio is conserved. One scenario that can explain this would be a massive inflow of pristine gas in to the galaxy \citep{amorin2010oxygen}.

%una columna

%una columna

%una columna

\section{Summary and conclusions}
\label{seccion_conclusiones}

We have presented physical and chemical properties of 24 GPs using MUSE/VLT data cubes. These galaxies are one of the best local analogs of high-redshift galaxies. Thus, their study is fundamental in order to understand the very first epoch of the formation and assembling of galaxies, and in particular the reionization.

For our set of GPs, we wanted to confirm the spatial extension of these sources. To do so, a study regarding the extension of all the sources within the FoV of the MUSE data cube was carried out. We established a criterion to determine whether a GP is resolved based on the comparison between the FWHM, $\mathrm{FW}\frac{1}{10}\mathrm{M,}$ and $\mathrm{FW}\frac{1}{100}\mathrm{M}$ of a stellar-like object and the GP itself. Retrieving seven spatially extended GPs in the core and the low surface brightness region, five GPs extended in the core, and four GPs extended in the low surface brightness region (see Table \ref{table:extension}).

We compare the emission line maps and the continuum maps. The only four emission line maps with no circular symmetry are the ones corresponding to GP06, GP07, GP13, and, marginally, GP01. Continuum maps present a much richer structure for all GPs that are proxies for the stellar underlying population. $[O\textsc{iii}]5007$ maps are the most extended ones, tracing the low surface brightness regions of ionized gas.

Regarding the ionization structure, $H\alpha / H\beta$ maps trace dusty regions and zones with low extinction where photons can travel without being absorbed; these maps present very different morphologies for the different GPs. 
The ionization parameter (as traced by $[O\textsc{iii}]/[O\textsc{ii}]$) tends to peak in the center of the galaxies, indicating that the highest ionization is near the star-forming region. GP20, GP06, GP15, and GP22 present the strongest ionization (see Table \ref{table:gas_properties} and Figs. \ref{GP15_color} and \ref{GP22_color}). The
$[O\textsc{iii}]/H\beta$, $[N\textsc{ii}]/H\alpha$, $[S\textsc{ii}]/H\alpha,$ and $[O\textsc{i}]/H\alpha$ ratios are studied in maps and in the BPT diagrams. These indicate  a tendency for higher excitation of the gas in the center of the galaxy (i.e., higher ratios of $[O\textsc{iii}]/H\beta$ and lower ratios of $[S\textsc{ii}]/H\alpha$ and $[O\textsc{i}]/H\alpha$ close to the $H\alpha$ peak). 
Still, the $[O\textsc{iii}]/H\beta$ and $[N\textsc{ii}]/H\alpha$ ratios do not present much spatial variation (with a maximum difference of 0.14 dex in all GPs, except for GP06 reaching over 0.4 dex), indicating uniformity in the gas excitation and metallicity. 
$[S\textsc{ii}]/H\alpha$ and $[O\textsc{i}]/H\alpha$ trace the boundaries of the ionized gas, and they present their lower values close to the center of the galaxies, suggesting a blister-type morphology (e.g., GP06, GP08, GP10, GP13, and GP23).

The BPT diagrams indicate hot massive stars are confirmed as the main source of ionizing photons. In particular, our GPs are located in the top left part of the diagram, where the most extreme galaxies reside (i.e., those with the lowest metallicity and highest excitation of the ionized gas). The absence of the $[FeX]\lambda 6374$ in all the spectra discards BHs with masses $>>10^5 M_\odot$  contributing to ionizing the gas.

We also produced the integrated spectrum of each GP by integrating the flux of a region defined by the $H\alpha$ map. 
The SFRs derived from the luminosity of the $H\alpha$ line indicate bursts of star-formation with mass-doubling timescales 2 dex lower than common star-forming galaxies.
Our study of the ionized gas properties using emission lines indicates low gas metallicity (i.e., $12+\log(O/H) = 7.6-8.4$), high electron temperatures ranging from $11500 \ \mathrm{K}$ to $15500 \ \mathrm{K,}$ and electron densities ranging from $30 \ \mathrm{cm^{-3}}$ to $530 \ \mathrm{cm^{-3}}$.
The nitrogen over oxygen ratio versus stellar mass of the GPs (see Fig. \ref{Nitrogeno}) generally follows the tendency of SDSS galaxies and ranges between $\log(N/O)=-1.5$ and $-0.85$, whereas these are clearly above the SDSS sequence in the $\log(N/O)$ versus $12 + \log(O/H)$ diagram (see Fig. \ref{Oxigeno y nitrogeno}), which possibly indicates the presence of an inflow of pristine gas into the galaxies.

We detected the nebular $HeII\lambda4686$ line in the galaxies GP06, GP15 (this particular galaxy being a confirmed LyC leaker \citep{izotov2016detection}), and
GP20, indicating the presence of very hard ionizing photons (E > 4Ry). We
checked that none of these GPs show WR features in their
spectra, which suggests that WR stars are not the main HeII excitation
contributors \citep[e.g.,][]{senchyna2017ultraviolet,kehrig2015extended,kehrig2018extended}. We also
note that two of these GPs present among the highest sSFR (> $8\times 10^8 yr^{-1}$),
suggesting that, besides a low metallicity, high sSFR can be a dominant factor to
determine the HeII emitting nature of a galaxy \citep[][]{kehrig2020mapping,perez2020photon}. A detailed analysis of the origin of the HeII
ionization, which keeps challenging up-to-date stellar models \citep[see e.g.,][]{eldridge2022new}, is beyond the scope of this work and will be
investigated in future work.

%GPs are characterised by having a strong star-burst that is by far the main source of electromagnetic radiation in the galaxy. The massive stars that are in the core ionize a large amount of gas that extends out of the place where stars are born. We can see this by looking at the $[O\textsc{iii}]{\lambda}5007\AA$ emission line maps, where the light extends several Kpc out from the center of the galaxy. Nevertheless, for most galaxies this emission present a high level of radial symmetry (ie. there is a lack of structure). Still, images of the continuum radiation indicates the presence of a richer structure, possibly giving us information about the subyacent stellar population and the circumstellar medium.

%GPs have low metallicity ($12+log(O/H)\sim 8$), have among the highest sSFR measured in the Universe, they are close to purely star-ionizing objects with no presence of massive BH and his extinction is mild, indicating that there's not a big presence of dust. For 3 GPs there is HeII emission, with is an indicator of really high ionization sources (ie. presence of massive stars). We have not found in any GP any tracer of WR stars.

%$[OIII]/H\beta$ and $[OIII]/[OII]$ maps indicate that the central part of the galaxy tend to have a peak in ionization. Tracers maps of metallicity ($[N\textsc{ii}]/H\alpha$) indicates pretty low gradients and density-bounded tracers ($[S\textsc{ii}]/H\alpha$) indicates a blister-type morphology where photons can escape through holes close to the center of emission.

\begin{acknowledgements}
We thanks the referee for their constructive comments.
Author Antonio Arroyo Polonio acknowledges financial support from the grant CEX2021-001131-S funded by MCIN/AEI/ 10.13039/501100011033. RA acknowledges support from ANID FONDECYT Regular Grant 1202007.
\end{acknowledgements}

\bibliographystyle{aa}
\bibliography{main.bib}

\begin{appendix}

\section{Emission line maps}
\label{appendix:emission line maps}

%dos columnas
\begin{figure*}[h!]
\centering
   \includegraphics[width=17cm]{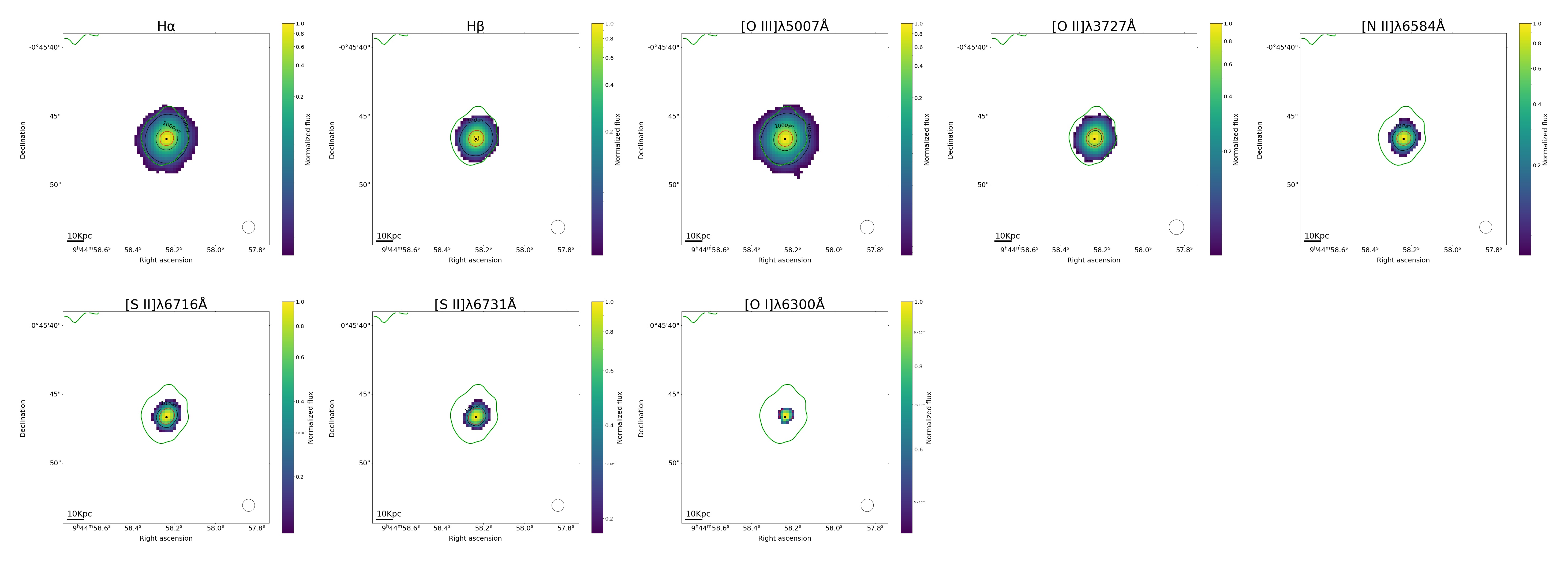}
     \caption{Emission line maps of GP01.}
     \label{1}
\end{figure*}

%dos columnas
\begin{figure*}[h!]
\centering
   \includegraphics[width=17cm]{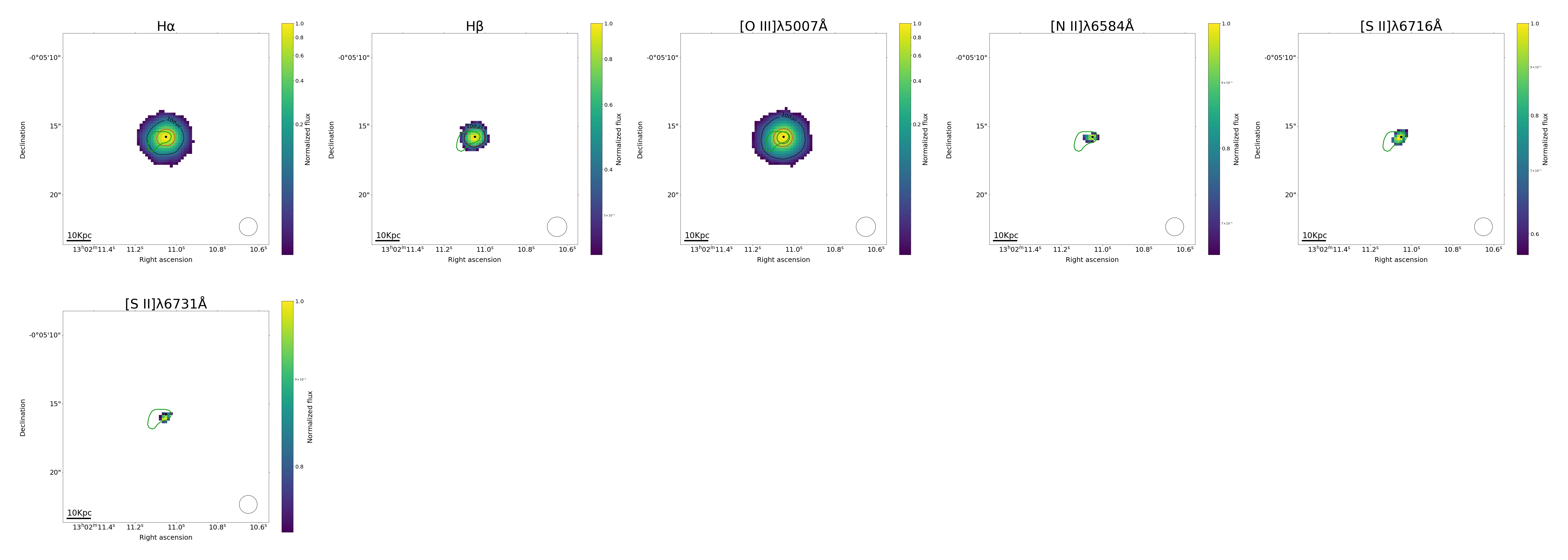}
     \caption{Emission line maps of GP02.}
     \label{hola}
\end{figure*}

%dos columnas
\begin{figure*}[h!]
\centering
   \includegraphics[width=17cm]{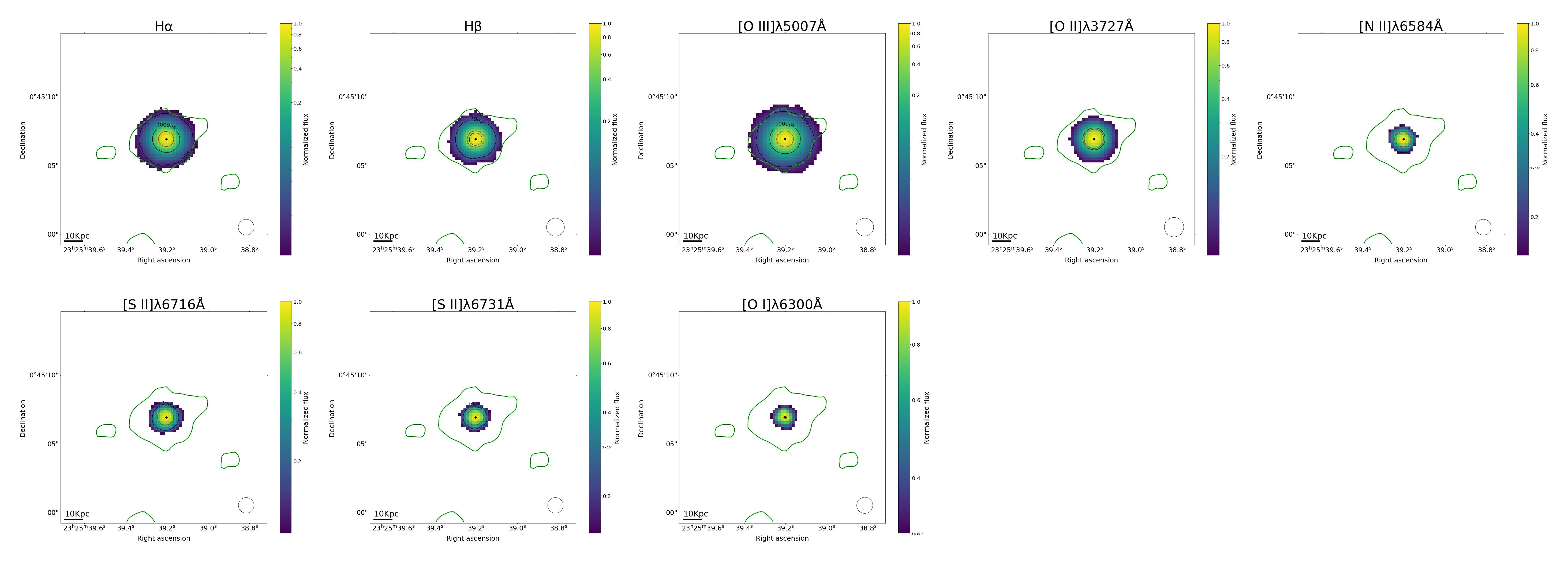}
     \caption{Emission line maps of GP03.}
     \label{hola}
\end{figure*}

%dos columnas
\begin{figure*}[h!]
\centering
   \includegraphics[width=17cm]{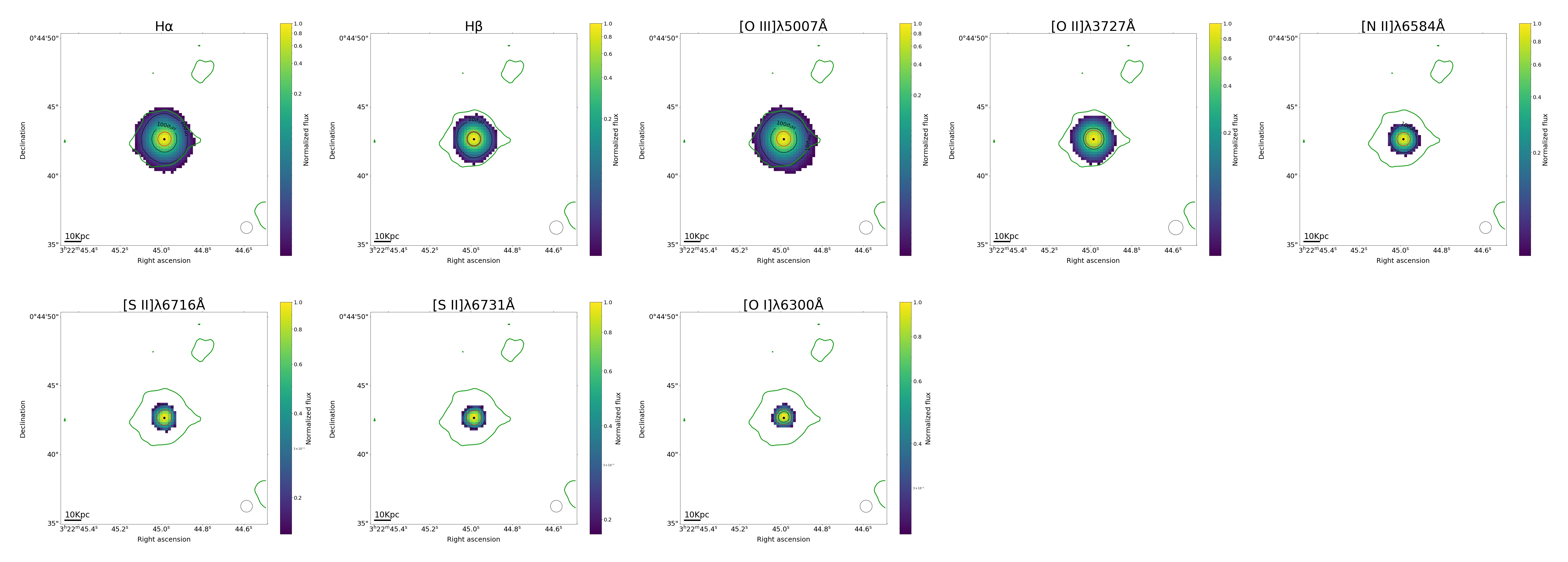}
     \caption{Emission line maps of GP04.}
     \label{hola}
\end{figure*}

%dos columnas
\begin{figure*}[h!]
\centering
   \includegraphics[width=17cm]{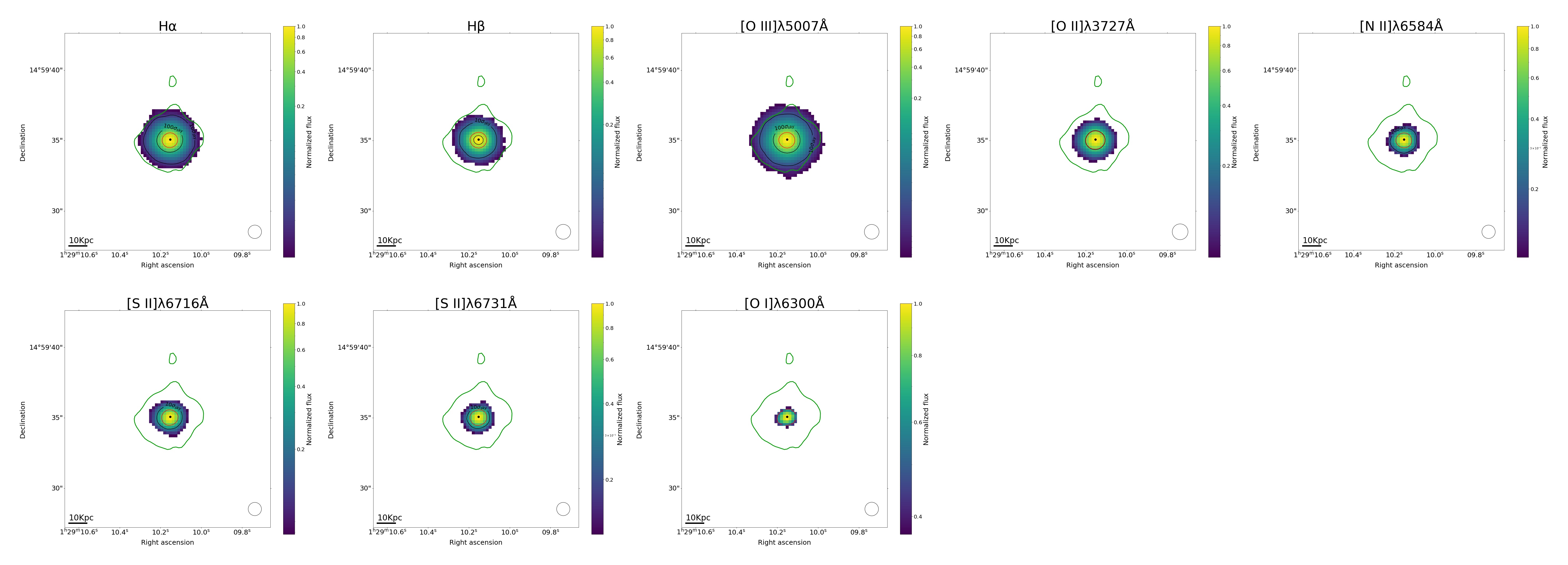}
     \caption{Emission line maps of GP05.}
     \label{hola}
\end{figure*}

%dos columnas
\begin{figure*}[h!]
\centering
   \includegraphics[width=17cm]{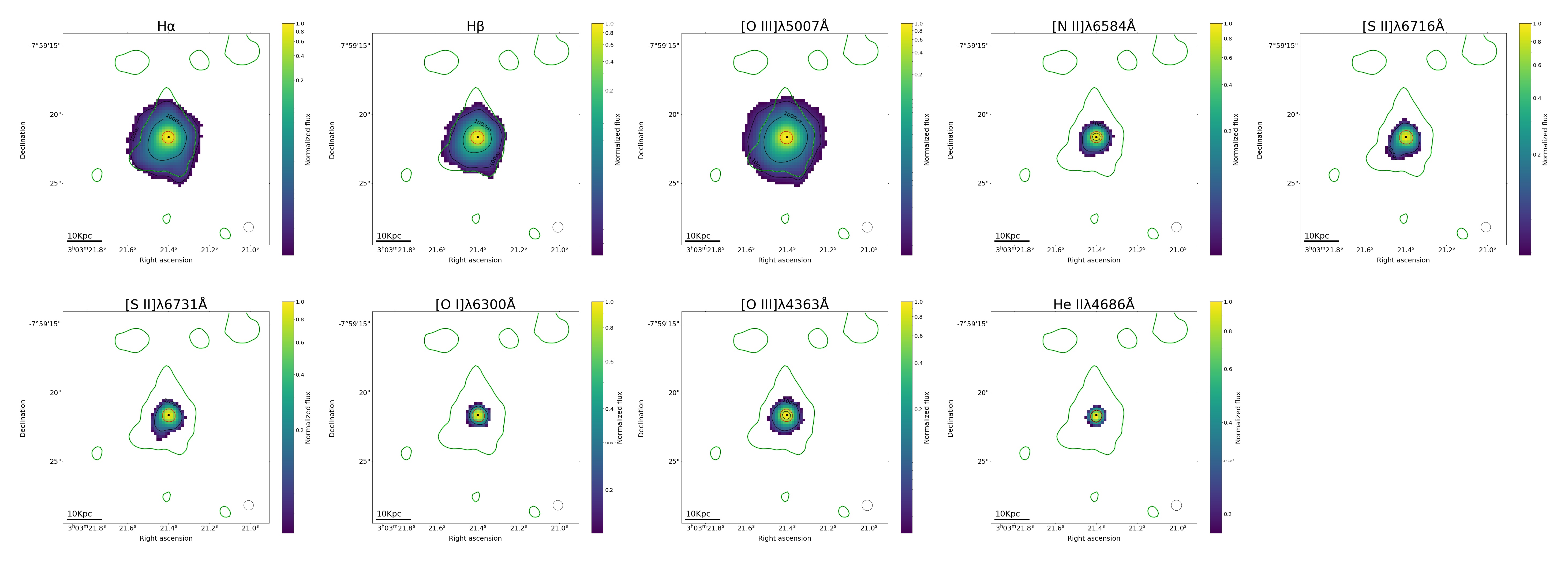}
     \caption{Emission line maps of GP06.}
     \label{06}
\end{figure*}

%dos columnas
\begin{figure*}[h!]
\centering
   \includegraphics[width=17cm]{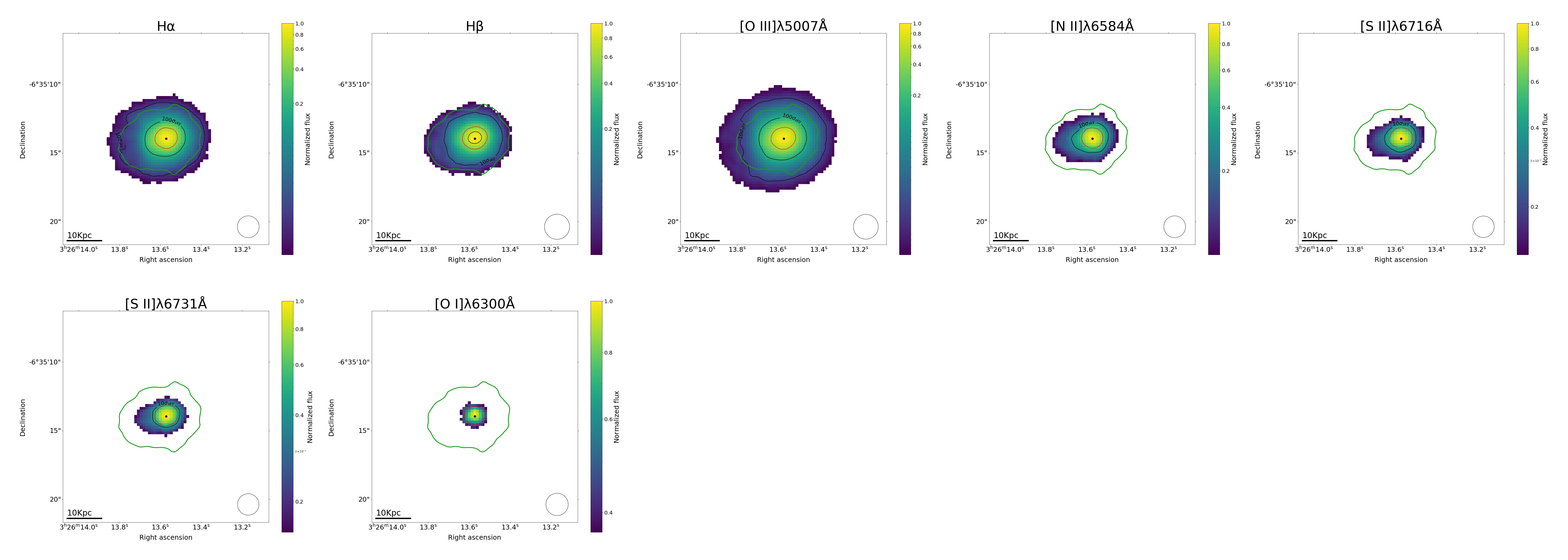}
     \caption{Emission line maps of GP07.}
     \label{07}
\end{figure*}

%dos columnas
\begin{figure*}[h!]
\centering
   \includegraphics[width=17cm]{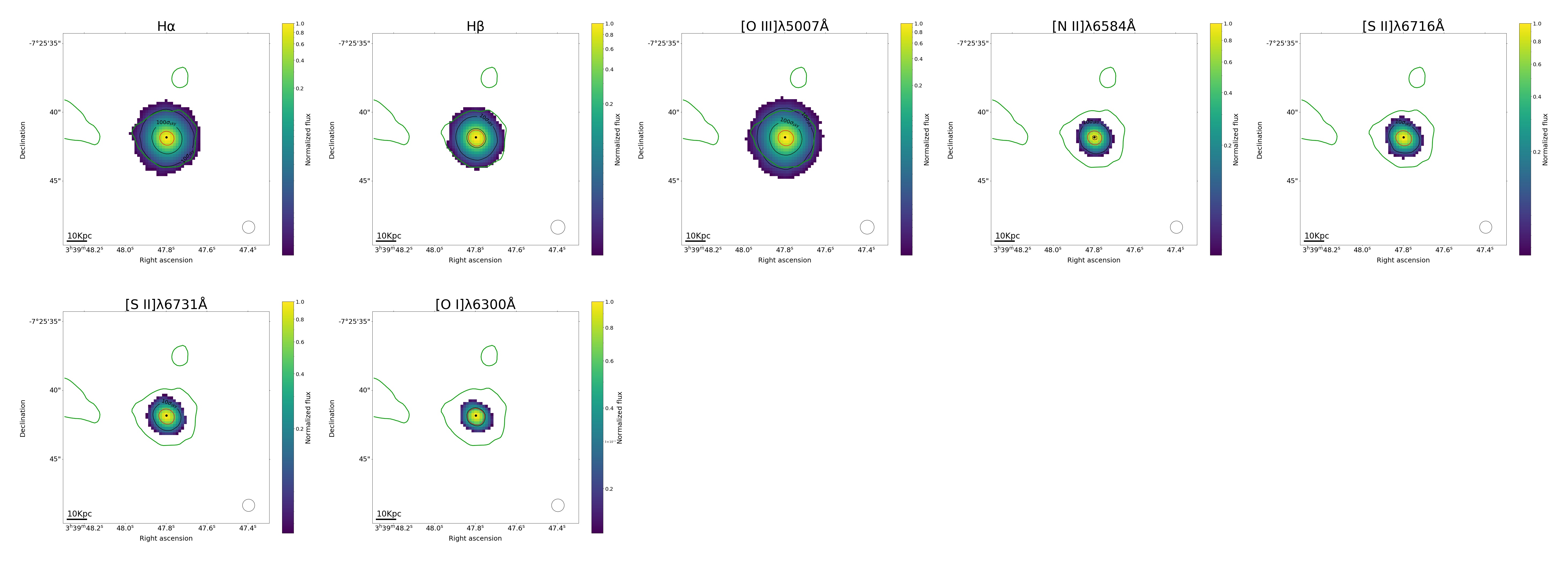}
     \caption{Emission line maps of GP08.}
     \label{hola}
\end{figure*}

%dos columnas
\begin{figure*}[h!]
\centering
   \includegraphics[width=17cm]{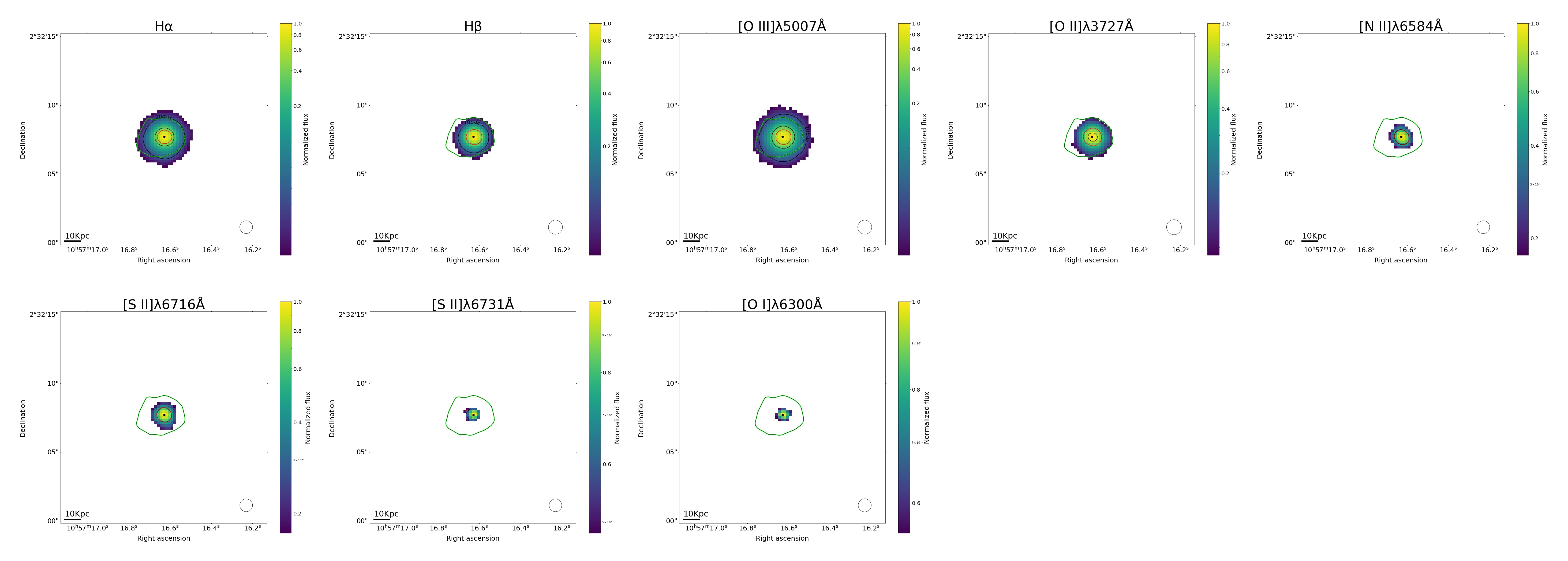}
     \caption{Emission line maps of GP09.}
     \label{hola}
\end{figure*}

%dos columnas
\begin{figure*}[h!]
\centering
   \includegraphics[width=17cm]{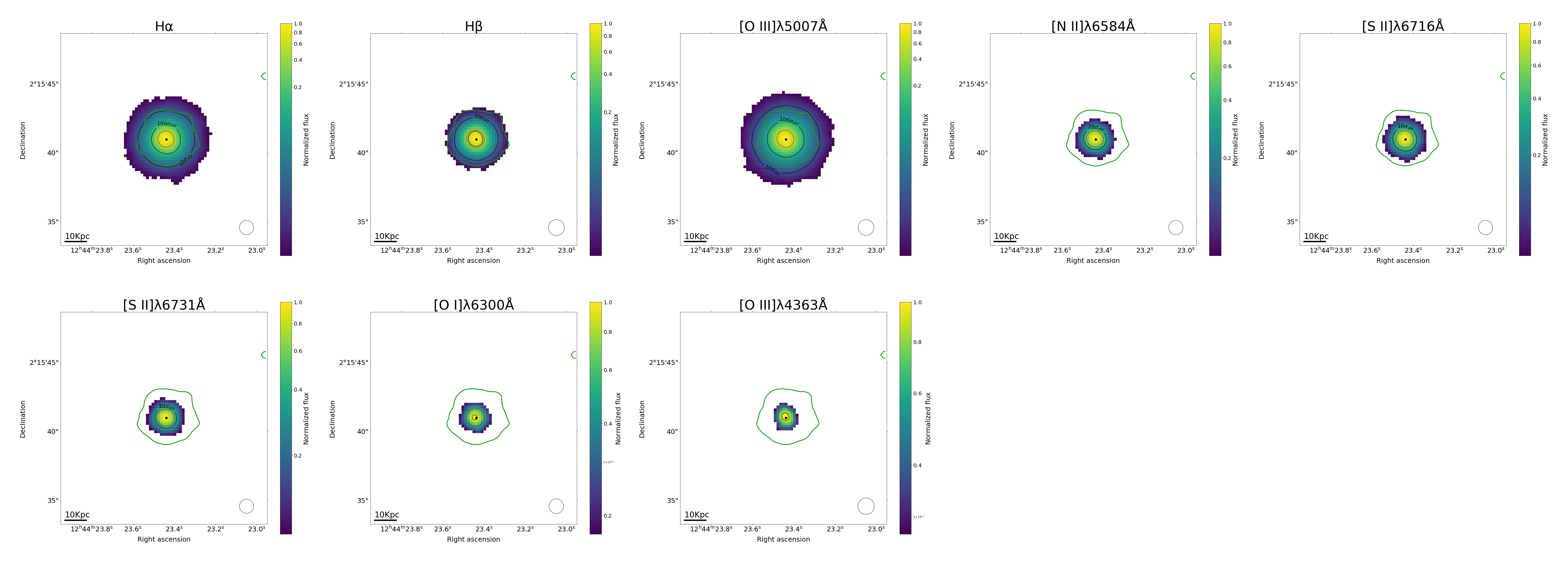}
     \caption{Emission line maps of GP10.}
     \label{hola}
\end{figure*}

%dos columnas
\begin{figure*}[h!]
\centering
   \includegraphics[width=17cm]{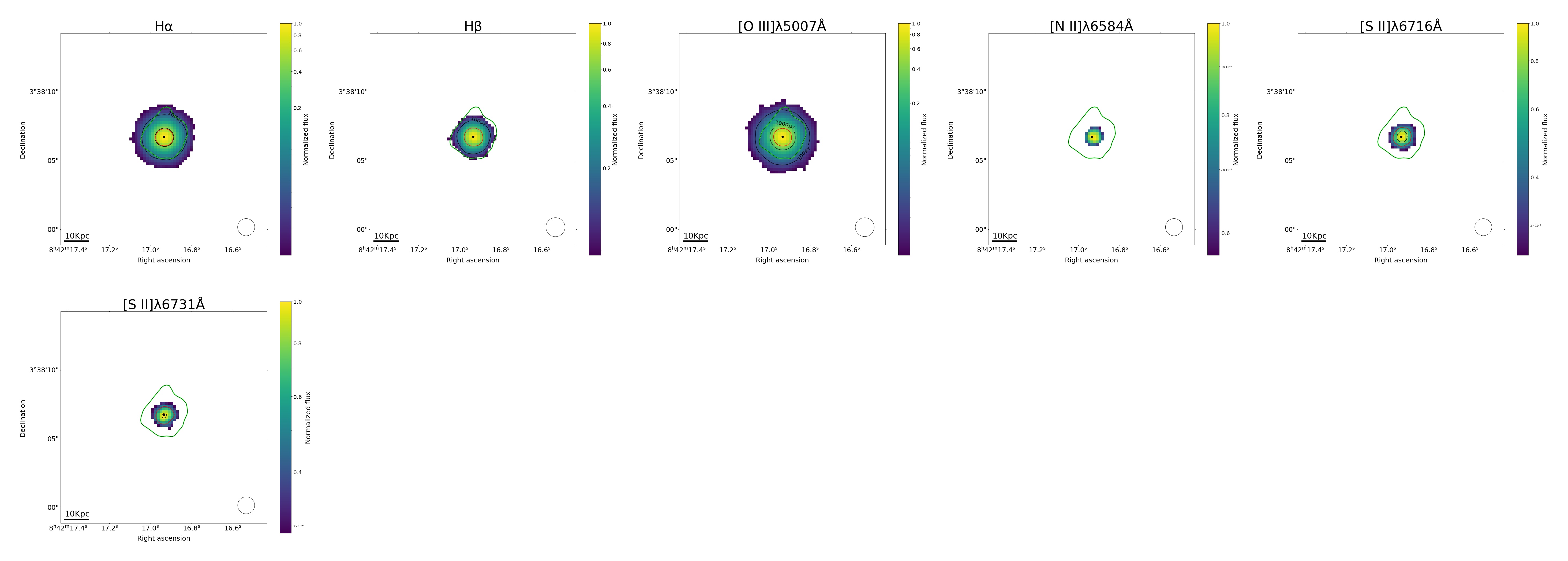}
     \caption{Emission line maps of GP11.}
     \label{hola}
\end{figure*}

%dos columnas
\begin{figure*}[h!]
\centering
   \includegraphics[width=17cm]{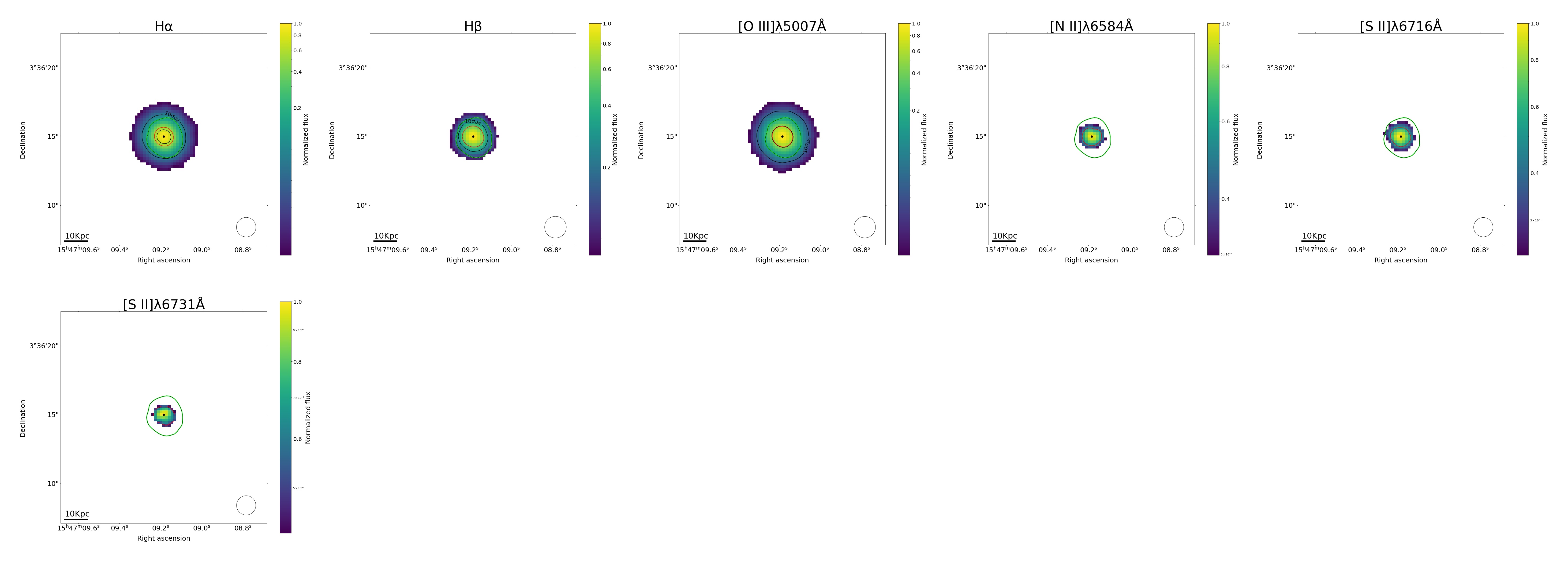}
     \caption{Emission line maps of GP12.}
     \label{hola}
\end{figure*}

%dos columnas
\begin{figure*}[h!]
\centering
   \includegraphics[width=17cm]{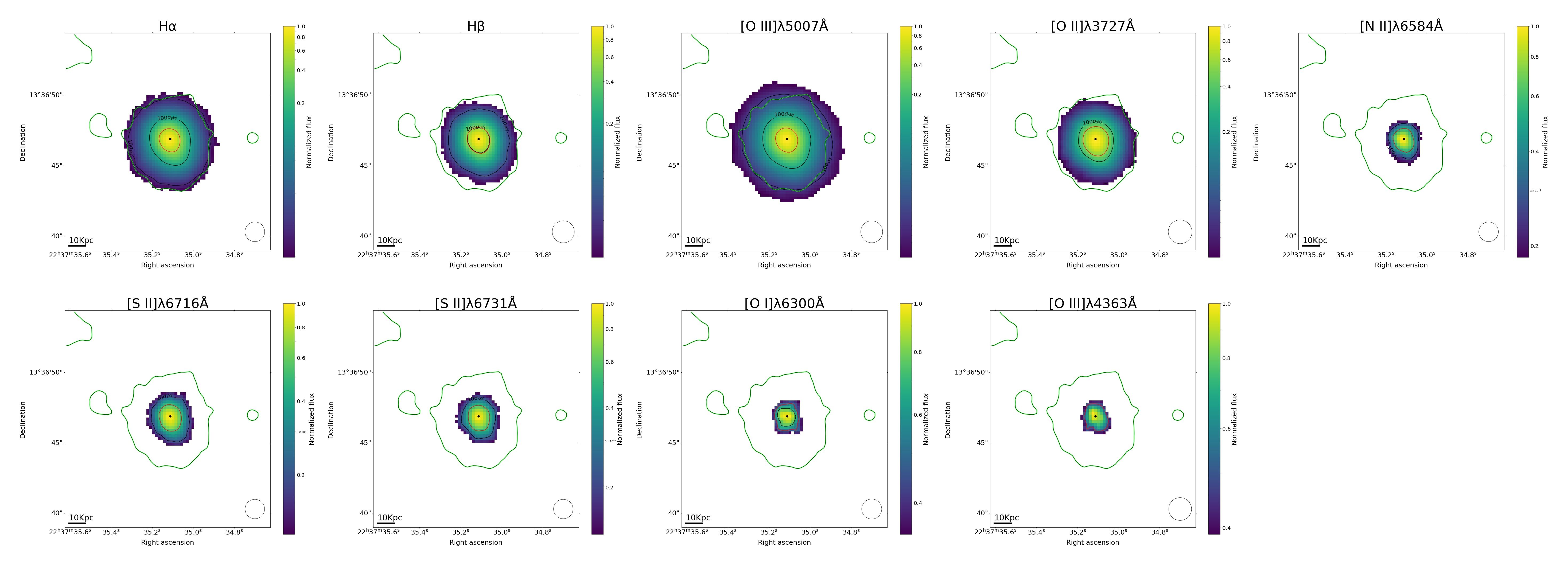}
     \caption{Emission line maps of GP13.}
     \label{13}
\end{figure*}

%dos columnas
\begin{figure*}[h!]
\centering
   \includegraphics[width=17cm]{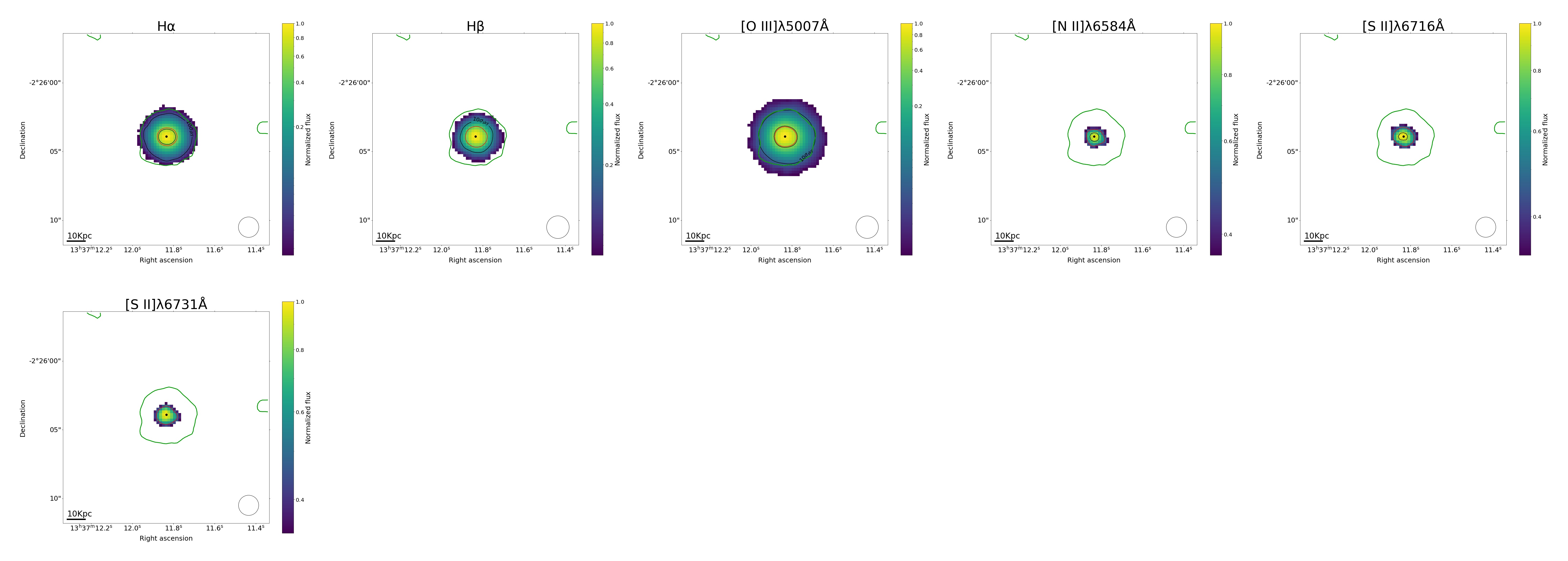}
     \caption{Emission line maps of GP14.}
     \label{hola}
\end{figure*}

%dos columnas
\begin{figure*}[h!]
\centering
   \includegraphics[width=17cm]{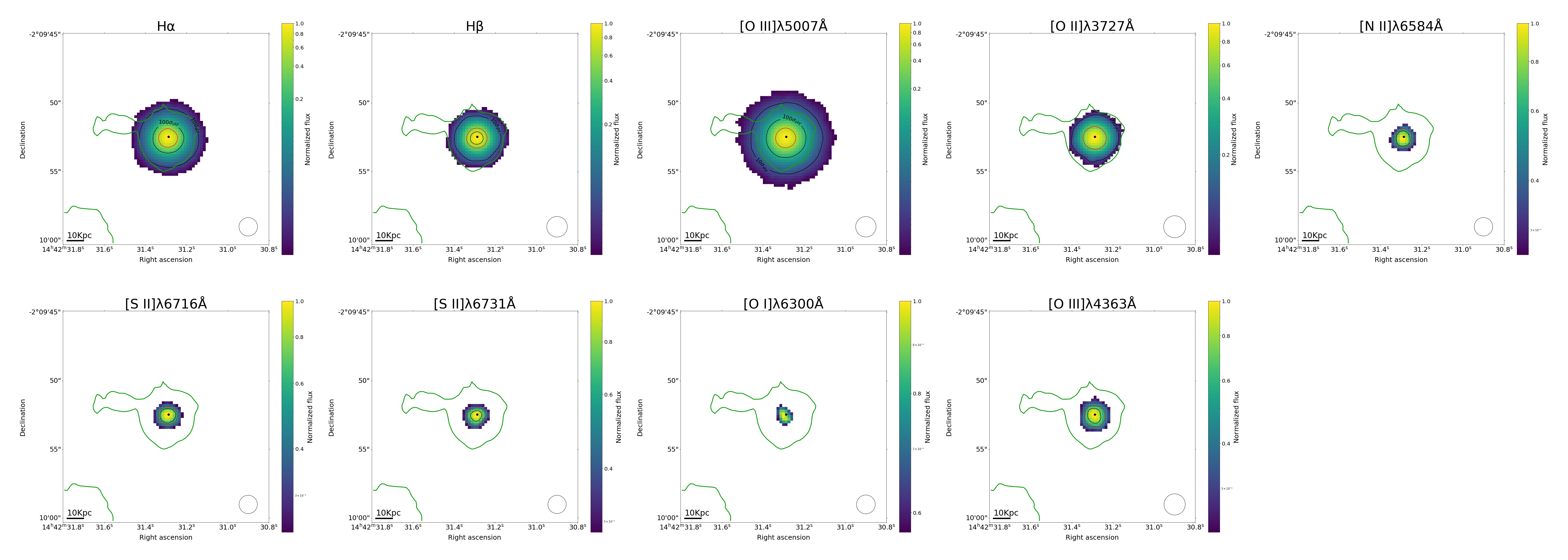}
     \caption{Emission line maps of GP15.}
     \label{hola}
\end{figure*}

%dos columnas
\begin{figure*}[h!]
\centering
   \includegraphics[width=17cm]{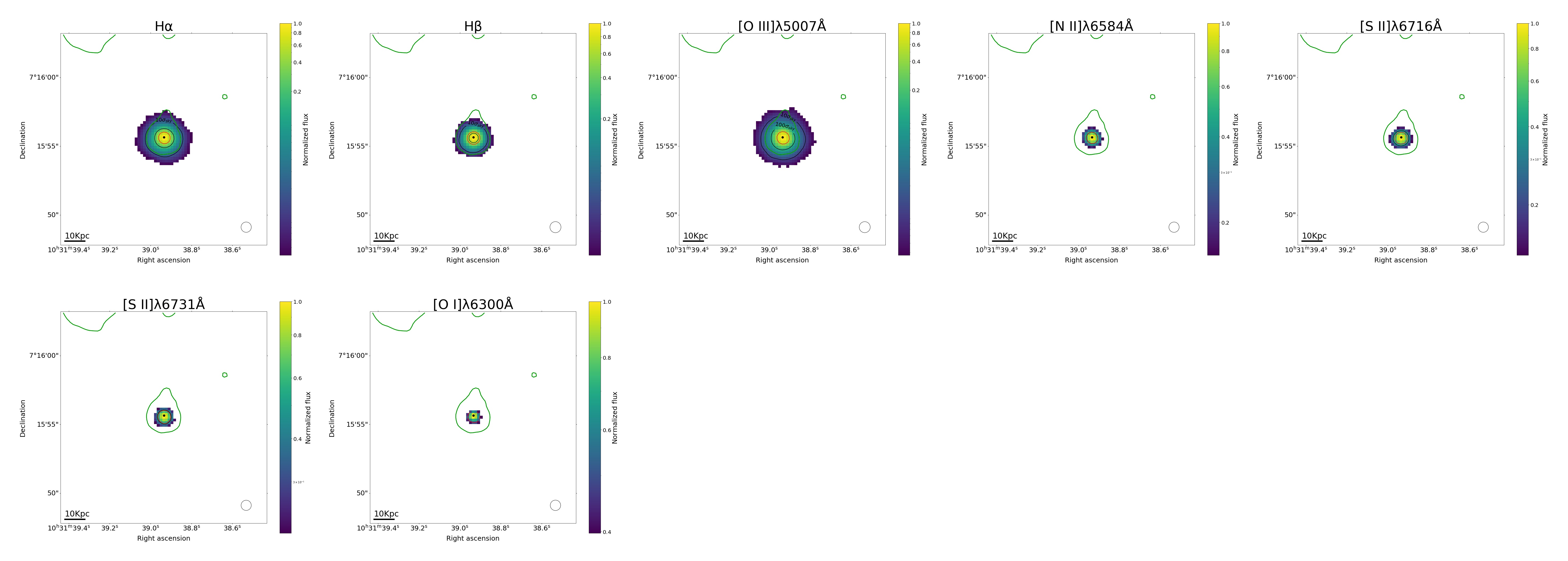}
     \caption{Emission line maps of GP16.}
     \label{hola}
\end{figure*}

%dos columnas
\begin{figure*}[h!]
\centering
   \includegraphics[width=17cm]{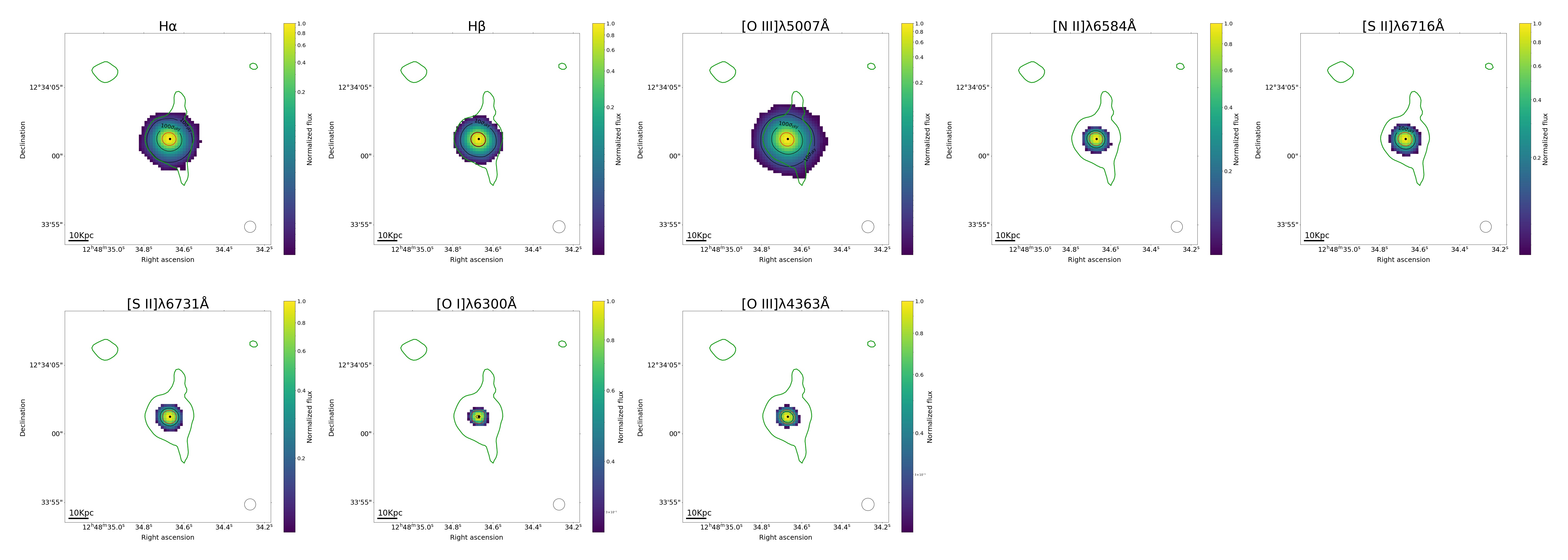}
     \caption{Emission line maps of GP17.}
     \label{hola}
\end{figure*}

%dos columnas
\begin{figure*}[h!]
\centering
   \includegraphics[width=17cm]{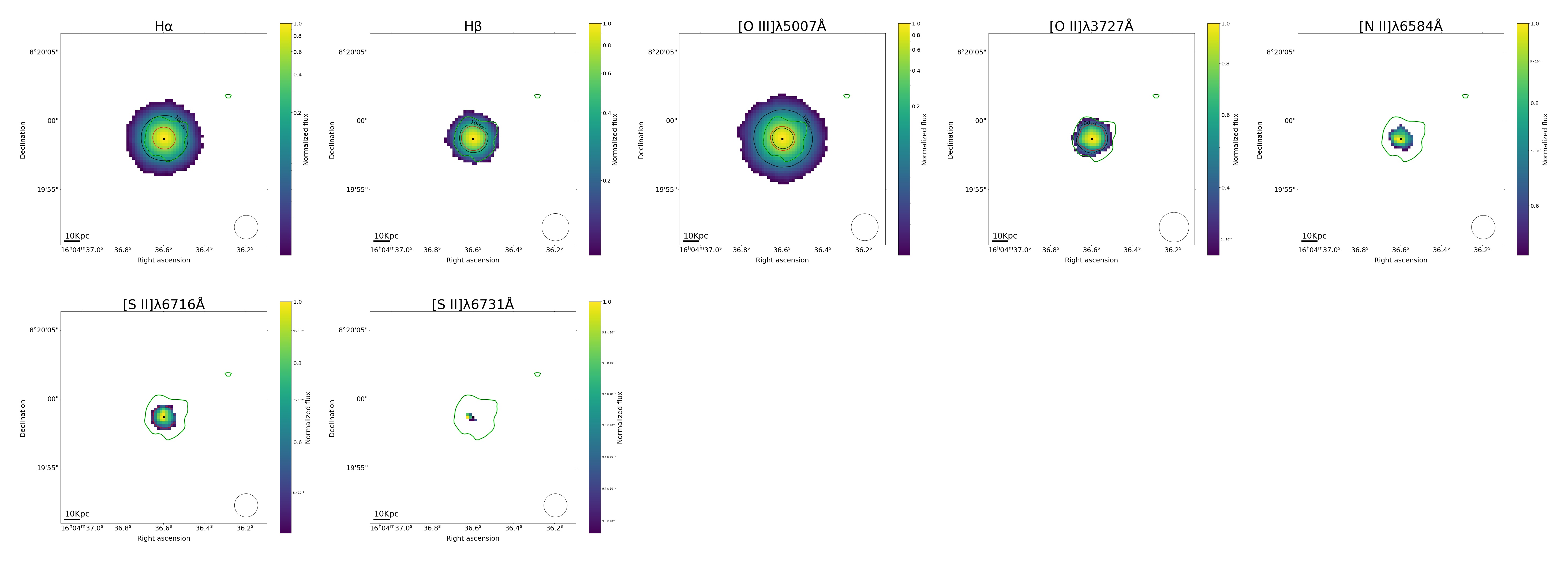}
     \caption{Emission line maps of GP18.}
     \label{hola}
\end{figure*}

%dos columnas
\begin{figure*}[h!]
\centering
   \includegraphics[width=17cm]{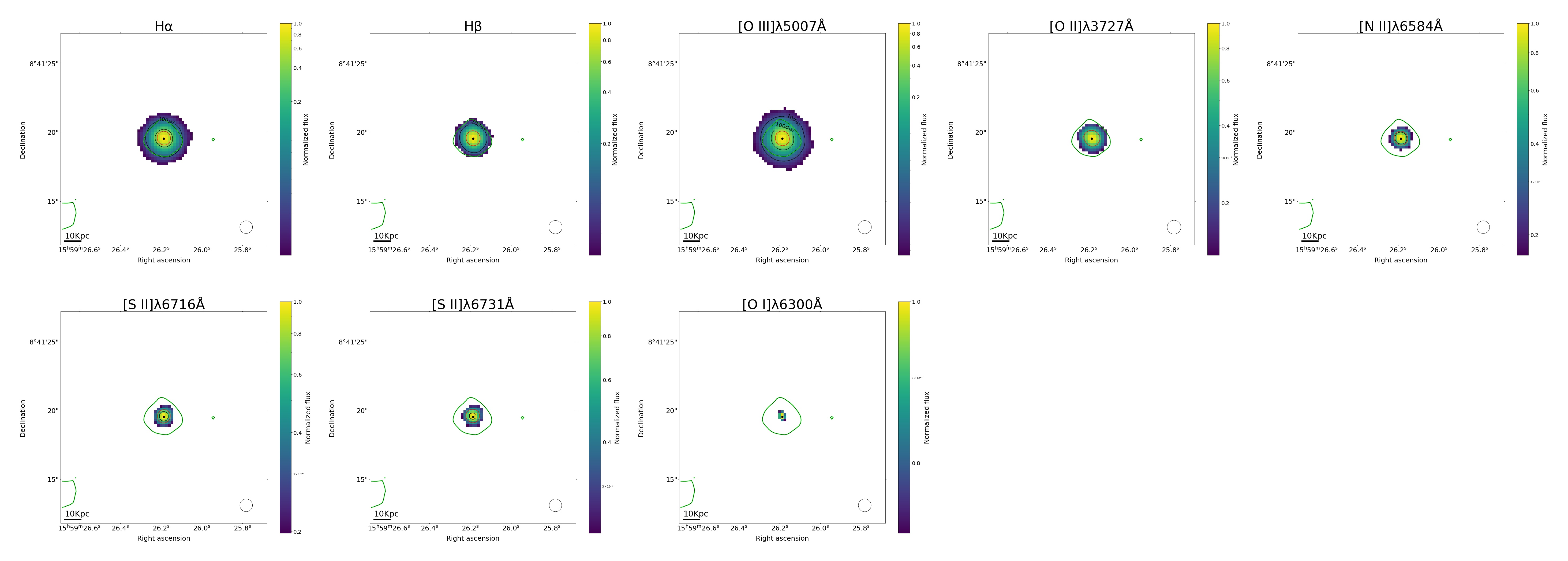}
     \caption{Emission line maps of GP19.}
     \label{hola}
\end{figure*}

%dos columnas
\begin{figure*}[h!]
\centering
   \includegraphics[width=17cm]{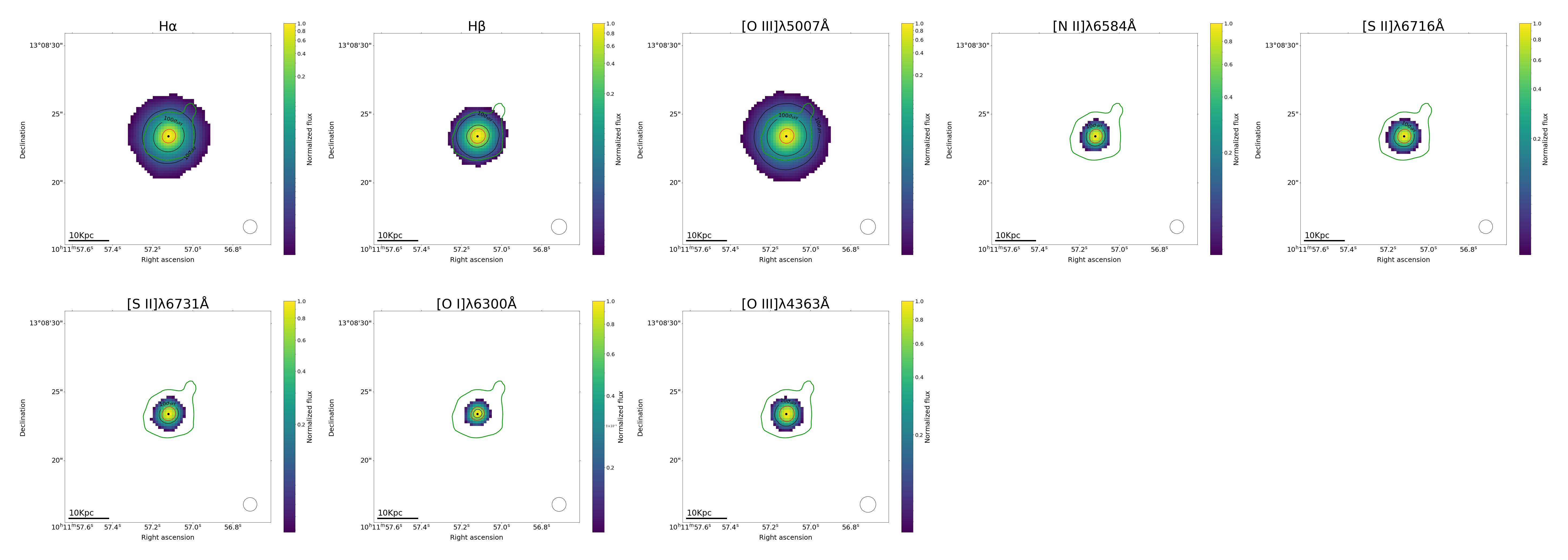}
     \caption{Emission line maps of GP20.}
     \label{hola}
\end{figure*}

%dos columnas
\begin{figure*}[h!]
\centering
   \includegraphics[width=17cm]{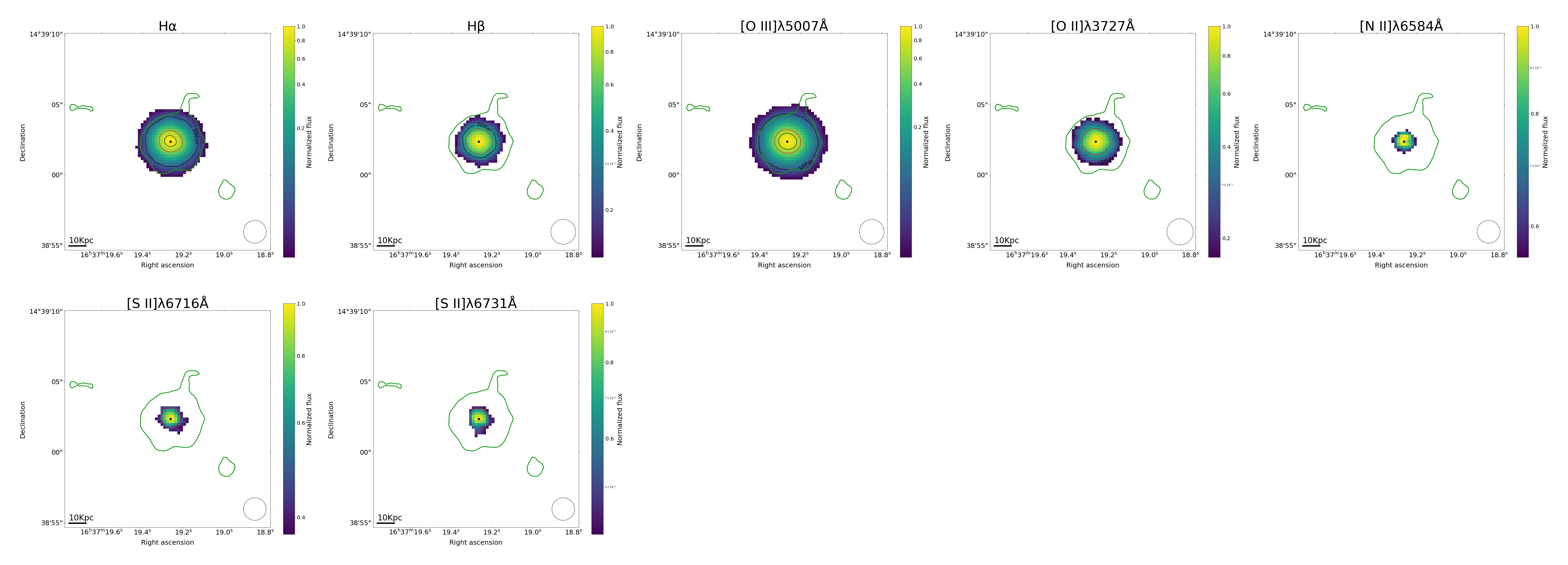}
     \caption{Emission line maps of GP21.}
     \label{hola}
\end{figure*}

%dos columnas
\begin{figure*}[h!]
\centering
   \includegraphics[width=17cm]{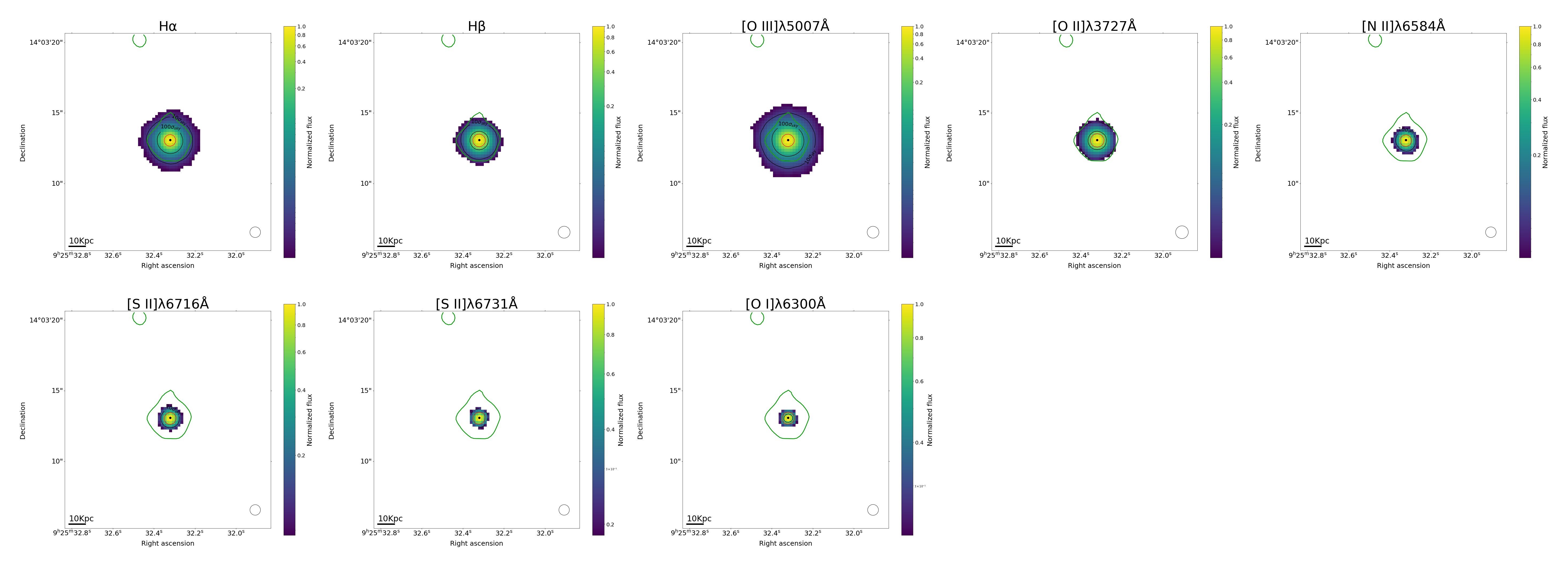}
     \caption{Emission line maps of GP22.}
     \label{hola}
\end{figure*}

%dos columnas
\begin{figure*}[h!]
\centering
   \includegraphics[width=17cm]{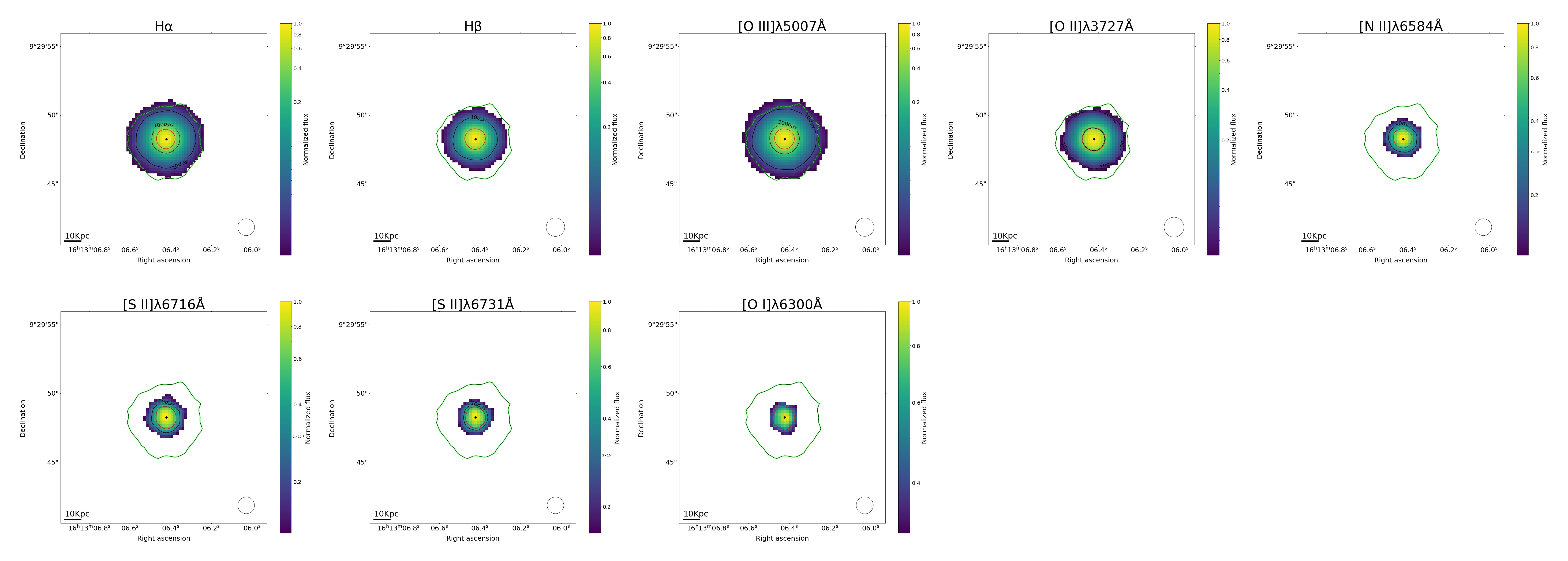}
     \caption{Emission line maps of GP23.}
     \label{hola}
\end{figure*}

%dos columnas
\begin{figure*}[h!]
\centering
   \includegraphics[width=17cm]{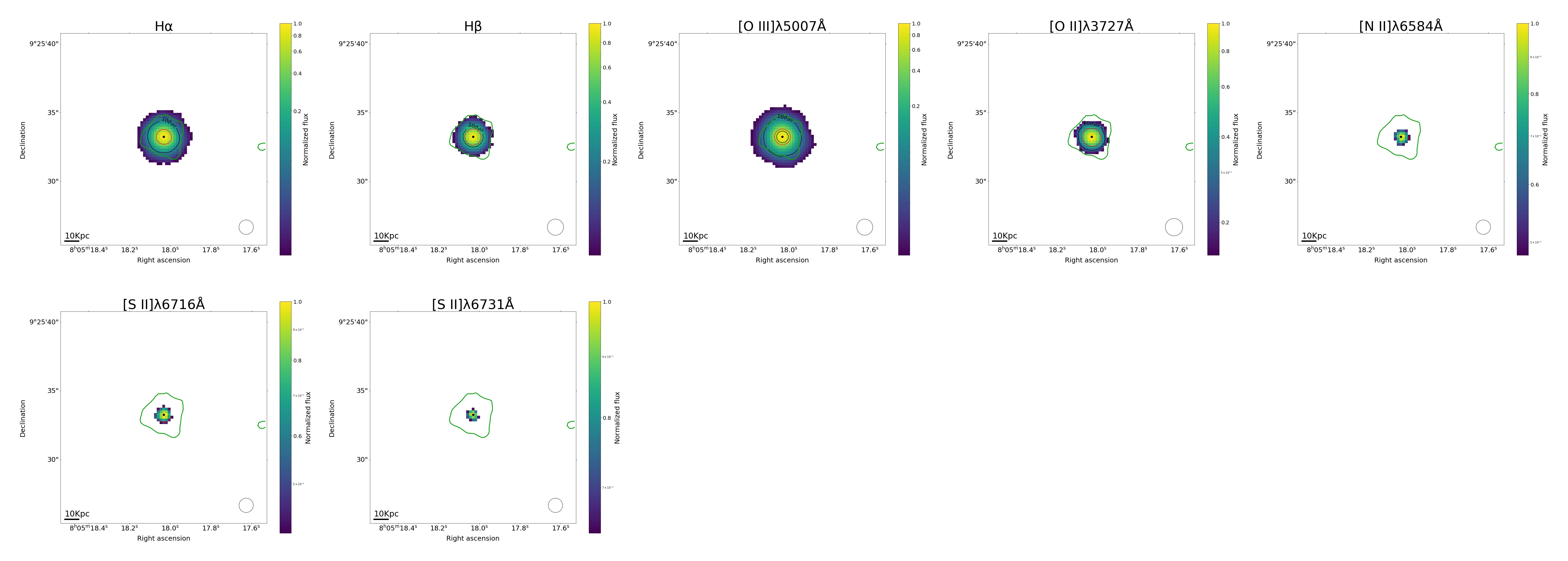}
     \caption{Emission line maps of GP24.}
     \label{hola}
\end{figure*}

\clearpage
\section{Line ratio maps}
\label{appendix:cocient maps}

%dos columnas
\begin{figure*}[h!]
\centering
   \includegraphics[width=17cm]{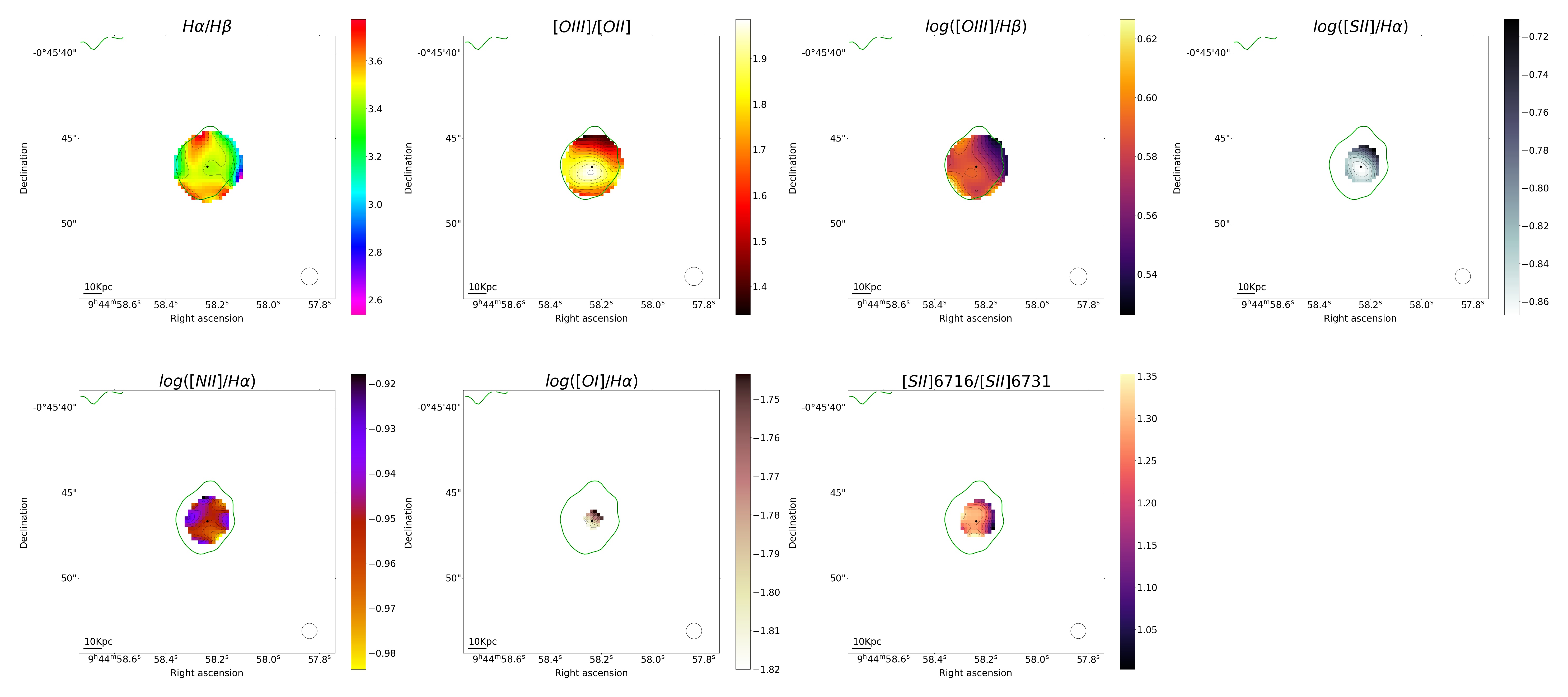}
     \caption{Line ratio maps for GP01.}
     \label{GP01_color}
\end{figure*}

%dos columnas
\begin{figure*}[h!]
\centering
   \includegraphics[width=17cm]{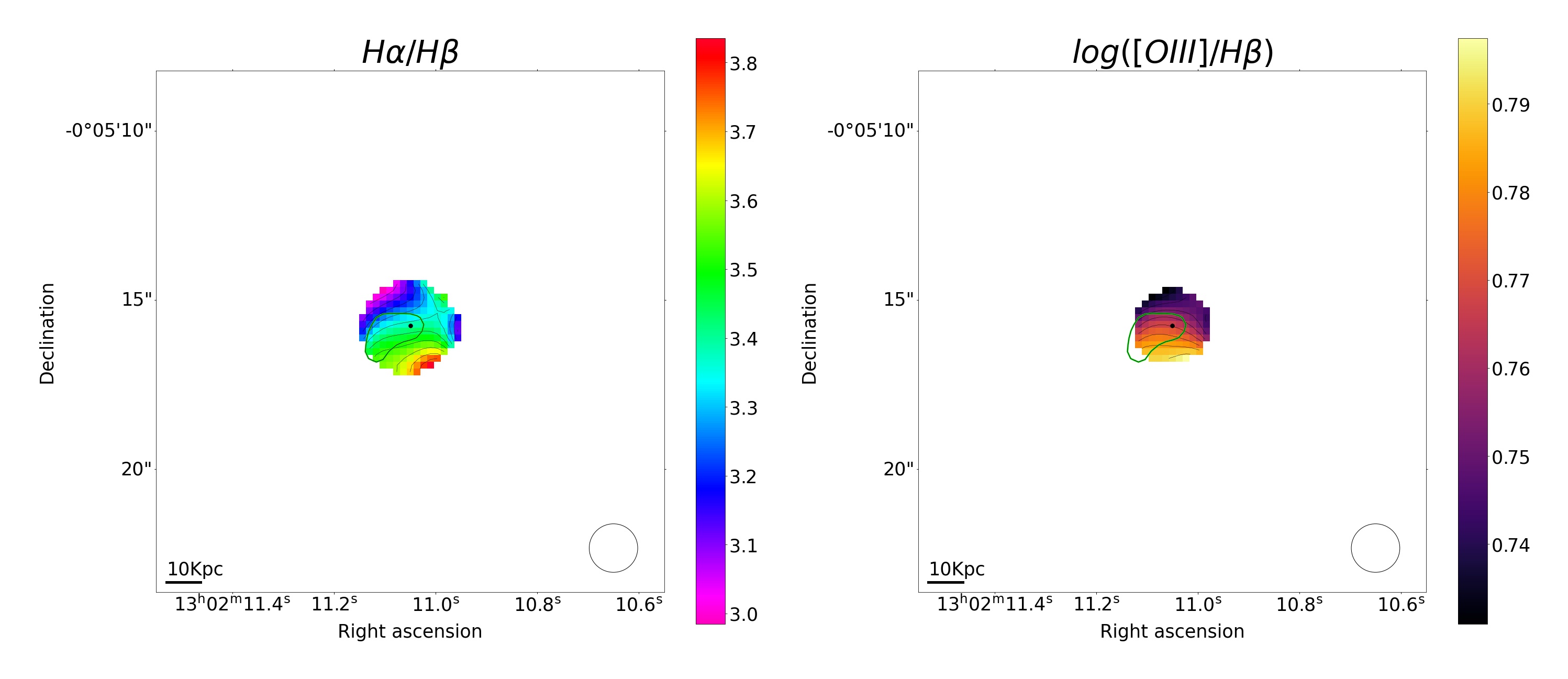}
     \caption{Line ratio maps for GP02.}
     \label{GP02_color}
\end{figure*}

%dos columnas
\begin{figure*}[h!]
\centering
   \includegraphics[width=17cm]{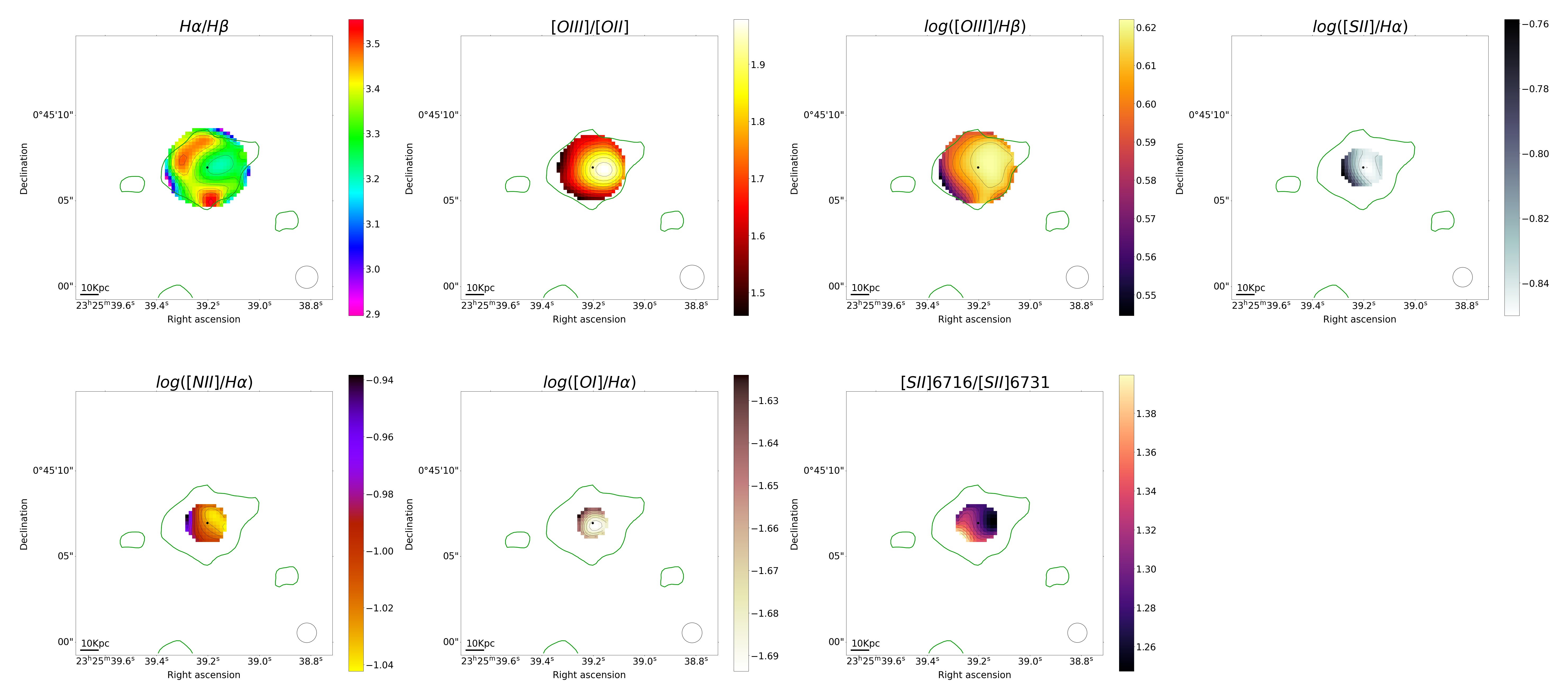}
     \caption{Line ratio maps for GP03.}
     \label{GP03_color}
\end{figure*}

%dos columnas
\begin{figure*}[h!]
\centering
   \includegraphics[width=17cm]{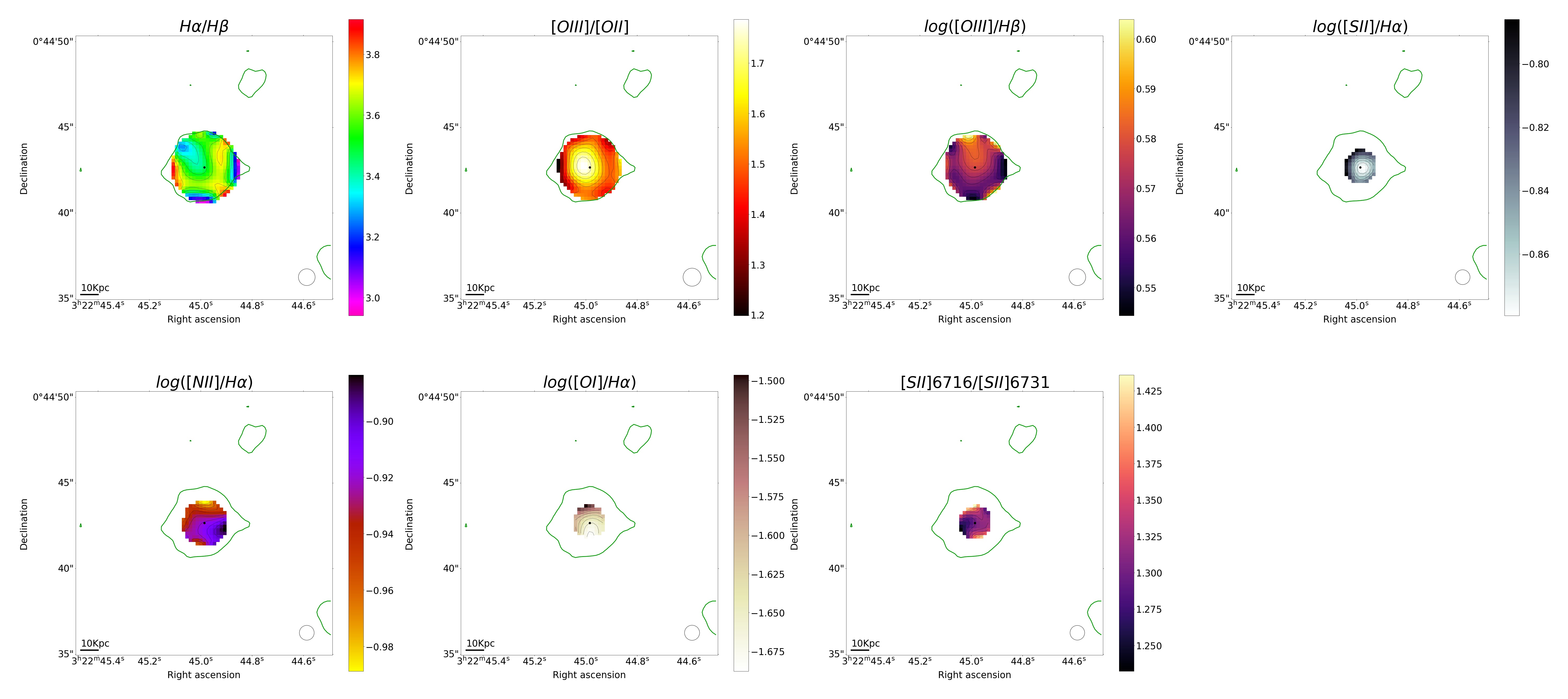}
     \caption{Line ratio maps for GP04.}
     \label{GP04_color}
\end{figure*}

%dos columnas
\begin{figure*}[h!]
\centering
   \includegraphics[width=17cm]{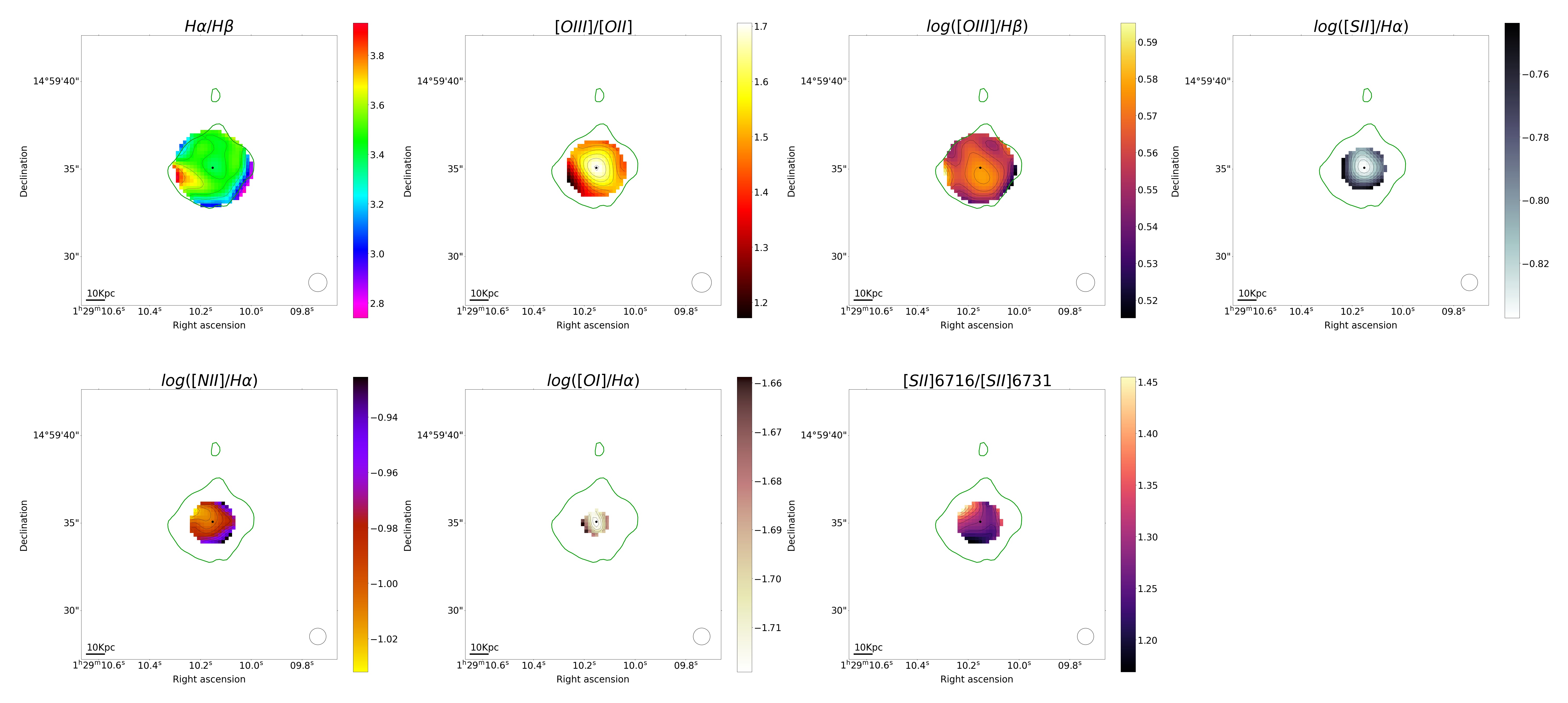}
     \caption{Line ratio maps for GP05.}
     \label{GP05_color}
\end{figure*}

%dos columnas
\begin{figure*}[h!]
\centering
   \includegraphics[width=17cm]{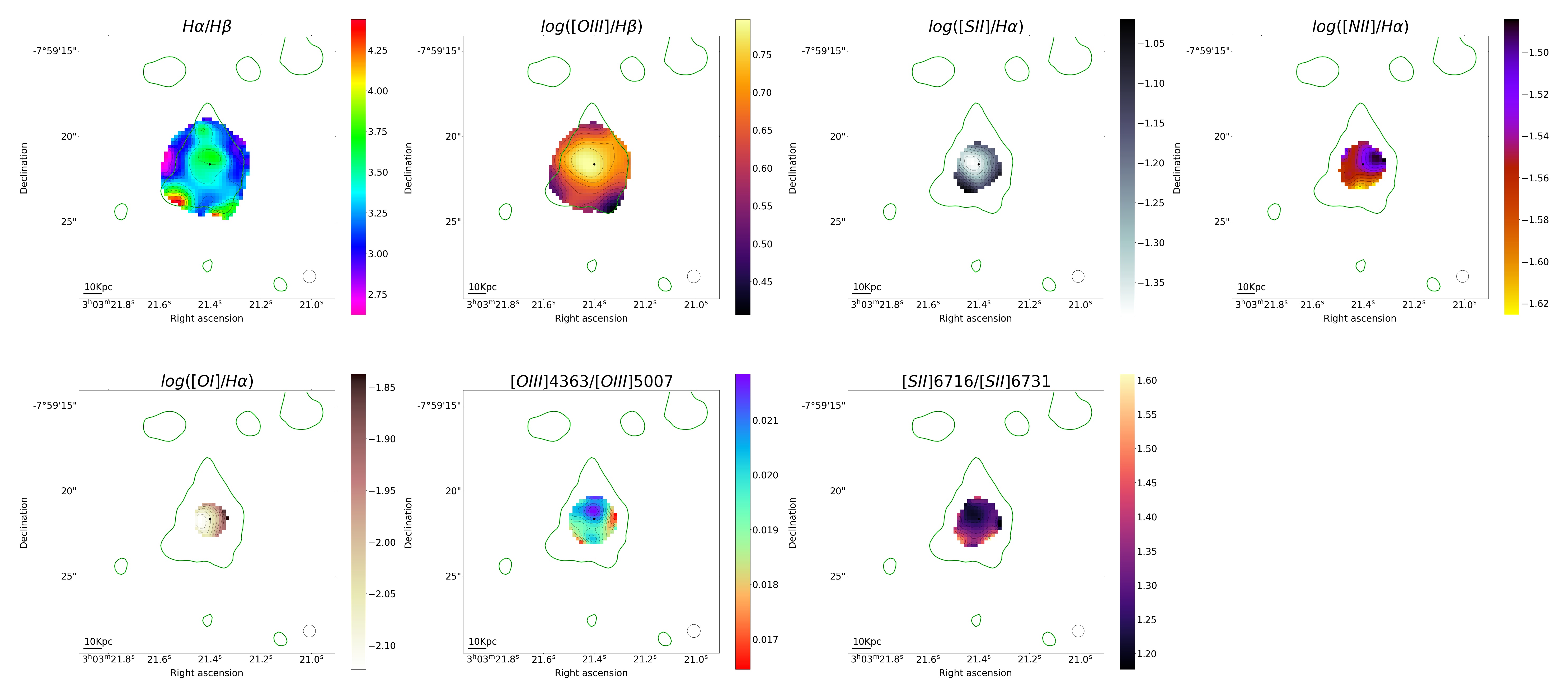}
     \caption{Line ratio maps for GP06.}
     \label{GP06_color}
\end{figure*}

%dos columnas
\begin{figure*}[h!]
\centering
   \includegraphics[width=17cm]{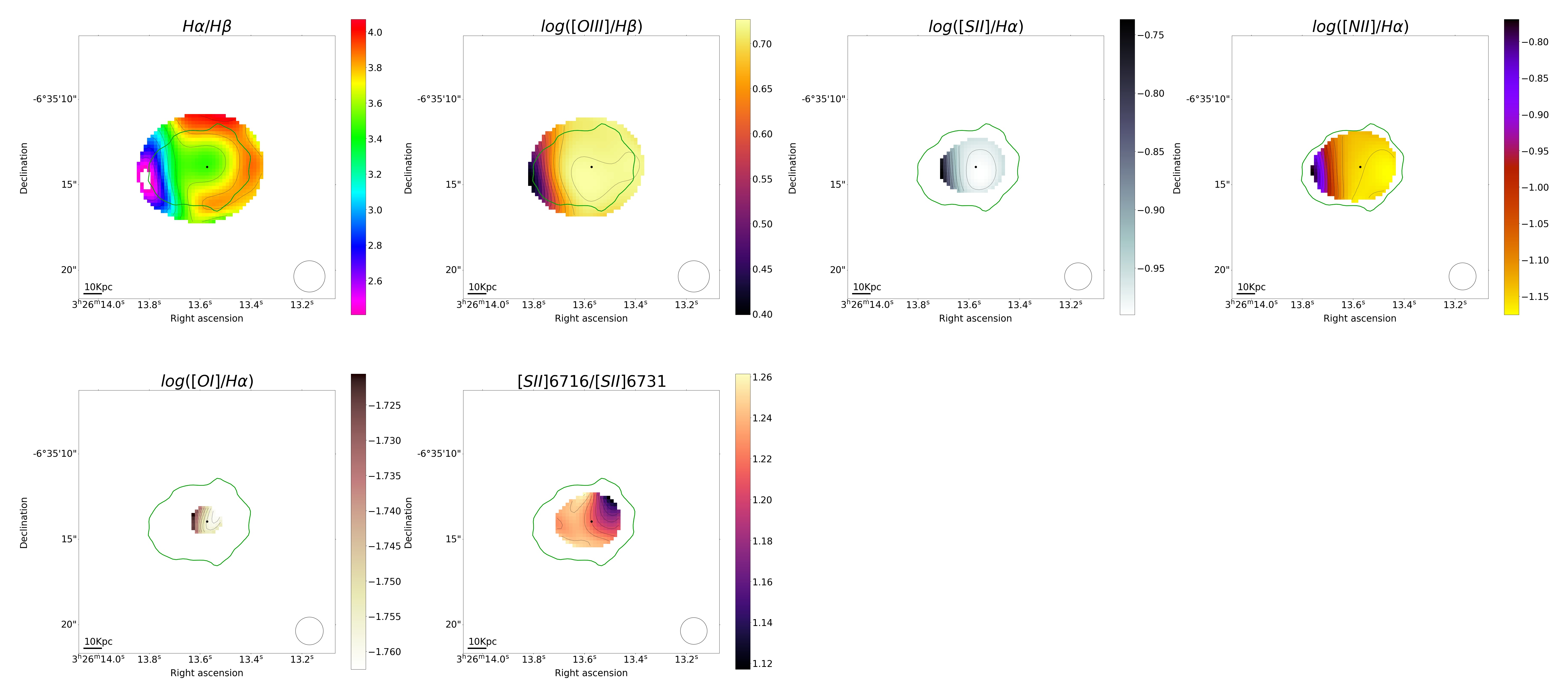}
     \caption{Line ratio maps for GP07. Spaxels > 20 kpc to the west from the H$\alpha$ peak are not reliable due to sky contamination.}
     \label{GP07_color}
\end{figure*}

%dos columnas
\begin{figure*}[h!]
\centering
   \includegraphics[width=17cm]{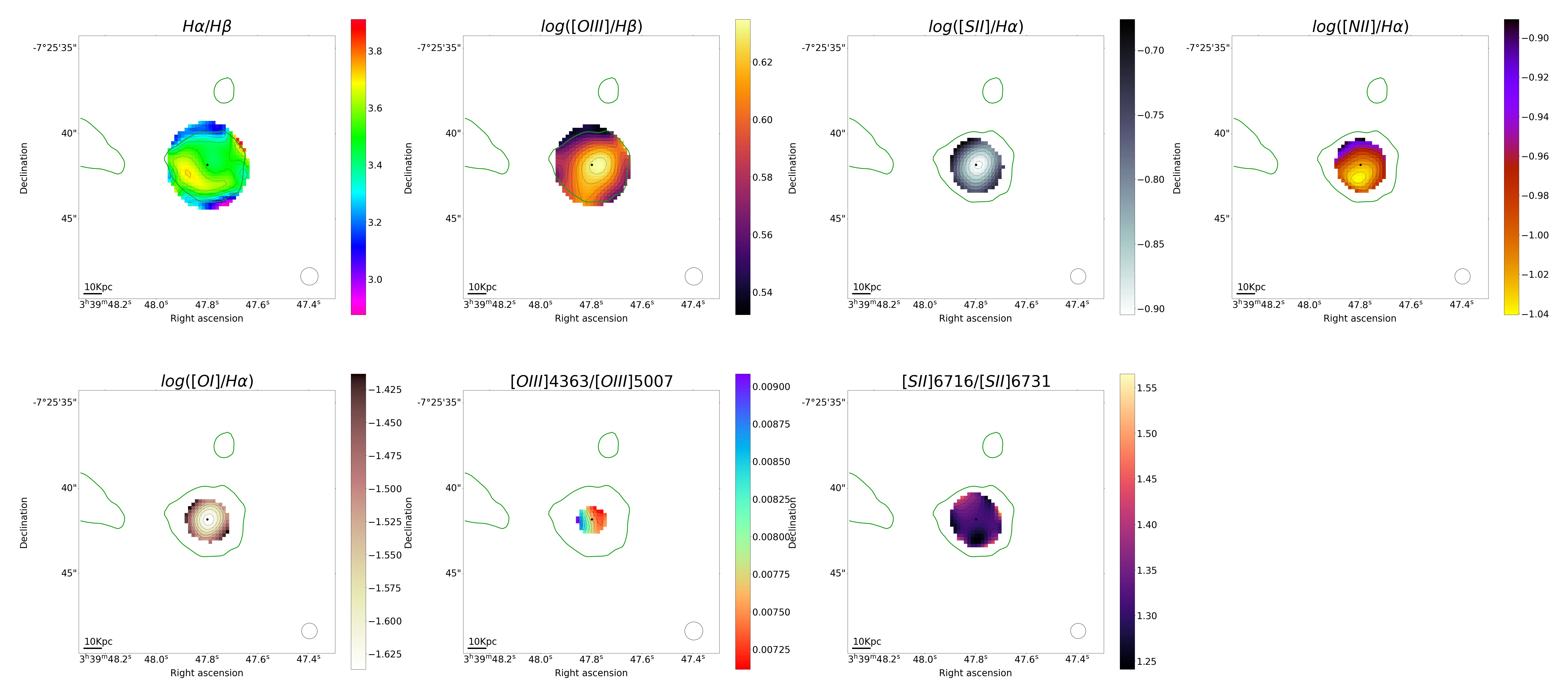}
     \caption{Line ratio maps for GP08.}
     \label{GP08_color}
\end{figure*}

%dos columnas
\begin{figure*}[h!]
\centering
   \includegraphics[width=17cm]{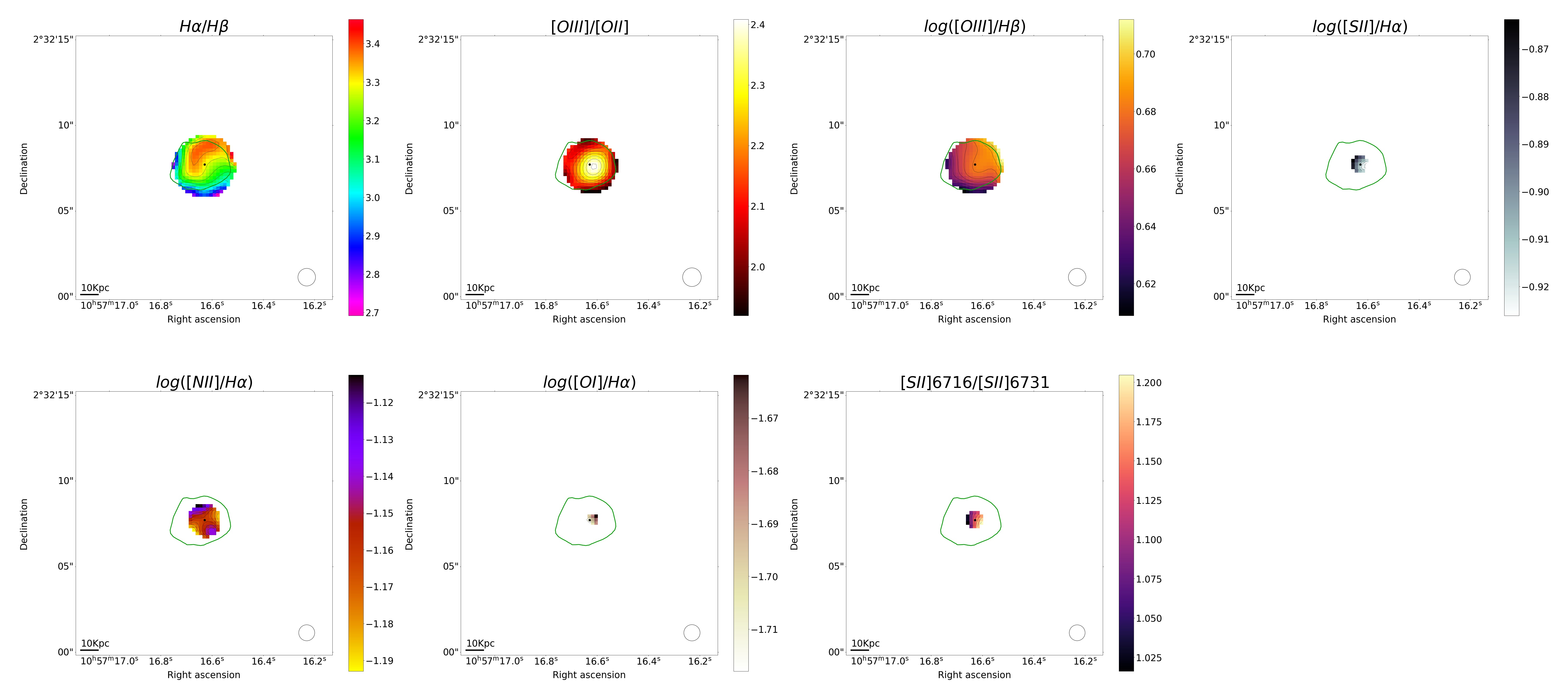}
     \caption{Line ratio maps for GP09.}
     \label{GP09_color}
\end{figure*}

%dos columnas
\begin{figure*}[h!]
\centering
   \includegraphics[width=17cm]{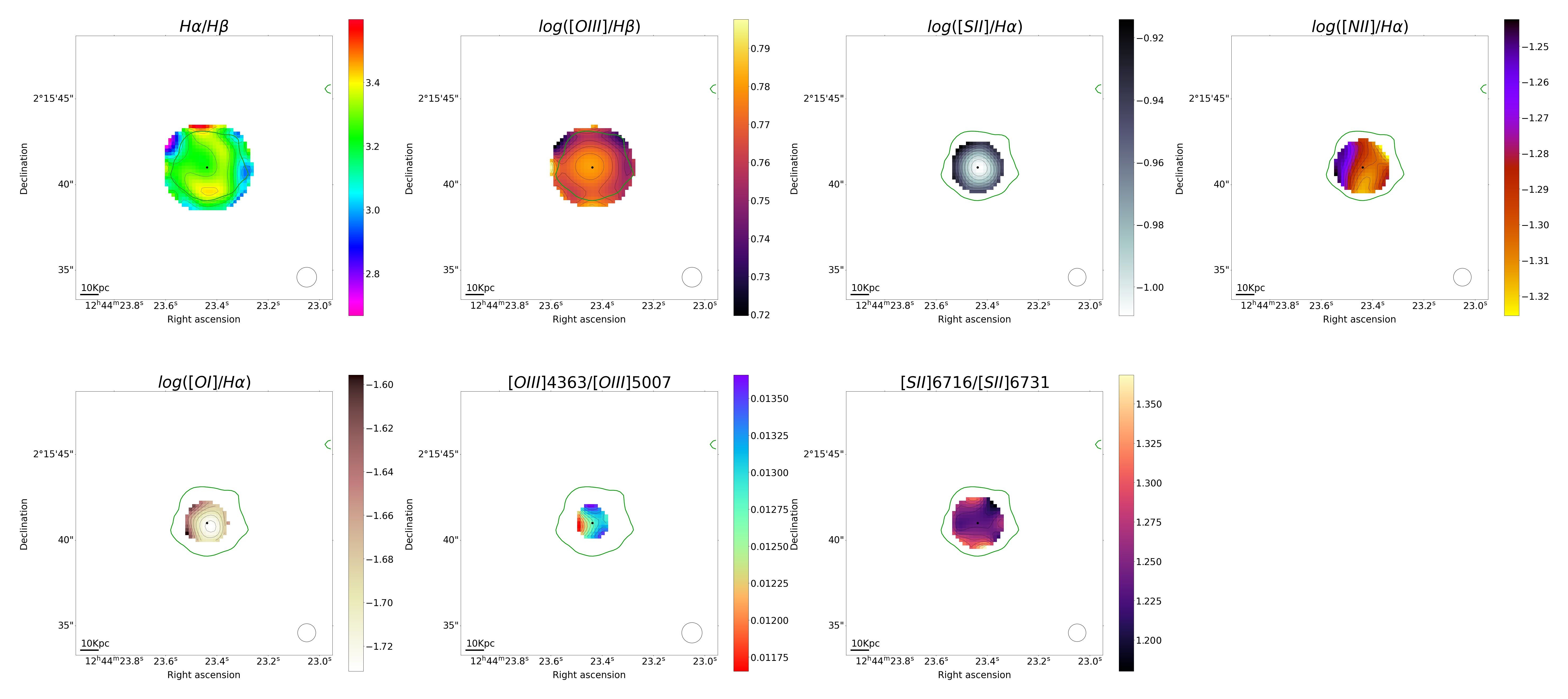}
     \caption{Line ratio maps for GP10.}
     \label{GP10_color}
\end{figure*}

%dos columnas
\begin{figure*}[h!]
\centering
   \includegraphics[width=17cm]{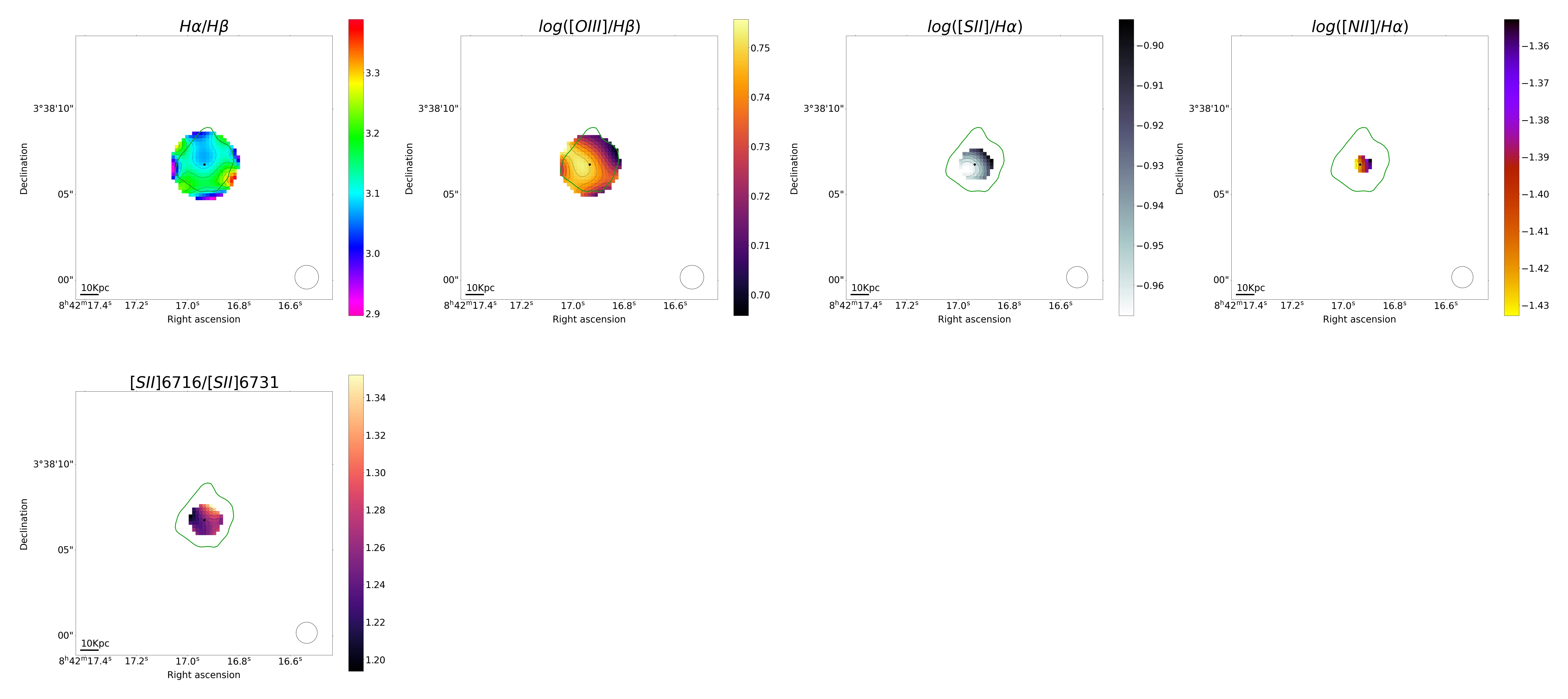}
     \caption{Line ratio maps for GP11.}
     \label{GP11_color}
\end{figure*}

%dos columnas
\begin{figure*}[h!]
\centering
   \includegraphics[width=17cm]{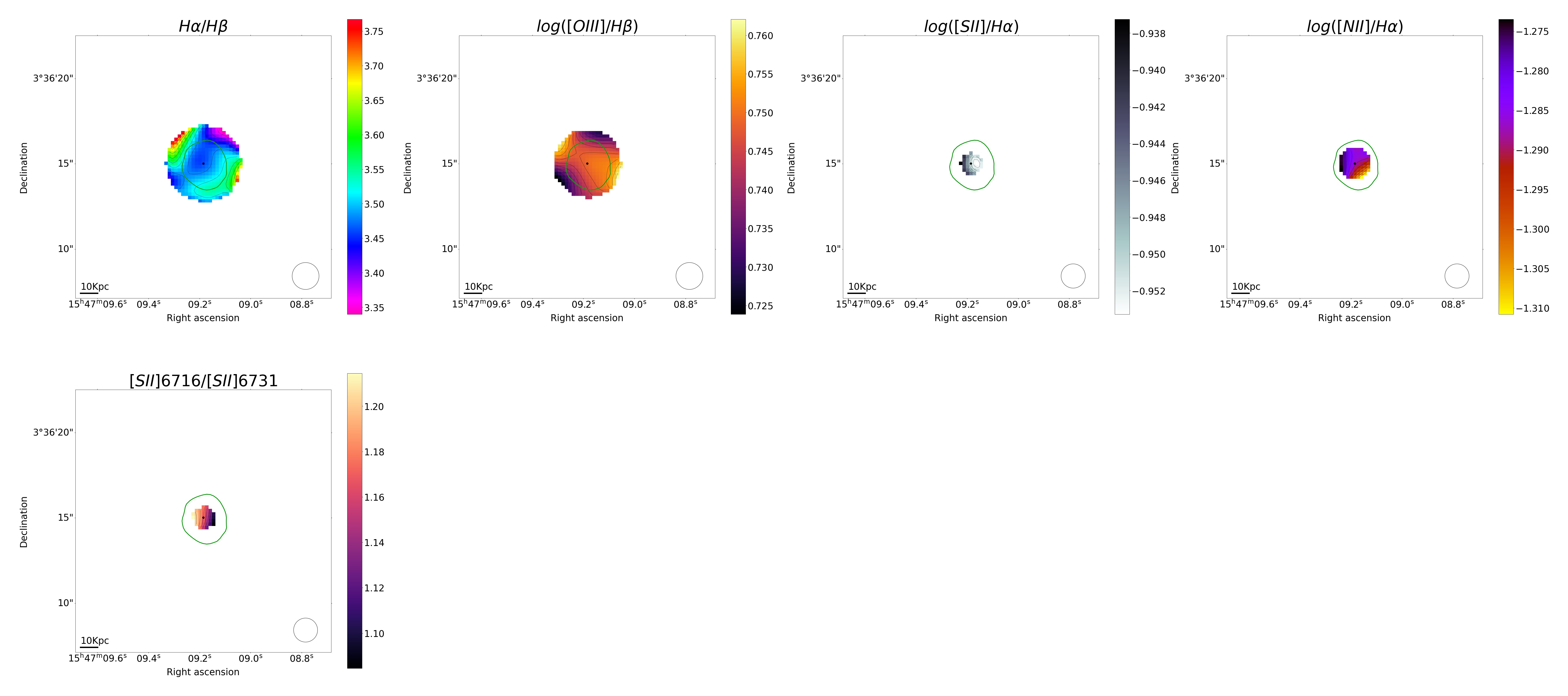}
     \caption{Line ratio maps for GP12.}
     \label{GP12_color}
\end{figure*}

%dos columnas
\begin{figure*}[h!]
\centering
   \includegraphics[width=17cm]{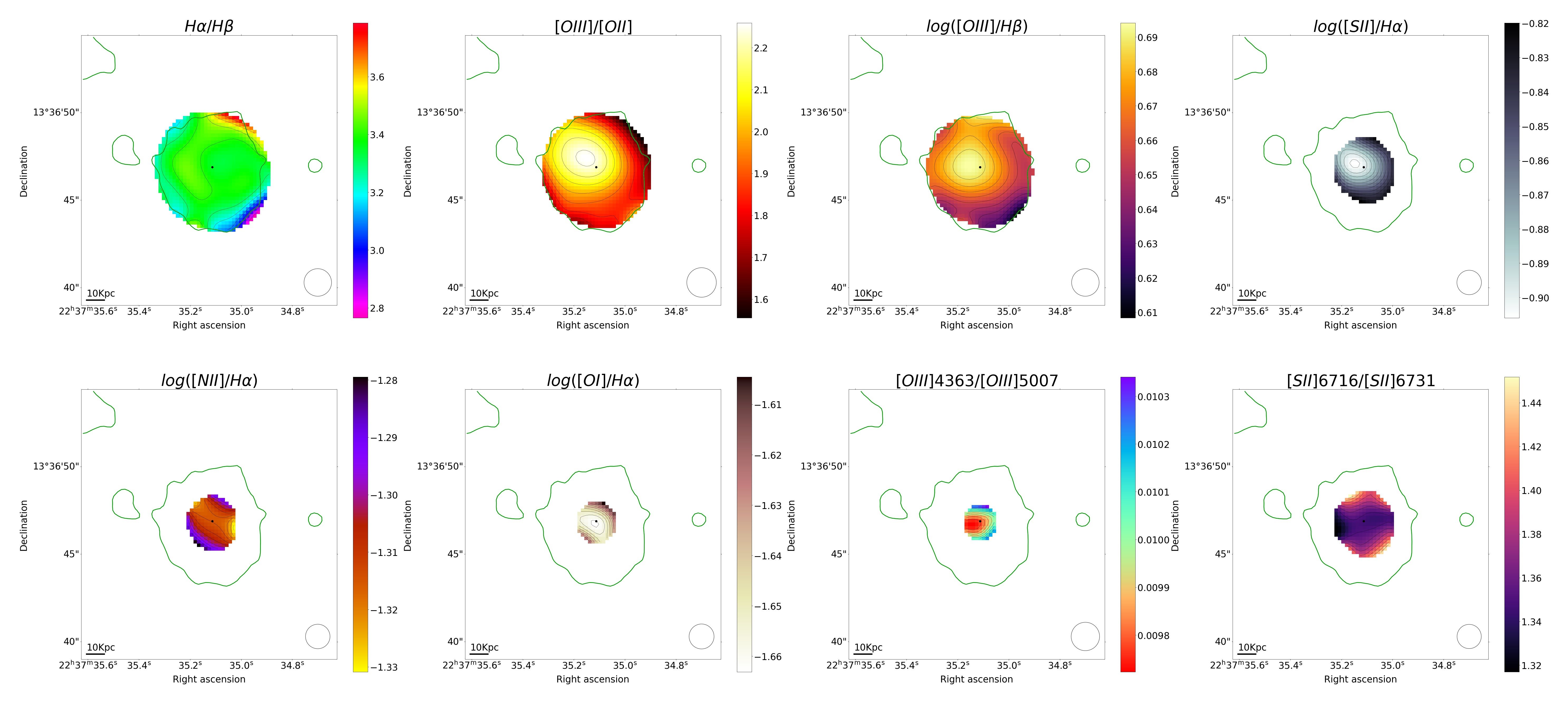}
     \caption{Line ratio maps for GP13.}
     \label{GP13_color}
\end{figure*}

%dos columnas
\begin{figure*}[h!]
\centering
   \includegraphics[width=17cm]{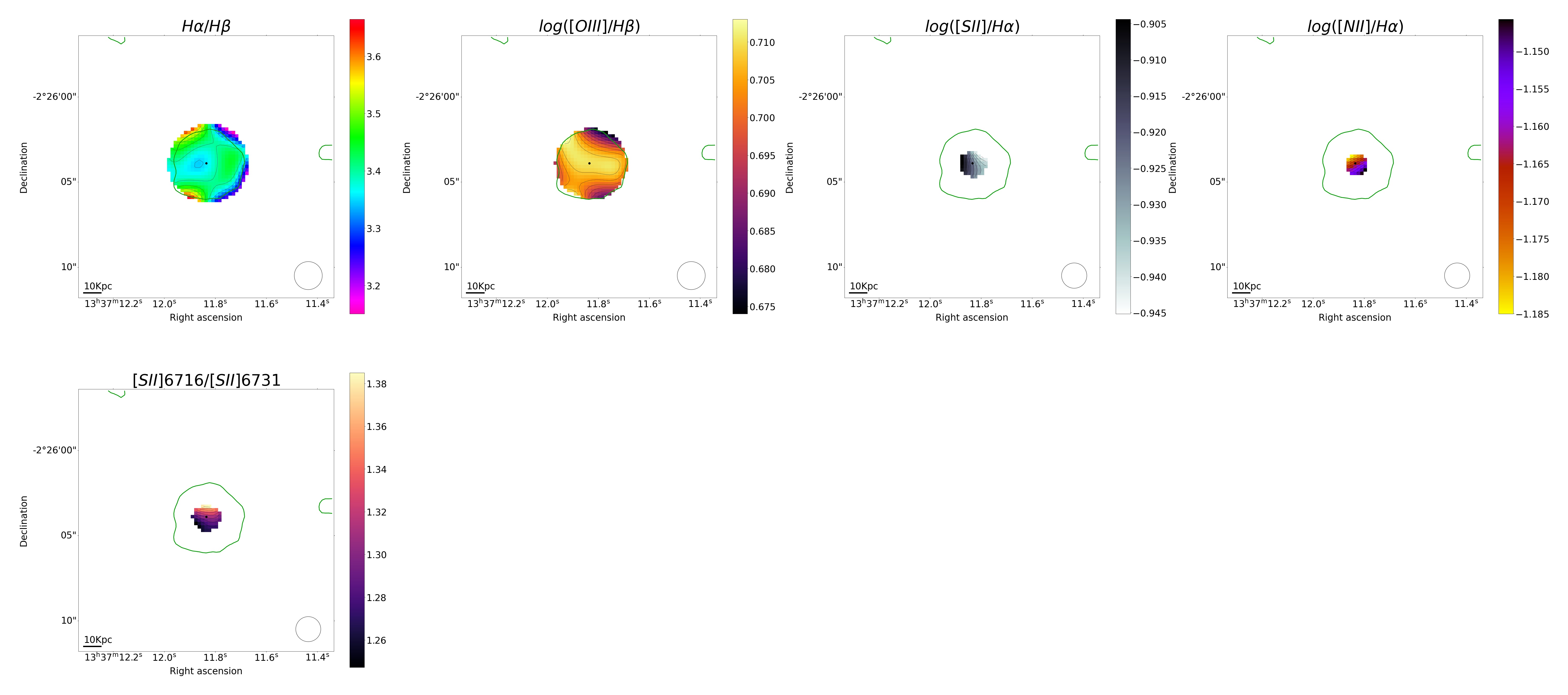}
     \caption{Line ratio maps for GP14.}
     \label{GP14_color}
\end{figure*}

%dos columnas
\begin{figure*}[h!]
\centering
   \includegraphics[width=17cm]{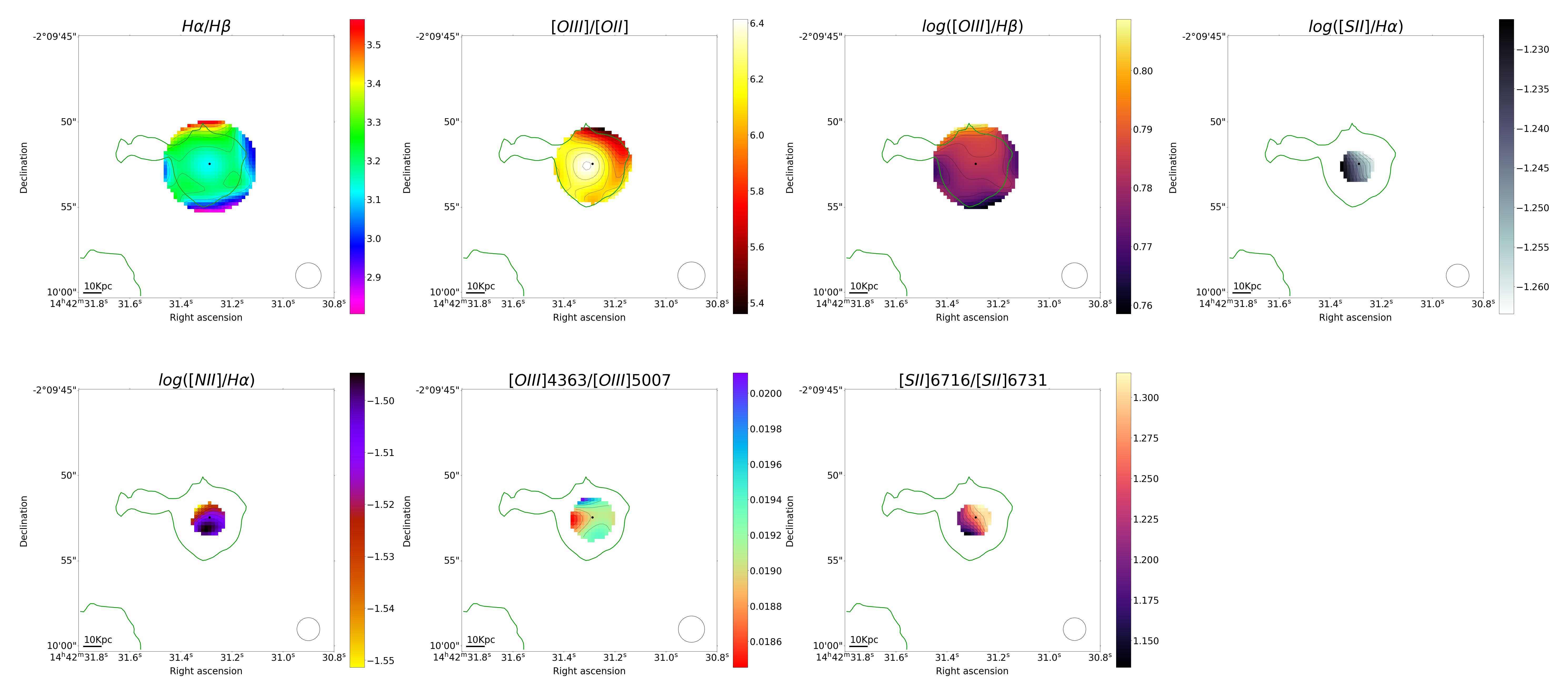}
     \caption{Line ratio maps for GP15.}
     \label{GP15_color}
\end{figure*}

%dos columnas
\begin{figure*}[h!]
\centering
   \includegraphics[width=17cm]{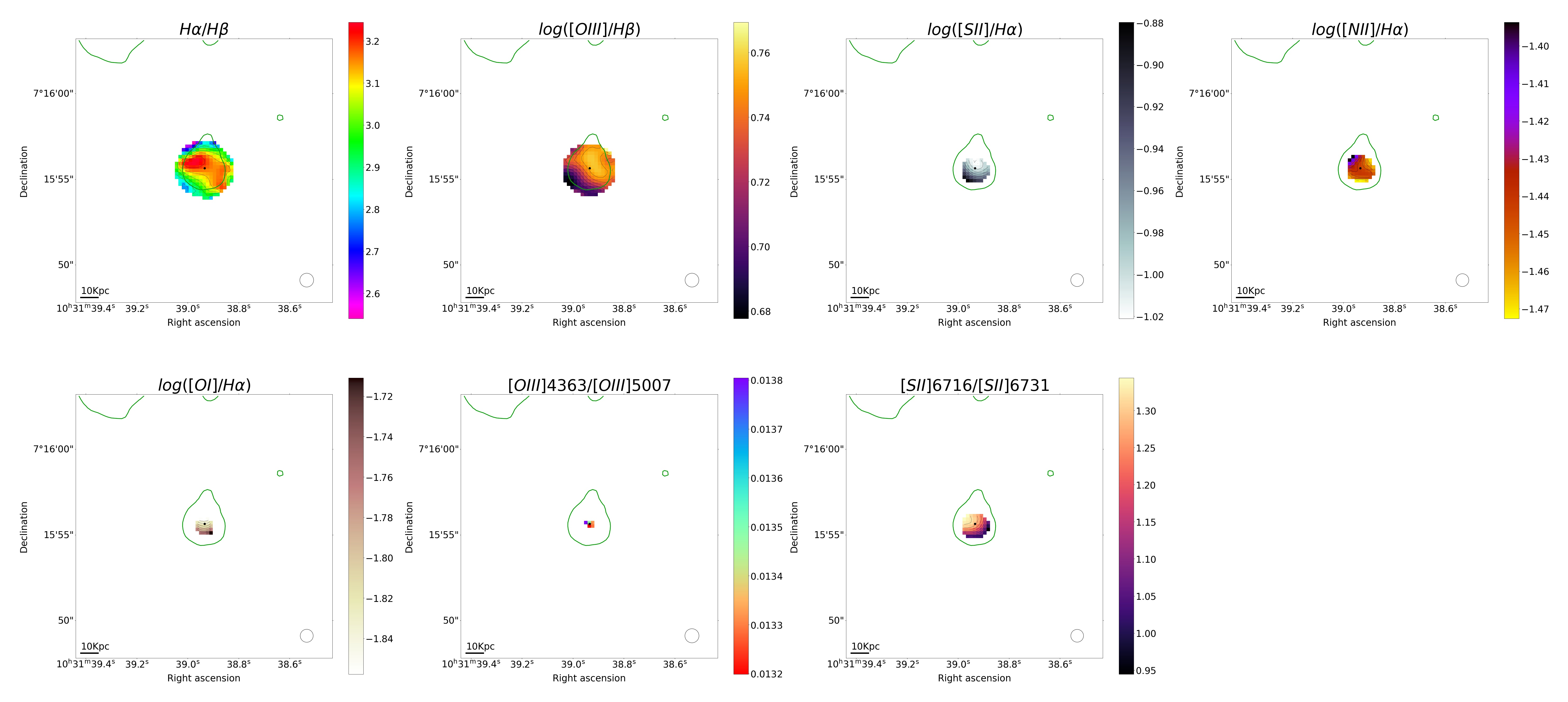}
     \caption{Line ratio maps for GP16.}
     \label{GP16_color}
\end{figure*}

%dos columnas
\begin{figure*}[h!]
\centering
   \includegraphics[width=17cm]{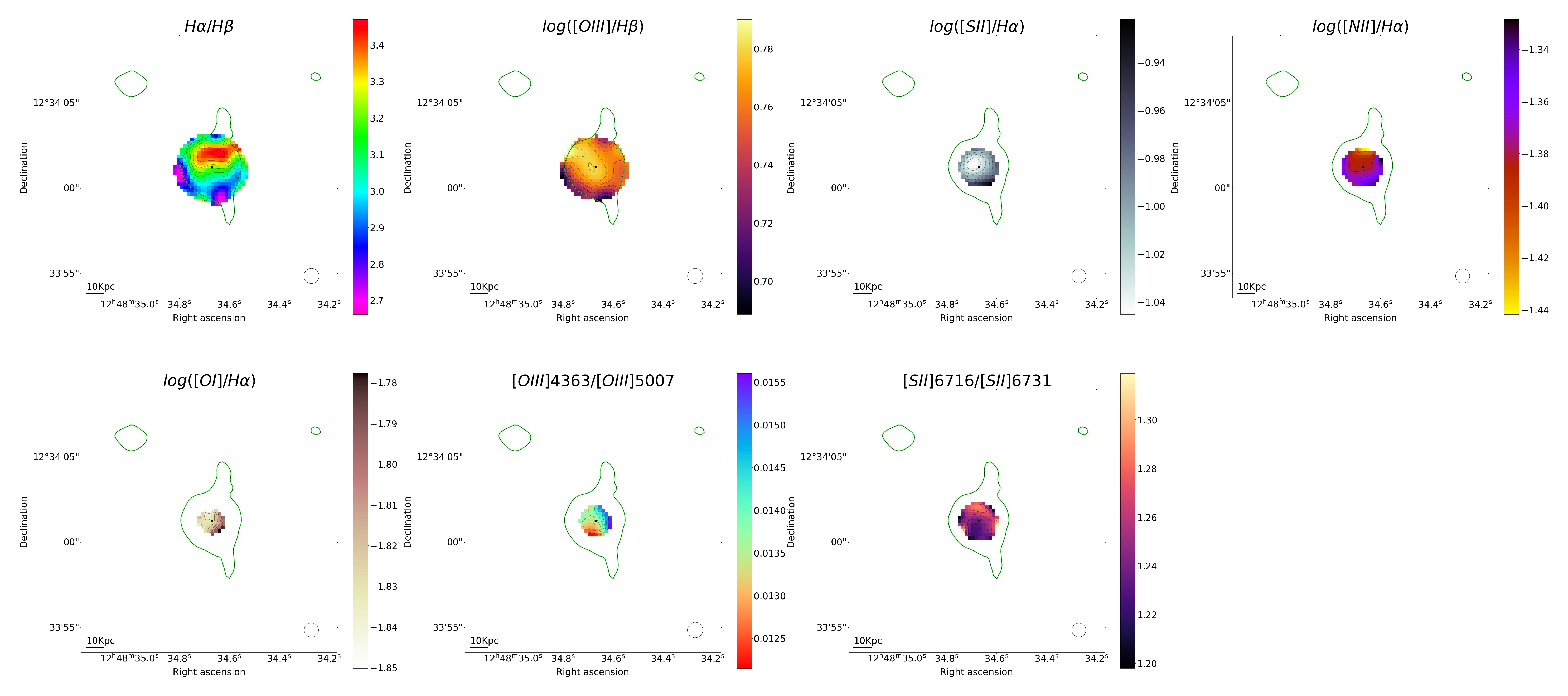}
     \caption{Line ratio maps for GP17.}
     \label{GP17_color}
\end{figure*}

%dos columnas
\begin{figure*}[h!]
\centering
   \includegraphics[width=17cm]{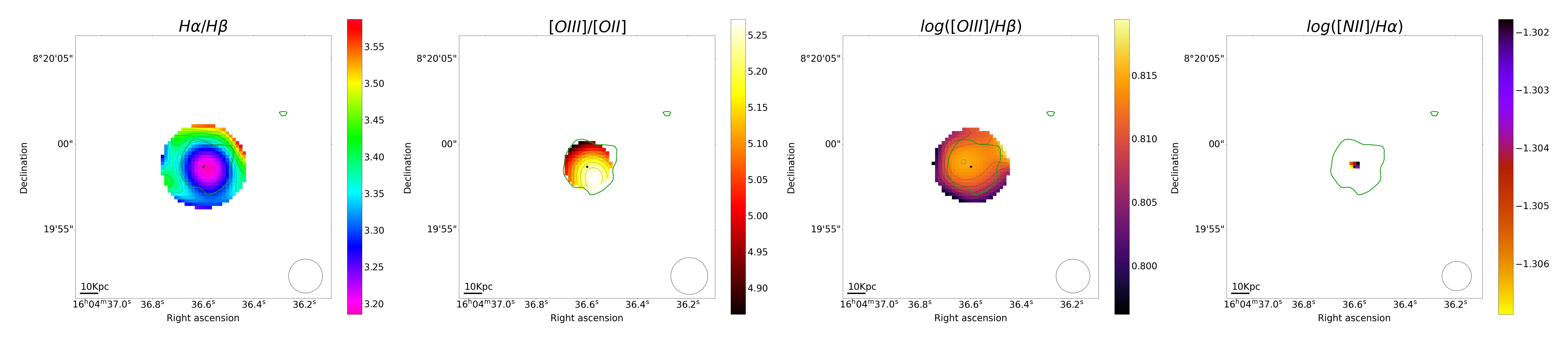}
     \caption{Line ratio maps for GP18.}
     \label{GP18_color}
\end{figure*}

%dos columnas
\begin{figure*}[h!]
\centering
   \includegraphics[width=17cm]{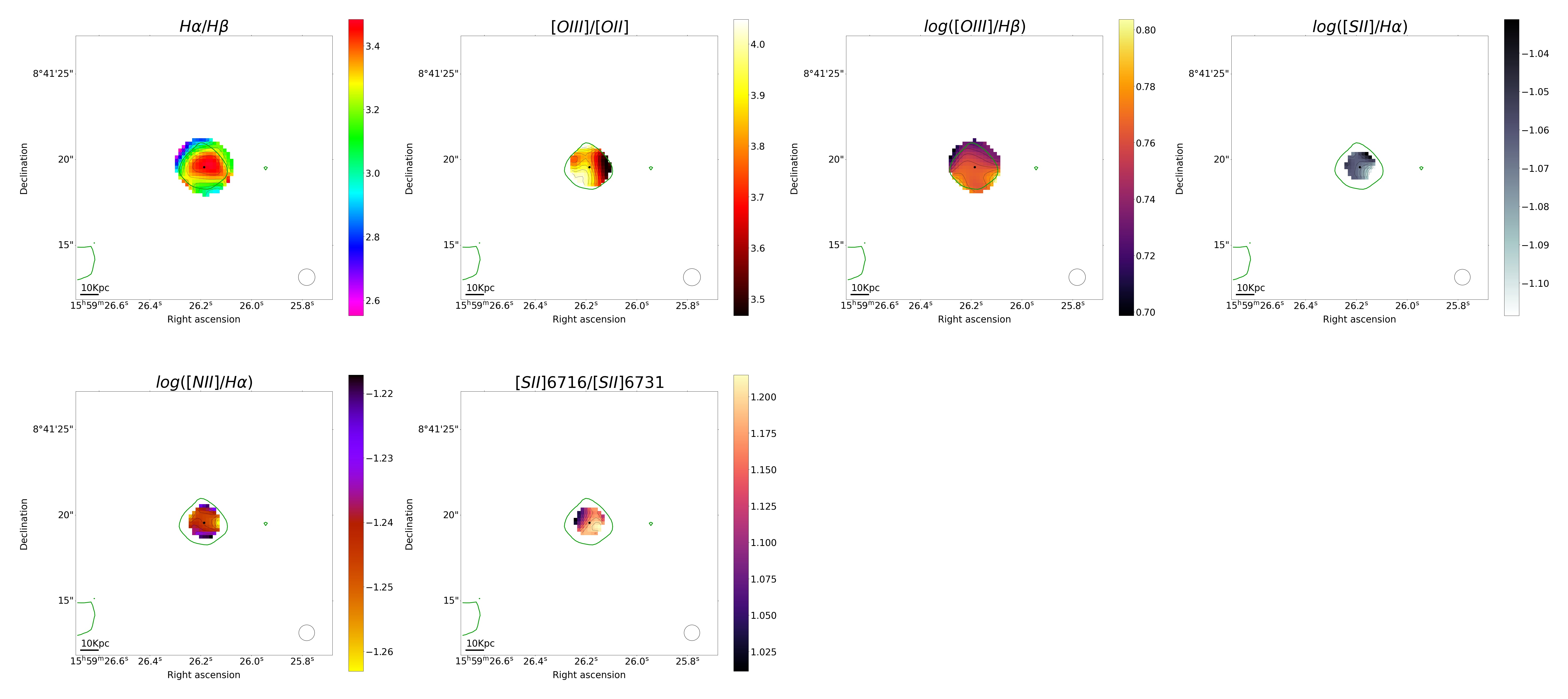}
     \caption{Line ratio maps for GP19.}
     \label{GP19_color}
\end{figure*}

%dos columnas
\begin{figure*}[h!]
\centering
   \includegraphics[width=17cm]{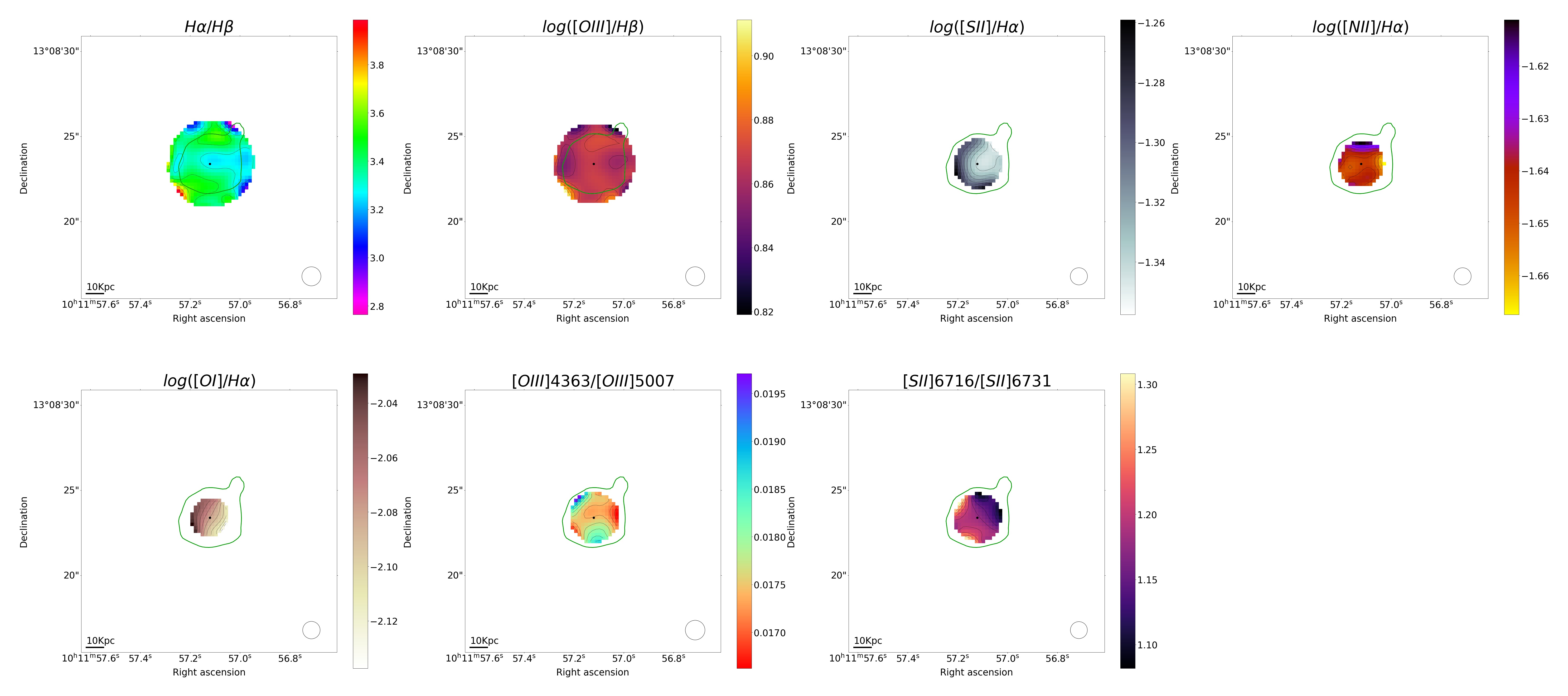}
     \caption{Line ratio maps for GP20.}
     \label{GP20_color}
\end{figure*}

%dos columnas
\begin{figure*}[h!]
\centering
   \includegraphics[width=17cm]{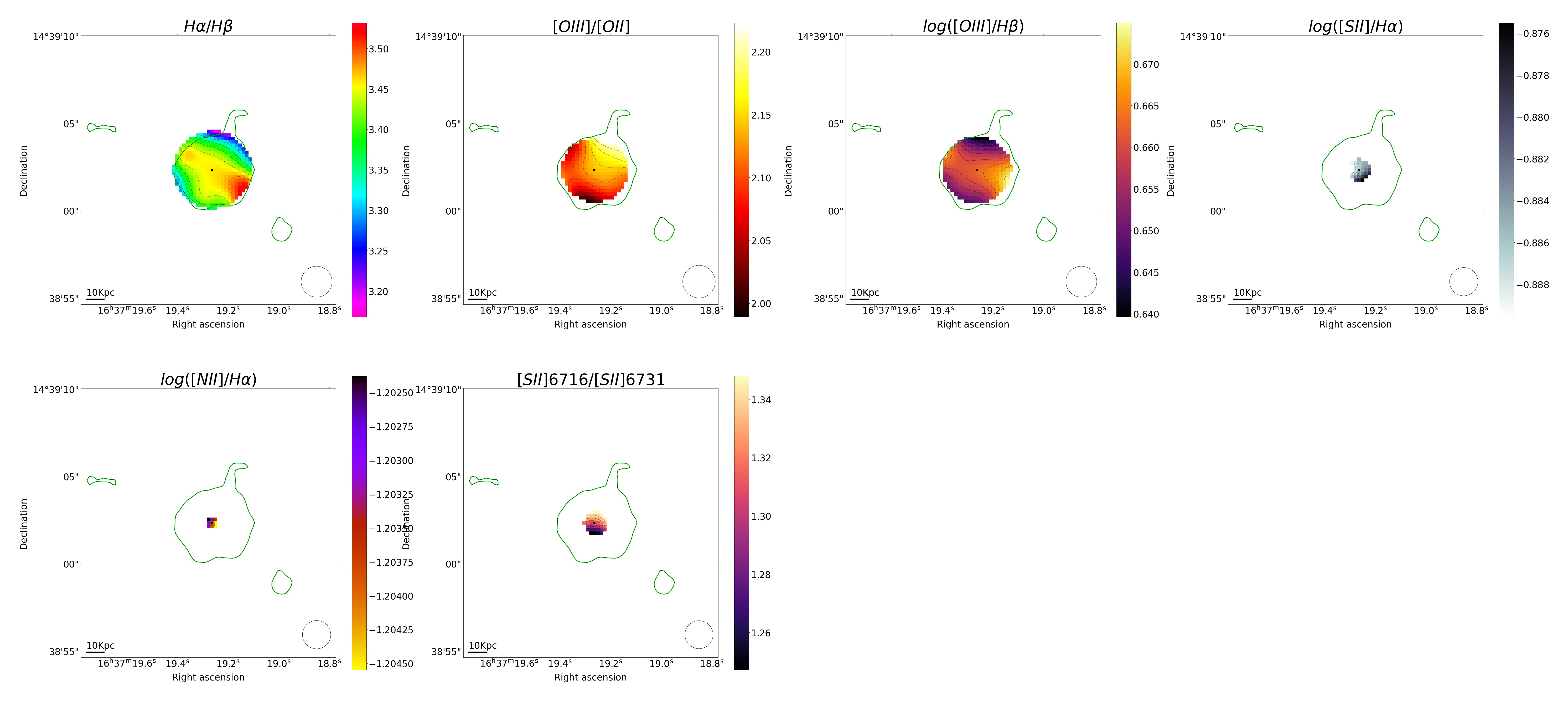}
     \caption{Line ratio maps for GP21.}
     \label{GP21_color}
\end{figure*}

%dos columnas
\begin{figure*}[h!]
\centering
   \includegraphics[width=17cm]{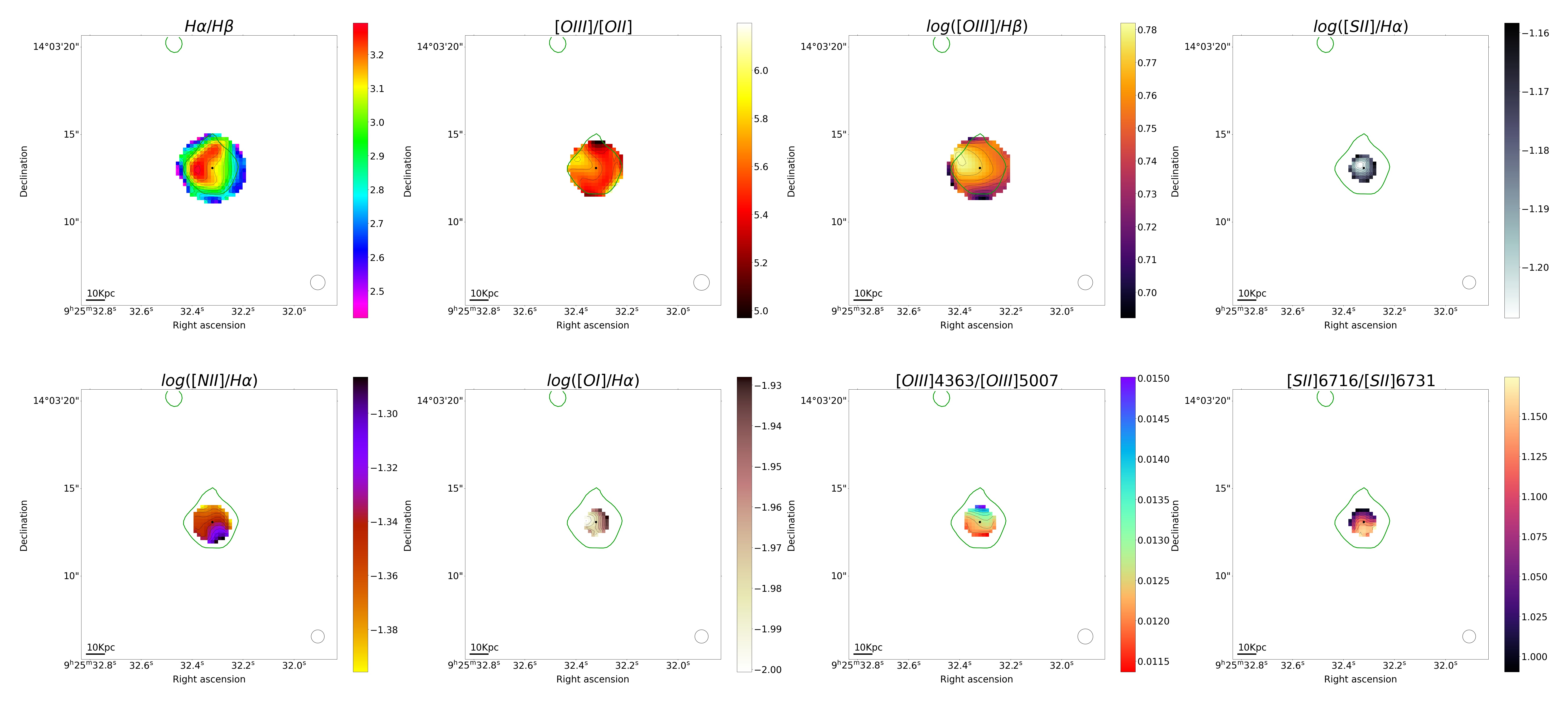}
     \caption{Line ratio maps for GP22.}
     \label{GP22_color}
\end{figure*}

%dos columnas
\begin{figure*}[h!]
\centering
   \includegraphics[width=17cm]{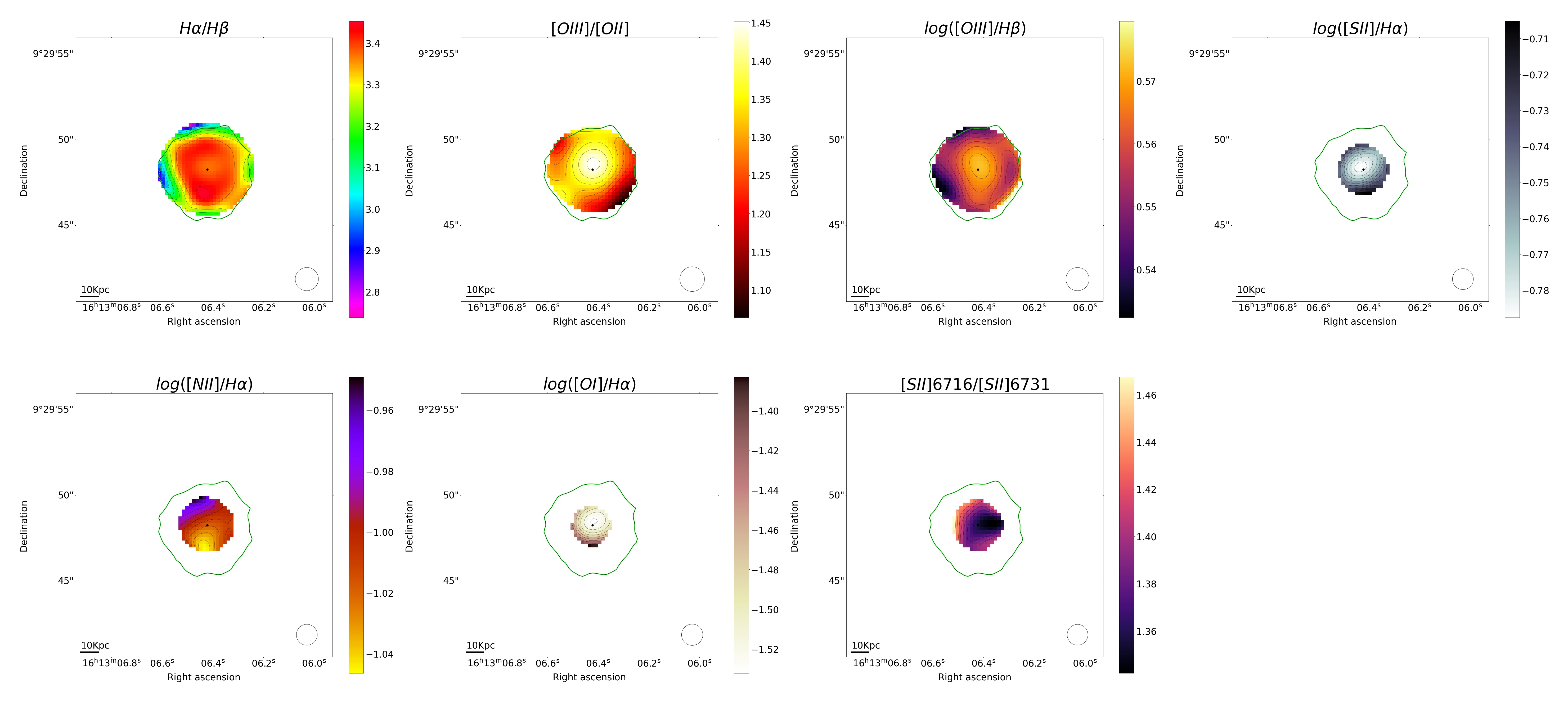}
     \caption{Line ratio maps for GP23.}
     \label{GP23_color}
\end{figure*}

%dos columnas
\begin{figure*}[h!]
\centering
   \includegraphics[width=17cm]{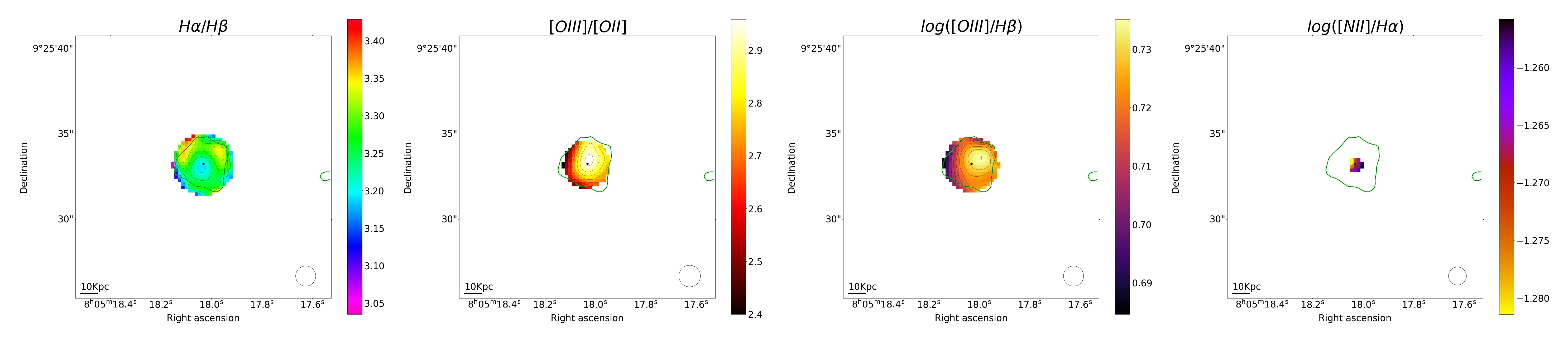}
     \caption{Line ratio maps for GP24.}
     \label{GP24_color}
\end{figure*}

\clearpage
\section{Continuum maps}
\label{appendix:continuum maps}

%dos columnas
\begin{figure*}[h!]
\centering
   \includegraphics[width=17cm]{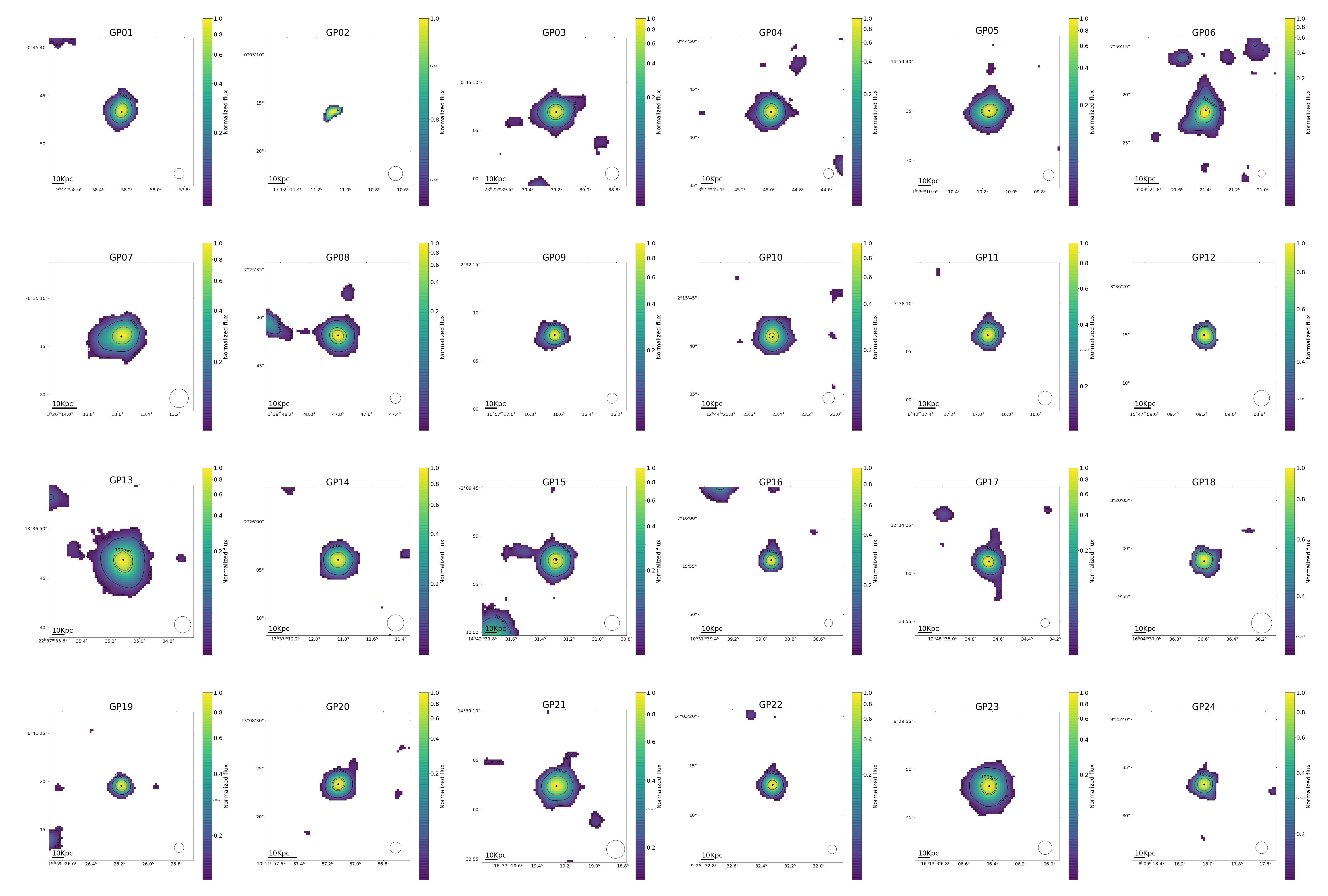}
     \caption{Continuum maps of all GPs.}
     \label{continuo}
\end{figure*}

\clearpage
\section{Radial profiles}
\label{apendix_radial_profiles}

%dos columnas
\begin{figure*}[h!]
\centering
   \includegraphics[width=15cm]{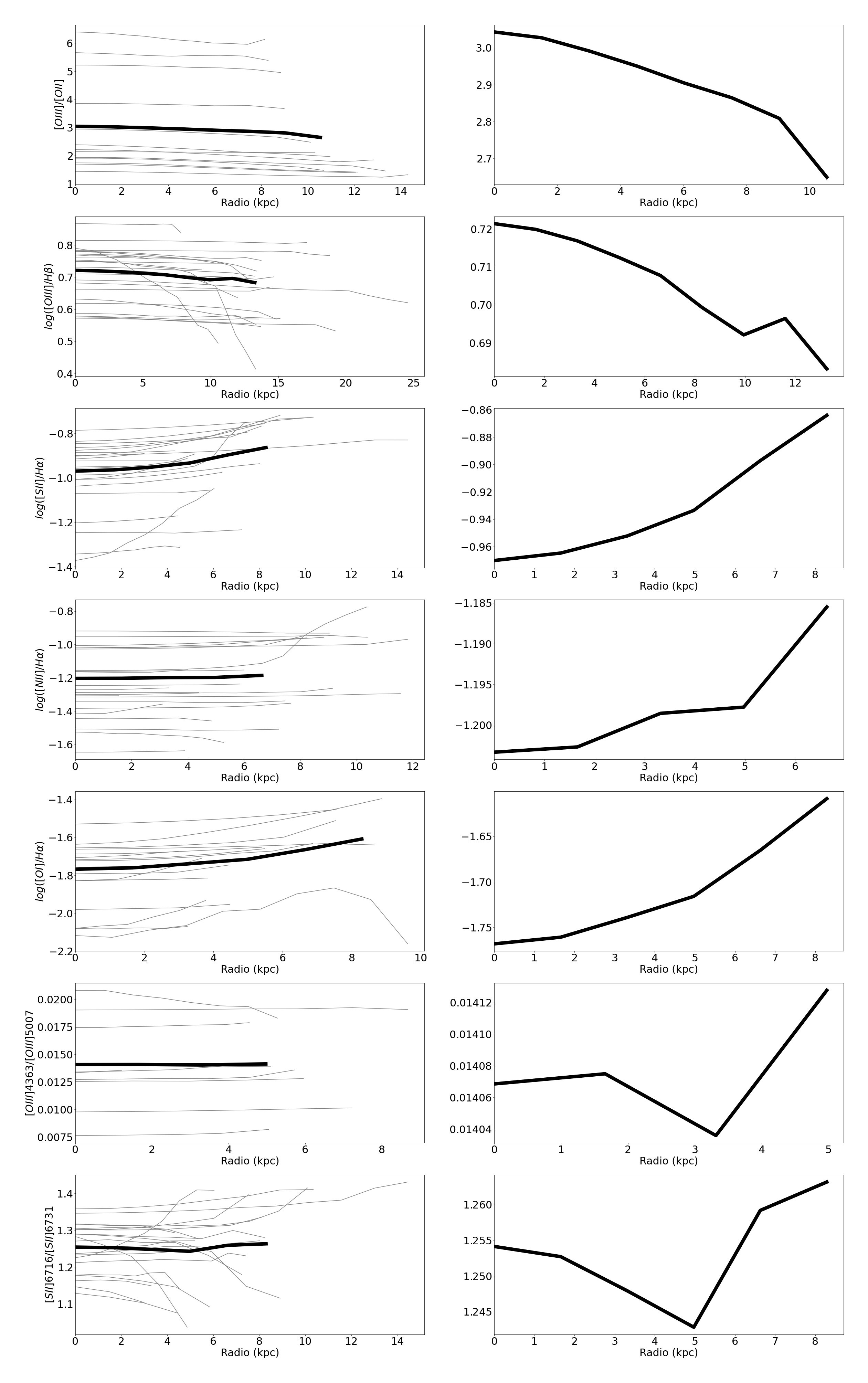}
     \caption{Radial profiles of remaining emission line ratios.}
     \label{radial_profiles}
\end{figure*}

\clearpage
\section{Spectra}
\label{appendix:Spectra}

%dos columnas
\begin{figure*}[h!]
\centering
   \includegraphics[width=15cm]{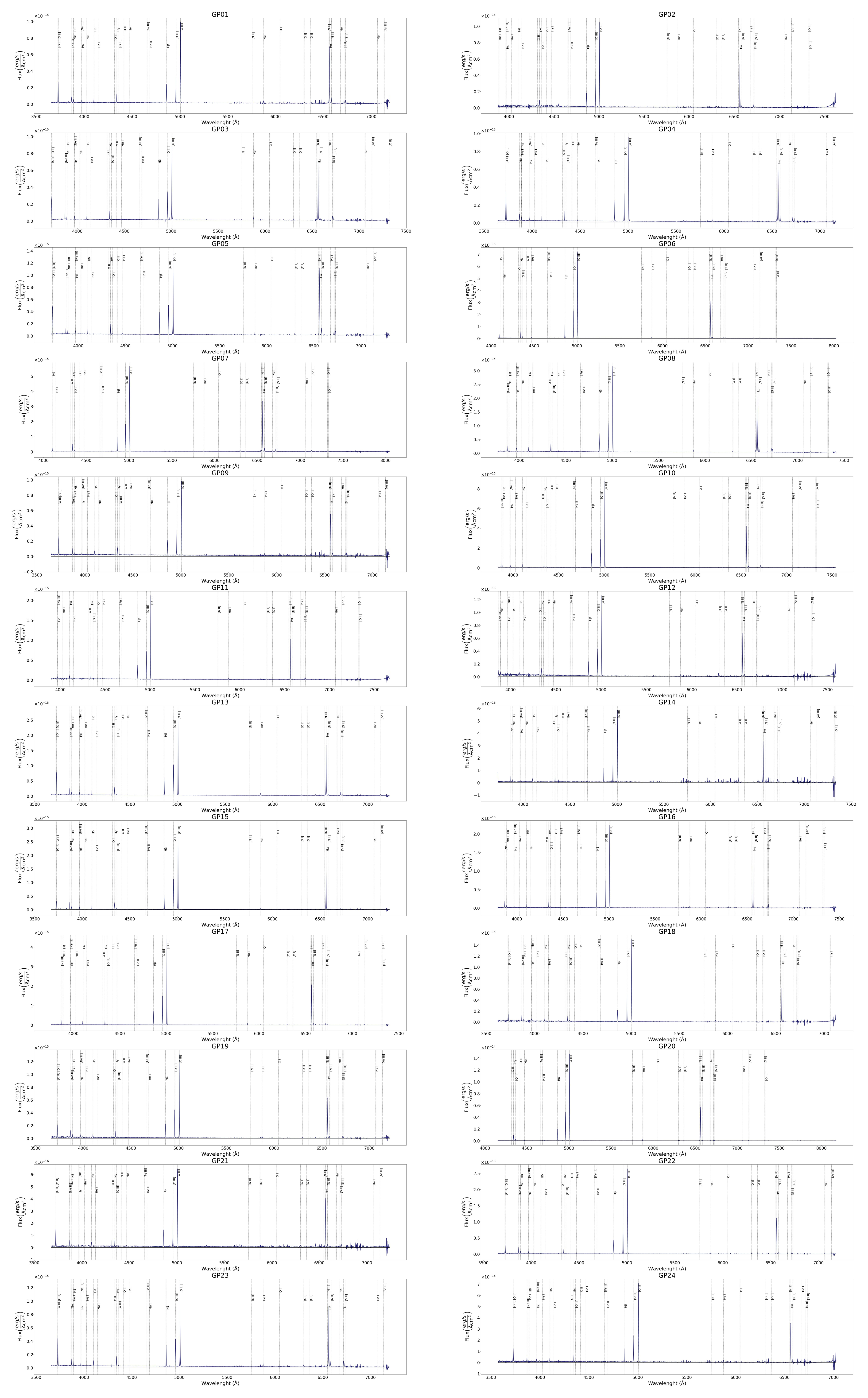}
     \caption{Integrated spectra.}
     \label{espectros}
\end{figure*}

% %dos columnas
% \begin{figure*}[h!]
% \centering
%    \includegraphics[width=0.0001cm]{Fotos/Espectros/espectros.jpg}
%      \caption{Integrated spectra}
%      \label{espectros}
% \end{figure*}

\clearpage
\section{Emission line data and properties of the ionized gas.}
\label{a1}

\begin{landscape}
\begin{table}[h!]                                  % used for centering table
\begin{threeparttable}
\label{table:Emission_line_fluxes}
\caption{Emission line fluxes, extinction and EW of $[OIII]$ line.} %Redshifts have been calculated using the position of the most prominent lines ie. $H\alpha$ and $[O\textsc{iii}]$. } 
\label{table:emission line fluxes}
\centering   
\begin{tabular}{lllllllllll}       % centered columns (4 columns)
\hline\hline   
  Name & \boldmath{$[OII]3727$} & \boldmath{$[OIII]4363$} & \boldmath{$[OIII]5007$} & \boldmath{$[OI]6300$} & \boldmath{$[NII]6584$} & \boldmath{$[SII]6716$} & \boldmath{$[SII]6731$} & \boldmath{$C_{H\beta}$} & \boldmath{$H\alpha \ flux$} & \boldmath{$[OIII] \ EW$}   \\      \textbf{(1)} & \textbf{(2)} & \textbf{(3)} & \textbf{(4)} & \textbf{(5)} & \textbf{(6)} & \textbf{(7)} & \textbf{(8)} & \textbf{(9)} & \textbf{(10)} & \textbf{(11)}   \\
  
\hline 
    \csvreader[tabular = c]{Tablas/GPs.csv}{}{\csvcolix}
  & \begin{filecontents*}{Tablas/Tabla_flujos2.csv}
Flux OII,Flux OIII4363,Flux OIII5007,Flux OI,Flux NII,Flux SII6716,Flux SII6731,Chbeta,Halpha luminosidad corregida,Halpha EW
223.4±1.4, ,386.0±0.8,4.9±1.2,27.3±1.2,19.9±1.2,14.4±1.3,0.3,265.3±1.6,181.27
190.3±4.0, ,555.8±1.5, ,14.9±0.6,20.0±0.6,14.2±0.6,0.206,92.7±0.4,382.38
235.3±1.3, ,416.2±3.3,6.3±0.5,18.6±0.8,20.6±0.7,15.1±0.8,0.239,167.1±0.7,240.11
219.6±1.1, ,373.4±0.7,7.2±0.6,29.3±0.5,21.2±0.9,18.9±1.1,0.277,280.4±0.8,204.96
225.5±1.3, ,368.2±0.8,6.3±0.4,26.3±0.5,22.8±0.6,18.8±0.6,0.215,285.5±0.9,239.1
89.2±1.0,11.21±0.12,583.9±0.6,2.8±0.18,6.32±0.23,7.98±0.17,6.6±0.16,0.221,263.6±0.5,682.41
207.1±1.3, ,514.1±0.7,5.3±0.5,21.09±0.35,16.7±0.5,14.4±0.5,0.235,231.6±0.5,529.68
221.2±1.3, ,407.8±0.6,7.3±0.2,26.78±0.32,23.4±0.3,14.2±0.3,0.252,576.4±1.2,262.73
203.5±1.6, ,473.2±1.0,5.1±1.6,17.0±2.1,14.8±2.0,13.4±2.1,0.171,193.8±1.6,253.81
179.3±1.2,6.5±0.5,590.5±0.7,4.8±0.3,11.82±0.15,14.78±0.18,11.92±0.16,0.181,664.3±0.9,747.48
187.3±1.2, ,498.0±1.2,4.5±0.9,11.6±0.5,18.4±0.6,13.5±0.6,0.142,125.5±0.38,352.65
206.3±2.7, ,551.6±1.3,4.2±0.9,12.21±0.47,15.3±0.8,11.0±0.9,0.285,170.8±0.5,534.84
231.4±1.0,8.8±1.2,473.0±0.7,6.4±0.5,13.1±0.9,20.4±1.1,14.5±1.2,0.237,570.3±2.7,340.42
229.9±1.8, ,503.3±1.1,6.1±1.0,17.2±1.0,17.8±1.0,14.8±1.0,0.321,132.4±0.7,351.05
95.2±1.0,11.6±0.5,622.3±0.8,2.2±0.5,6.8±1.0,8.5±1.1,6.8±1.3,0.148,371.9±1.7,945.78
185.4±2.1, ,549.3±1.0,10.7±1.0,10.3±1.0,16.9±1.1,14.3±1.1,0.142,155.4±0.8,562.67
179.5±1.6,8.1±0.3,602.1±0.8,4.4±0.3,11.17±0.28,15.54±0.29,11.3±0.3,0.162,325.8±0.7,703.58
124.6±2.1, ,675.7±1.2,4.5±0.9,12.0±0.8,12.9±2.6,10.6±2.5,0.204,320.4±1.5,782.27
149.5±1.8, ,573.6±1.0,3.98±0.4,13.28±0.3,12.1±0.7,10.2±0.7,0.201,194.3±0.5,637.66
80.8±1.1,12.6±0.2,736.8±0.8,2.01±0.16,5.75±0.12,7.1±0.16,5.97±0.16,0.151,209.0±0.3,1999.99
223.0±2.0, ,458.1±1.7,6.5±0.7,18.9±1.0,20.9±1.4,16.1±1.6,0.279,176.3±0.9,281.3
103.8±0.7, ,590.7±0.7,2.7±0.5,10.4±0.5,9.4±0.5,8.1±0.5,0.123,328.7±1.0,978.47
277.8±1.3, ,379.5±1.0,9.5±1.0,28.5±1.2,26.9±2.0,17.9±2.4,0.276,297.0±1.9,179.29
189.8±2.1, ,525.8±1.2, ,13.5±2.0,26.6±2.0,21.2±2.1,0.218,176.8±1.4,412.39
\end{filecontents*}
\csvreader[tabular = c]{Tablas/Tabla_flujos2.csv}{}{\csvcoli}
  & \csvreader[tabular = c]{Tablas/Tabla_flujos2.csv}{}{\csvcolii}
  & \csvreader[tabular = c]{Tablas/Tabla_flujos2.csv}{}{\csvcoliii}
  & \csvreader[tabular = c]{Tablas/Tabla_flujos2.csv}{}{\csvcoliv}
  & \csvreader[tabular = c]{Tablas/Tabla_flujos2.csv}{}{\csvcolv}
  & \csvreader[tabular = c]{Tablas/Tabla_flujos2.csv}{}{\csvcolvi}
  & \csvreader[tabular = c]{Tablas/Tabla_flujos2.csv}{}{\csvcolvii}
  & \csvreader[tabular = c]{Tablas/Tabla_flujos2.csv}{}{\csvcolviii}
  & \csvreader[tabular = c]{Tablas/Tabla_flujos2.csv}{}{\csvcolix}
  & \csvreader[tabular = c]{Tablas/Tabla_flujos2.csv}{}{\csvcolx}

  \\
\hline

\end{tabular}
\begin{tablenotes}
      \small
      \item Column (1): Name of the galaxy. Columns (2), (3), (4), (5), (6), (7) and (8): Extinction corrected flux of several lines normalized to $H\beta$ ($H\beta$ flux = 100). Column (9): Extinction coefficient. Column (10): $H\alpha$ luminosity ($\times 10^{40} \frac{erg}{s}$). Column (11): $H\alpha$ equivalent width ($\AA$).
    \end{tablenotes}
\end{threeparttable}
\end{table}
\end{landscape}

%%%%%%%%%%%%%%%%%%%%%%%%%%
\begin{landscape}
\begin{table}[h!]                                  
\addtocounter{table}{-1}  % decrease table counter by 1
\ContinuedFloat
\begin{threeparttable}
\caption{Continued.} 
\label{table:emission line fluxes_continued}
\centering   
\begin{tabular}{llllllllll}       
\hline\hline   
  Name & \boldmath{$[NeIII]3869$} & \boldmath{$[FeIII]4658$} & \boldmath{$[HeII]4686$} & \boldmath{$[ArIV]4711$} & \boldmath{$[ArIV]4740$} & \boldmath{$[HeI]4922$} & \boldmath{$[ArIII]7136$} & \boldmath{$[OII]7319$} & \boldmath{$[ArIII]7751$}    \\      \textbf{(1)} & \textbf{(2)} & \textbf{(3)} & \textbf{(4)} & \textbf{(5)} & \textbf{(6)} & \textbf{(7)} & \textbf{(8)} & \textbf{(9)} & \textbf{(10)}    \\
  
\hline 
    \csvreader[tabular = c]{Tablas/GPs.csv}{}{\csvcolix}
  & \begin{filecontents*}{Tablas/Tabla_flujos_lime2.csv}
H1_3704A,O2_3726A_b,H1_3750A,H1_3771A,H1_3798A,H1_3835A,Ne3_3869A,H1_3889A_m,H1_3970A,He1_4026A,S2_4069A,H1_4102A,H1_4341A,O3_4363A,He1_4471A,Fe3_4658A,He2_4686A,Ar4_4711A_m,Ar4_4740A,H1_4861A,He1_4922A,O3_4959A,O3_5007A_b,He1_5048A,He1_5876A,O1_6300A_b,H1_6563A_b,He1_6678A,S2_6716A_b,He1_7065A,Ar3_7136A,O2_7319A_m,Ar3_7751A
 , , , , ,3.5±0.6,32.4±0.6,16.3±0.5,21.9±1.0, , ,23.5±0.4,46.5±0.4,4.4±0.5,3.97±0.31, , , , ,100.0±0.7, ,128.3±1.0, , ,11.1±0.5, , , , , ,7.6±1.5, , 
 , , , , , , ,15.3±1.5,27.1±2.1, , ,27.5±1.2,45.5±1.1,15.4±1.3,4.4±0.8, , , , ,100.0±0.9, ,195.0±1.3, , ,11.74±0.48, , , , ,2.49±0.35,7.6±0.5, , 
 , , , , ,3.5±0.5,35.8±0.5,16.05±0.4,22.5±0.8, , ,23.7±0.4,45.06±0.4, ,3.15±0.23, , , , ,100.0±0.8, ,135.5±2.2, , ,11.46±0.19, , ,2.65±0.33, , ,7.44±0.24, , 
 , , , , ,2.6±0.5,31.1±0.5,15.6±0.4,18.9±0.6, , ,22.6±0.4,45.4±0.24, , , , , , ,100.0±0.5, ,127.8±0.9, , ,11.2±0.28, , , , , , , , 
 , , , , ,4.8±0.5,29.7±0.5,17.2±0.4,22.3±0.7, , ,23.9±0.4,45.8±0.4, ,3.75±0.2, , , , ,100.0±0.7, ,122.1±1.3, , ,11.4±0.21, , ,3.08±0.19, , ,6.02±0.2, , 
 , , , , , , , , , , ,25.5±0.1,45.22±0.14,11.94±0.09,3.53±0.05, ,1.85±0.06,1.77±0.05,,100.0±0.4, ,192.6±0.7, , ,10.17±0.09, , ,2.606±0.046, ,5.11±0.09,3.92±0.09,1.57±0.21, 
 , , , , , , , , , , ,24.5±0.5,47.1±0.4,5.5±0.3,4.07±0.18, , , , ,100.0±0.7, ,174.5±1.6, , ,11.09±0.15, , ,2.82±0.15, ,3.0±0.08,7.44±0.12,2.8±0.3,1.9±0.3
 , , , , ,3.9±0.3,34.1±0.4,15.9±0.3,22.5±0.5, , ,23.5±0.4,45.6±0.3, ,3.45±0.16, , , , ,100.0±0.6, ,138.6±1.3, , ,11.68±0.13, , , , , ,6.55±0.19, , 
 , , , , ,3.5±1.0,42.4±0.8,17.7±0.7,23.8±0.9, , ,24.9±0.6,48.1±0.5, , , , , , ,100.0±0.6, ,160.1±1.0, , ,12.3±0.5, , , , , , , , 
 , , , , ,3.6±0.5,44.2±0.4,16.8±0.3,30.2±0.6,1.27±0.18, ,25.05±0.18,46.8±0.4,7.75±0.22,3.9±0.2, , , , ,100.0±0.7, ,199.1±1.5, , ,11.64±0.16, , ,3.18±0.06, ,3.42±0.08,6.55±0.1,1.9±0.3, 
 , , , , , , , ,28.1±1.2, , ,23.8±0.6,46.1±0.5,8.0±0.5,4.2±0.5, , , , ,100.0±0.4, ,191.9±0.7, , ,10.87±0.17, , ,2.61±0.11, ,2.48±0.11,7.27±0.15,1.6±0.7, 
 , , , , , ,50.6±2.0,23.5±1.7,27.4±2.0,6.1±2.1, ,25.9±1.1,46.7±1.0,14.9±1.1,4.3±0.6, , , , ,100.0±0.8, ,185.0±1.4, , ,11.77±0.31, , ,3.07±0.22, ,2.91±0.24,7.4±0.6, , 
 , , , ,2.35±0.22,3.9±0.4,40.4±0.4,16.86±0.22,26.2±0.6, , ,24.55±0.21,45.3±0.9,5.1±0.8,2.91±0.13, , , , ,100.0±0.6, ,161.8±1.2, , ,14.15±0.35, , ,2.55±0.14, , ,5.84±0.11, , 
 , , , , ,3.6±0.6,40.7±0.6,15.6±0.6,22.5±0.8, , ,21.6±0.6,44.6±0.7,5.3±0.5, , , , , ,100.0±0.4, ,169.4±0.6, , ,11.68±0.26, , ,4.27±0.46, ,4.7±1.2,5.7±0.5, , 
 , ,2.1±0.2,2.82±0.15,4.09±0.14,5.56±0.16,44.2±0.4,16.42±0.19,28.5±0.4, , ,22.67±0.2,43.4±0.5,11.2±0.3,3.86±0.1, , , , ,100.0±0.7, ,211.3±1.8, , ,15.28±0.29, , ,2.94±0.22, ,4.77±0.22,6.2±0.17, , 
 , , , ,5.8±0.5,5.2±0.5,44.5±0.4,17.1±0.4,28.4±0.8,0.9±0.3, ,24.23±0.31,43.45±0.25,6.47±0.24,3.2±0.3, , , , ,100.0±0.4, ,189.1±0.7, , ,10.76±0.22, , ,3.3±0.19, ,2.87±0.33,7.1±0.5,2.7±0.4, 
 , , ,2.3±0.2,3.78±0.24,5.81±0.2,48.1±0.3,17.59±0.19,31.0±0.8,1.65±0.12,1.47±0.15,25.06±0.21,47.1±0.3,7.94±0.19,3.64±0.12, , , , ,100.0±0.5, ,200.2±1.2, , ,11.21±0.12, , ,2.78±0.2, ,3.45±0.33,6.88±0.16,2.11±0.25, 
 , , , ,4.2±1.0,5.2±1.0,52.0±1.0,17.7±1.1,31.6±1.4, , ,24.6±0.8,45.7±0.6,11.7±0.7,4.1±0.6, , , , ,100.0±0.6, ,218.1±1.4, , ,12.29±0.34, , ,2.7±0.6, ,4.4±3.4, , , 
1.9±0.6, ,1.3±0.7,2.7±0.7,4.0±0.6,6.9±0.7,51.6±0.8,20.0±0.8,31.5±1.0, , ,25.5±0.6,48.7±0.4,9.8±0.6,4.36±0.44, , , , ,100.0±0.5, ,193.6±1.1, , ,12.68±0.33, , ,3.2±0.19, ,3.83±0.19,6.11±0.26, , 
 , , , , , , , , , , , ,47.83±0.22,13.06±0.08,4.14±0.06,0.97±0.06,1.08±0.06,2.18±0.06,1.24±0.05,100.0±0.4,0.95±0.05,246.4±1.3, , ,11.21±0.09, , ,3.2±0.06, ,3.951±0.035,5.06±0.06,1.48±0.2,1.28±0.09
 , , , ,1.8±0.8,4.6±1.0,35.9±1.1,15.1±1.0,23.6±1.4, , ,23.1±0.8,44.6±1.4,8.5±1.5,2.9±0.5, , , , ,100.0±0.9, ,152.8±0.6, , ,11.6±0.5, , ,2.57±0.33, , ,7.1±0.4, , 
 , ,1.82±0.18,3.01±0.15,3.93±0.14,6.12±0.17,46.5±0.4,17.8±0.2,30.7±0.5,1.76±0.14, ,25.13±0.27,45.4±0.5,7.49±0.18,4.18±0.12, , , , ,100.0±1.0, ,203.3±2.6, , ,12.64±0.39, , ,3.35±0.1, ,5.03±0.21,7.0±1.0, , 
 , , , , ,2.9±0.5,33.9±0.5,15.2±0.4,19.7±0.7, , ,22.6±0.4,45.8±0.4, , , , , , ,100.0±1.0, ,122.9±0.9, , ,12.7±0.7, , , , , ,7.0±1.2, , 
 , , ,3.9±0.9,3.9±0.8,5.9±0.8,43.2±0.8,16.5±0.8,31.0±1.4, , ,26.4±0.8,47.1±0.7,9.5±0.7,2.8±0.5, , , , ,100.0±0.7, ,180.0±1.2, , ,12.96±0.46, , , , , , , , 
\end{filecontents*}
\csvreader[tabular = c]{Tablas/Tabla_flujos_lime2.csv}{}{\csvcolvii}
  & \csvreader[tabular = c]{Tablas/Tabla_flujos_lime2.csv}{}{\csvcolxvi}
  & \begin{filecontents*}{Tablas/Tabla_flujos_lime_nueva.csv}
H1_3704A,O2_3726A_b,H1_3750A,H1_3771A,H1_3798A,H1_3835A,Ne3_3869A,H1_3889A_m,H1_3970A,He1_4026A,S2_4069A,H1_4102A,H1_4341A,O3_4363A,He1_4471A,Fe3_4658A,He2_4686A,Ar4_4711A_m,Ar4_4740A,H1_4861A,He1_4922A,O3_4959A,O3_5007A_b,He1_5048A,He1_5876A,O1_6300A_b,H1_6563A_b,He1_6678A,S2_6716A_b,He1_7065A,Ar3_7136A,O2_7319A_m,Ar3_7751A
 , , , , ,3.5±0.6,32.4±0.6,16.3±0.5,21.9±1.0, , ,23.5±0.39,46.47±0.38,4.41±0.47,3.97±0.31, , , , ,100.0±0.7, ,128.3±1.0, , ,11.1±0.5, , , , , ,7.6±1.5, , 
 , , , , , , ,15.3±1.5,27.1±2.1, , ,27.5±1.2,45.5±1.1,15.4±1.3,4.4±0.8, , , , ,100.0±0.9, ,195.0±1.3, , ,11.74±0.48, , , , ,2.49±0.35,7.59±0.48, , 
 , , , , ,3.5±0.5,35.78±0.48,16.05±0.4,22.5±0.8, , ,23.69±0.32,45.06±0.4, ,3.15±0.23, , , , ,100.0±0.8, ,135.5±2.2, , ,11.46±0.19, , ,2.65±0.33, , ,7.44±0.24, , 
 , , , , ,2.59±0.49,31.14±0.47,15.57±0.39,18.9±0.6, , ,22.63±0.42,45.4±0.24, , , , , , ,100.0±0.5, ,127.8±0.9, , ,11.2±0.28, , , , , , , , 
 , , , , ,4.84±0.44,29.72±0.47,17.16±0.35,22.3±0.7, , ,23.92±0.38,45.83±0.39, ,3.75±0.2, , , , ,100.0±0.7, ,122.1±1.3, , ,11.4±0.21, , ,3.08±0.19, , ,6.02±0.2, , 
 , , , , , , , , , , ,25.5±0.1,45.22±0.14,11.94±0.09,3.53±0.05, ,1.85±0.06,1.768±0.049, ,100.0±0.39, ,192.6±0.7, , ,10.17±0.09, , ,2.606±0.046, ,5.11±0.09,3.92±0.09,1.57±0.21, 
 , , , , , , , , , , ,24.48±0.46,47.14±0.38,5.49±0.26,4.07±0.18, , , , ,100.0±0.7, ,174.5±1.6, , ,11.09±0.15, , ,2.82±0.15, ,3.0±0.08,7.44±0.12,2.79±0.34,1.9±0.28
 , , , , ,3.9±0.34,34.06±0.37,15.92±0.31,22.5±0.44, , ,23.46±0.38,45.55±0.26, ,3.45±0.16, , , , ,100.0±0.6, ,138.6±1.3, , ,11.68±0.13, , , , , ,6.55±0.19, , 
 , , , , ,3.5±1.0,42.4±0.8,17.7±0.7,23.8±0.9, , ,24.9±0.6,48.13±0.48, , , , , , ,100.0±0.6, ,160.1±1.0, , ,12.3±0.5, , , , , , , , 
 , , , , ,3.56±0.45,44.15±0.37,16.8±0.31,30.2±0.6,1.27±0.18, ,25.05±0.18,46.78±0.41,7.75±0.22,3.9±0.14, , , , ,100.0±0.7, ,199.1±1.5, , ,11.64±0.16, , ,3.18±0.06, ,3.42±0.08,6.55±0.1,1.88±0.29, 
 , , , , , , , ,28.1±1.2, , ,23.8±0.6,46.07±0.49,8.0±0.5,4.21±0.46, , , , ,100.0±0.41, ,191.9±0.7, , ,10.87±0.17, , ,2.61±0.11, ,2.48±0.11,7.27±0.15,1.6±0.7, 
 , , , , , ,50.6±2.0,23.5±1.7,27.4±2.0,6.1±2.1, ,25.9±1.1,46.7±1.0,14.9±1.1,4.3±0.6, , , , ,100.0±0.8, ,185.0±1.4, , ,11.77±0.31, , ,3.07±0.22, ,2.91±0.24,7.4±0.6, , 
 , , , ,2.35±0.22,3.92±0.35,40.44±0.34,16.86±0.22,26.2±0.6, , ,24.55±0.21,45.3±0.9,5.1±0.8,2.91±0.13, , , , ,100.0±0.6, ,161.8±1.2, , ,14.15±0.35, , ,2.55±0.14, , ,5.84±0.11, , 
 , , , , ,3.6±0.6,40.7±0.6,15.6±0.6,22.5±0.8, , ,21.6±0.6,44.6±0.7,5.31±0.45, , , , , ,100.0±0.4, ,169.4±0.6, , ,11.68±0.26, , ,4.27±0.46, ,4.7±1.2,5.65±0.47, , 
 , ,2.1±0.2,2.82±0.15,4.09±0.14,5.56±0.16,44.18±0.35,16.42±0.19,28.51±0.42, , ,22.67±0.2,43.38±0.45,11.22±0.31,3.86±0.1, ,1.34±0.07, , ,100.0±0.7, ,211.3±1.8, , ,15.28±0.29, , ,2.94±0.22, ,4.77±0.22,6.2±0.17, , 
 , , , ,5.78±0.47,5.19±0.43,44.51±0.38,17.13±0.36,28.4±0.8,0.93±0.27, ,24.23±0.31,43.45±0.25,6.47±0.24,3.19±0.29, , , , ,100.0±0.38, ,189.1±0.7, , ,10.76±0.22, , ,3.3±0.19, ,2.87±0.33,7.13±0.49,2.67±0.39, 
 , , ,2.3±0.24,3.78±0.24,5.81±0.2,48.12±0.26,17.59±0.19,31.0±0.8,1.65±0.12,1.47±0.15,25.06±0.21,47.06±0.27,7.94±0.19,3.64±0.12, , , , ,100.0±0.5, ,200.2±1.2, , ,11.21±0.12, , ,2.78±0.2, ,3.45±0.33,6.88±0.16,2.11±0.25, 
 , , , ,4.2±1.0,5.2±1.0,52.0±1.0,17.7±1.1,31.6±1.4, , ,24.6±0.8,45.7±0.6,11.7±0.7,4.1±0.6, , , , ,100.0±0.6, ,218.1±1.4, , ,12.29±0.34, , ,2.7±0.6, ,4.4±3.4, , , 
1.9±0.6, ,1.3±0.7,2.7±0.7,4.0±0.6,6.9±0.7,51.6±0.8,20.0±0.8,31.5±1.0, , ,25.5±0.6,48.74±0.41,9.8±0.6,4.36±0.44, , , , ,100.0±0.5, ,193.6±1.1, , ,12.68±0.33, , ,3.2±0.19, ,3.83±0.19,6.11±0.26, , 
 , , , , , , , , , , , ,47.83±0.22,13.06±0.08,4.14±0.06,0.97±0.06,1.08±0.06,2.18±0.06,1.239±0.046,100.0±0.41,0.945±0.042,246.4±1.3, , ,11.21±0.09, , ,3.2±0.06, ,3.951±0.035,5.06±0.06,1.48±0.2,1.28±0.09
 , , , ,1.8±0.8,4.6±1.0,35.9±1.1,15.1±1.0,23.6±1.4, , ,23.1±0.8,44.6±1.4,8.5±1.5,2.9±0.5, , , , ,100.0±0.9, ,152.8±0.6, , ,11.6±0.5, , ,2.57±0.33, , ,7.06±0.36, , 
 , ,1.82±0.18,3.01±0.15,3.93±0.14,6.12±0.17,46.53±0.39,17.83±0.18,30.72±0.41,1.76±0.14, ,25.13±0.27,45.37±0.47,7.49±0.18,4.18±0.12, , , , ,100.0±1.0, ,203.3±2.6, , ,12.64±0.39, , ,3.35±0.1, ,5.03±0.21,7.0±1.0, , 
 , , , , ,2.9±0.5,33.92±0.49,15.23±0.38,19.7±0.7, , ,22.6±0.39,45.87±0.36, , , , , , ,100.0±1.0, ,122.9±0.9, , ,12.7±0.7, , , , , ,7.0±1.2, , 
 , , ,3.9±0.9,3.9±0.8,5.9±0.8,43.2±0.8,16.5±0.8,31.0±1.4, , ,26.4±0.8,47.1±0.7,9.5±0.7,2.8±0.5, , , , ,100.0±0.7, ,180.0±1.2, , ,12.96±0.46, , , , , , , , 
\end{filecontents*}
\csvreader[tabular = c]{Tablas/Tabla_flujos_lime_nueva.csv}{}{\csvcolxvii}
  & \csvreader[tabular = c]{Tablas/Tabla_flujos_lime2.csv}{}{\csvcolxviii}
  & \csvreader[tabular = c]{Tablas/Tabla_flujos_lime2.csv}{}{\csvcolxix}
  & \csvreader[tabular = c]{Tablas/Tabla_flujos_lime2.csv}{}{\csvcolxxi}
  & \csvreader[tabular = c]{Tablas/Tabla_flujos_lime2.csv}{}{\csvcolxxxi}
  & \csvreader[tabular = c]{Tablas/Tabla_flujos_lime2.csv}{}{\csvcolxxxii}
  & \csvreader[tabular = c]{Tablas/Tabla_flujos_lime2.csv}{}{\csvcolxxxiii}

  \\
\hline
                            
\end{tabular}
\begin{tablenotes}
      \small
      \item All columns represent extinction corrected flux of several lines normalized to $H\beta$ ($H\beta$ flux = 100).
    \end{tablenotes}
\end{threeparttable}
\end{table}
\end{landscape}

\begin{landscape}
\begin{table}[h!]                                  % used for centering table
\begin{threeparttable}
\caption{Properties of the ionized gas.} %Redshifts have been calculated using the position of the most prominent lines ie. $H\alpha$ and $[O\textsc{iii}]$. } 
\label{table:gas_properties}
\centering   
\begin{tabular}{llllllllll}       % centered columns (4 columns)
\hline\hline   
  \textbf{Name} & \textbf{O32} & \textbf{log([OIII]/H\boldmath{$\beta$})} & \textbf{log([NII]/H\boldmath{$\alpha$})} & \textbf{log([SII]/H\boldmath{$\alpha$})} & \textbf{log([OI]/H\boldmath{$\alpha$})} & \boldmath{$T_e$} & \boldmath{$n_e$} & \boldmath{$12 + \log(O/H)$} & \boldmath{$\log(N/O)$}    \\      \textbf{(1)} & \textbf{(2)} & \textbf{(3)} & \textbf{(4)} & \textbf{(5)} & \textbf{(6)} & \textbf{(7)} & \textbf{(8)} & \textbf{(9)} & \textbf{(10)}    \\
  
\hline 
    \csvreader[tabular = c]{Tablas/GPs.csv}{}{\csvcolix}
  & \begin{filecontents*}{Tablas/PROPIEDADES_DEL_GAS_IONIZADO2.csv}
R23 OIII5007+OIII4959/OII,O32 OIII5007/OII,log(OIII_Hbeta),log(NII_Halpha),log(SII_Halpha),log(OI_Halpha),Temperatura,Densidad,Metalicidad,log(N/O)
7.28±0.02,1.81±0.02,0.587±0.004,-1.02±0.02,-1.02±0.02,-1.76±0.09, ,71.0±41.0,8.03±0.03,-0.95±0.06
9.31±0.05,2.93±0.06,0.745±0.005,-1.28±0.02,-1.28±0.02,-1.77±0.07, ,210.0±70.0,8.05±0.04,-1.24±0.03
7.87±0.05,1.81±0.02,0.62±0.01,-1.18±0.02,-1.18±0.02,-1.66±0.03, ,86.0±27.0,8.21±0.05,-1.13±0.02
7.21±0.02,1.67±0.01,0.572±0.003,-0.992±0.007,-0.992±0.007,-1.6±0.03, ,351.0±42.0,8.05±0.02,-1.02±0.03
7.12±0.02,1.63±0.01,0.563±0.002,-1.034±0.009,-1.034±0.009,-1.65±0.03, ,260.0±70.0,8.06±0.02,-1.10±0.04
8.69±0.03,6.56±0.08,0.766±0.002,-1.65±0.02,-1.65±0.02,-2.01±0.03,15059.0±21.0,302.0±20.0,7.902±0.008,-1.23±0.04
8.92±0.04,2.50±0.02,0.711±0.003,-1.131±0.007,-1.131±0.007,-1.74±0.04, ,373.0±24.0,8.16±0.02,-1.00±0.03
7.65±0.02,1.82±0.02,0.609±0.002,-1.042±0.005,-1.042±0.005,-1.60±0.02, ,200.0±70.0,8.13±0.06,-1.01±0.01
8.33±0.03,2.32±0.02,0.674±0.004,-1.23±0.05,-1.23±0.05,-1.74±0.12, ,520.0±170.0,8.17±0.02,-1.08±0.07
9.62±0.07,3.33±0.03,0.770±0.002,-1.382±0.006,-1.382±0.006,-1.78±0.03,11910.0±90.0,246.0±9.0,8.24±0.01,-1.27±0.03
8.56±0.02,2.62±0.02,0.698±0.004,-1.38±0.02,-1.38±0.02,-1.80±0.08, ,87.0±22.0,8.05±0.03,-1.30±0.05
9.35±0.04,2.79±0.04,0.742±0.004,-1.36±0.02,-1.36±0.02,-1.83±0.08, ,64.0±38.0,8.13±0.04,-1.20±0.05
8.65±0.02,2.02±0.01,0.675±0.002,-1.33±0.03,-1.336±0.03,-1.648±0.03,14880.0±260.0,48.0±28.0,7.95±0.02,-1.29±0.05
9.02±0.03,2.18±0.02,0.702±0.004,-1.22±0.03,-1.22±0.03,-1.67±0.06, ,340.0±60.0,8.26±0.03,-1.14±0.05
9.27±0.02,6.38±0.07,0.794±0.003,-1.62±0.06,-1.62±0.06,-2.11±0.09,14880.0±70.0,230.0±120.0,7.94±0.01,-1.21±0.07
9.18±0.03,2.96±0.03,0.739±0.003,-1.44±0.05,-1.44±0.05,-1.42±0.04, ,350.0±60.0,8.06±0.02,-1.37±0.05
9.83±0.02,3.30±0.03,0.778±0.002,-1.39±0.02,-1.39±0.01,-1.82±0.03,12970.0±39.0,280.0±70.0,8.148±0.009,-1.26±0.02
10.18±0.03,5.53±0.10,0.827±0.003,-1.37±0.03,-1.37±0.03,-1.80±0.08, ,320.0±180.0,8.04±0.05,-1.2±0.1
9.13±0.03,3.87±0.05,0.758±0.002,-1.33±0.01,-1.33±0.01,-1.86±0.03, ,350.0±50.0,8.13±0.03,-1.07±0.06
10.67±0.02,8.73±0.12,0.867±0.002,-1.695±0.009,-1.695±0.009,-2.15±0.04,14233.0±26.0,345.0±22.0,8.048±0.004,-1.25±0.02
8.33±0.03,2.06±0.02,0.661±0.006,-1.18±0.03,-1.18±0.03,-1.65±0.05, ,150.0±50.0,8.27±0.03,-1.14±0.05
8.73±0.01,5.67±0.05,0.762±0.002,-1.43±0.02,-1.43±0.02,-2.03±0.07, ,390.0±50.0,7.97±0.04,-1.09±0.05
7.95±0.02,1.31±0.01,0.579±0.007,-0.99±0.02,-0.99±0.02,-1.47±0.05, ,180.0±70.0,8.14±0.02,-1.06±0.06
8.88±0.03,2.77±0.04,0.719±0.004,-1.32±0.06,-1.32±0.06,-1.79±0.06, ,300.0±70.0,8.12±0.03,-1.46±0.08
\end{filecontents*}
\csvreader[tabular = c]{Tablas/PROPIEDADES_DEL_GAS_IONIZADO2.csv}{}{\csvcolii}
  & \csvreader[tabular = c]{Tablas/PROPIEDADES_DEL_GAS_IONIZADO2.csv}{}{\csvcoliii}
  & \csvreader[tabular = c]{Tablas/PROPIEDADES_DEL_GAS_IONIZADO2.csv}{}{\csvcoliv}
  & \csvreader[tabular = c]{Tablas/PROPIEDADES_DEL_GAS_IONIZADO2.csv}{}{\csvcolv}
  & \csvreader[tabular = c]{Tablas/PROPIEDADES_DEL_GAS_IONIZADO2.csv}{}{\csvcolvi}
  & \csvreader[tabular = c]{Tablas/PROPIEDADES_DEL_GAS_IONIZADO2.csv}{}{\csvcolvii}
  & \csvreader[tabular = c]{Tablas/PROPIEDADES_DEL_GAS_IONIZADO2.csv}{}{\csvcolviii}
  & \csvreader[tabular = c]{Tablas/PROPIEDADES_DEL_GAS_IONIZADO2.csv}{}{\csvcolix}
  & \csvreader[tabular = c]{Tablas/PROPIEDADES_DEL_GAS_IONIZADO2.csv}{}{\csvcolx}

  \\
\hline
                            
\end{tabular}
\begin{tablenotes}
      \small
      \item Column (1): Name of the galaxy. Column (2): Ionization parameter ($[OIII]/[OII]$). Columns (3), (4) and (5): BPT ratios. Column (6): Electron temperature (K). Column(7): Electron density $(cm^{-3})$. Column (8): Metallicity and Column (9): Nitrogen over oxygen abundance.
    \end{tablenotes}
\end{threeparttable}
\end{table}
\end{landscape}

\end{appendix}

\end{document}